\documentclass[]{aastex631}
\usepackage{amsmath}
\usepackage{soul}

\newcommand{\threeml}{\texttt{3ML}}

\newcommand{\core}{\textbf{core}}
\newcommand{\rone}{\textbf{r$_1$}}
\newcommand{\rtwo}{\textbf{r$_2$}}

\begin{document}

\title{The IXPE view of GRB~221009A}

\author[0000-0002-6548-5622]{Michela Negro}
\correspondingauthor{Michela Negro}
\email{mnegro1@umbc.edu}
\affiliation{University of Maryland, Baltimore County, Baltimore, MD 21250, USA}
\affiliation{NASA Goddard Space Flight Center, Greenbelt, MD 20771, USA}
\affiliation{Center for Research and Exploration in Space Science and Technology, NASA/GSFC, Greenbelt, MD 20771, USA}

\author[0000-0002-7574-1298]{Niccol\`o Di Lalla}
\affiliation{Department of Physics and Kavli Institute for Particle Astrophysics and Cosmology, Stanford University, Stanford, California 94305, USA}

\author[0000-0002-5448-7577]{Nicola Omodei}
\affiliation{Department of Physics and Kavli Institute for Particle Astrophysics and Cosmology, Stanford University, Stanford, California 94305, USA}

\author[0000-0002-2149-9846]{P\'eter Veres}
\affil{Department of Space Science, University of Alabama in Huntsville, 320 Sparkman Drive, Huntsville, AL 35899}
\affil{Center for Space Plasma and Aeronomic Research, University of Alabama in Huntsville, 320 Sparkman Drive, Huntsville, AL 35899, USA}

\author[0000-0002-8665-0105]{Stefano Silvestri}
\affiliation{Istituto Nazionale di Fisica Nucleare, Sezione di Pisa, Largo B. Pontecorvo 3, 56127 Pisa, Italy}
\affiliation{Università di Pisa, Dipartimento di Fisica Enrico Fermi, Largo B. Pontecorvo 3, 56127 Pisa, Italy}

\author[0000-0002-0998-4953]{Alberto Manfreda}
\affiliation{Istituto Nazionale di Fisica Nucleare, Sezione di Napoli, Strada Comunale Cinthia, 80126 Napoli, Italy}
\affiliation{Istituto Nazionale di Fisica Nucleare, Sezione di Pisa, Largo B. Pontecorvo 3, 56127 Pisa, Italy}

\author[0000-0002-2942-3379]{Eric Burns}
\affiliation{Department of Physics \& Astronomy, Louisiana State University, Baton Rouge, LA 70803, USA}

\author[0000-0002-9785-7726]{Luca Baldini}
\affiliation{Istituto Nazionale di Fisica Nucleare, Sezione di Pisa, Largo B. Pontecorvo 3, 56127 Pisa, Italy}
\affiliation{Dipartimento di Fisica, Università di Pisa, Largo B. Pontecorvo 3, 56127 Pisa, Italy}

\author[0000-0003-4925-8523]{Enrico Costa}
\affiliation{INAF Istituto di Astrofisica e Planetologia Spaziali, Via del Fosso del Cavaliere 100, 00133 Roma, Italy}

\author[0000-0003-4420-2838]{Steven R. Ehlert}
\affiliation{NASA Marshall Space Flight Center, Huntsville, AL 35812, USA}

% ...

\begin{abstract}
We present the IXPE observation of GRB~221009A which includes upper limits on the linear polarization degree of both prompt and afterglow emission in the soft X-ray energy band.
GRB~221009A is an exceptionally bright gamma-ray burst (GRB) that reached Earth on 2022 October 9 after travelling through the dust of the Milky Way. The Imaging X-ray Polarimetry Explorer (IXPE) pointed at GRB~221009A on October 11 to observe, for the first time, the 2--8 keV X-ray polarization of a GRB afterglow. 
We set an upper limit to the polarization degree of the afterglow emission of 13.8\% at a 99\% confidence level. This result provides constraints on the jet opening angle and the viewing angle of the GRB, or alternatively, other properties of the emission region. Additionally, IXPE captured halo-rings of dust-scattered photons which are echoes of the GRB prompt emission. The 99\% confidence level upper limit to the prompt polarization degree depends on the background model assumption and it ranges between $\sim 55\%$ to $\sim 82\%$.
This single IXPE pointing provides both the first assessment of X-ray polarization of a GRB afterglow and the first GRB study with polarization observations of both the prompt and afterglow phases.

\end{abstract}


\section{Introduction} \label{sec:intro}

Gamma-Ray Bursts (GRBs) are among the most energetic events in the Universe. These events are characterized by a ``prompt'' gamma-ray emission, the most luminous phase of the burst, followed by a temporally decaying ``afterglow'' that can last for days or even years and is observed across the whole electromagnetic spectrum whenever the needed sensitivity is available. GRBs are conventionally classified by duration of the prompt phase into the short ($<2$ sec) and long ($>2$ sec) class, with distinct physical origins \citep{1998Natur.395..670G,GW170817-GRB170817A, 2015ApJ...815..102F,Burns2021}. Long GRBs originates from collapsars \citep{woosley2006supernova}, a rare sub-type of type Ic supernovae. In the standard model, a fast spinning core collapses into a rapidly spinning black hole which devours some of the massive star progenitor. This results in a hyper-accreting process that powers bipolar, ultrarelativistic, collimated jets which ultimately release the prompt GRB signature \citep{kumar2015physics}. Despite detecting more than 10,000 GRBs in their prompt phase, our understanding of these events and the underlying physical processes is still limited \citep{zhang2018physics}. We also notice that in the X-Ray band covered by IXPE (2$-$8 keV), measurements of the prompt emission are meager. In fact, our knowledge relies on less than 100 detections by BeppoSAX and HETE-2 \citep[see, e.g.,][]{2019RLSFN..30S.171F, 2011NCimR..34..585C,2003ICRC....5.2741T}.

Advances in understanding require new diagnostics, e.g. observations of polarization or multiple messengers. However, so far, only upper limits on neutrino emission from either prompt or early-afterglow emission have been set, suggesting a leptonic composition of the jet bulk \citep{ICgrbsearch}. Only a few GRBs (before GRB 221009A) have been detected in the very-high-energy regime and only one (short GRB) coincident with gravitational waves \citep{GW170817-GRB170817A}. Polarization measurements of the prompt emission of GRBs can represent a unique observable to constrain the outflow composition and dynamics, to determine the structure of the magnetic fields at the jet formation, and provide insights on the radiation mechanisms behind the observed GRB spectra as well as on our viewing angle within the jet opening angle \citep[see, e.g.,][for a recent overview]{Gill2021}. Thus far, GRB polarization observations in the prompt phase have only occurred in the hard X-ray / soft gamma-ray band, reporting generally high polarization degrees, but never an unambiguous detection \citep[see, e.g.,][for a critical review]{McConnell}. The largest catalog of prompt GRB gamma-ray polarization measurements comes from the POLAR mission, with 14 observations but no clear detection. The picture is further complicated because the time-integrated polarization seems to be affected by polarization angle swing in time \citep{kole2020polar}. The forthcoming POLAR-2 \citep{hulsman2020polar} and COSI \citep{tomsick2021compton} missions are designed for significantly larger detection catalogs, as is the proposed LEAP mission \citep{mcconnell2021large}.  

After the prompt emission the jet propagates and interacts with the ambient medium, developing a shock, where electrons are accelerated and produce synchrotron emission, referred to as afterglow, throughout the whole electromagnetic spectrum, from radio to very-high energy gamma-rays. Observations of polarization in the afterglow phase can also provide insight into jet physics and structure \citep{rossi2004polarization}.
Models in the literature \citep{kuwata2022synchrotron} predict a progressive loss of coherence of the propagating jet magnetic fields, which results in an expected low polarization degree (below 5--3\%) of the late-phases of the GRB afterglow emission. These predictions are largely consistent with results from time-resolved GRB afterglow measurements of optical polarization \citep{mundell2013highly, 2016A&AT...29..205C, 2020ApJ...892..131S}. Radio polarization measurements typically show a lower polarization degree than do the optical measurements \citep{urata2019first, Urata2022radiopol}, motivating an increasing interest in a multiwavelength modeling of late-afterglow polarization \citep{2021ApJ...913...58S, 2021MNRAS.506.4275B}. No observations of afterglow polarization have been reported so far at X-ray energies.

On 2022 October 9 an exceptionally bright transient event outshone the rest of the high-energy sky. The first trigger was recorded in the gamma-ray band by the \textit{Fermi} Gamma-ray Burst Monitor (GBM) at 13:16:59.988\,UTC \citep{GCNfermiGBMtrigger, GCNLesage},  and the same event was also strongly detected by the \textit{Fermi} Large Area Telescope (LAT) up to a hundred GeV \citep{ATelLATrefined}. The Large High Altitude Air Shower Observatory (LHAASO) also reported the detection of gamma-rays up to 18 TeV \citep{2022GCN.32677....1H}. After about an hour from the initial trigger, as soon as the source was observed, the Burst Alert Telescope on board of the Neil Gehrels \textit{Swift} Observatory triggered on the same event and the \textit{Swift} X-Ray Telescope (XRT)\citep{Swift-XRT} was on target 143 seconds later and the \textit{Swift} Ultraviolet/Optical Telescope (UVOT) located it at (RA(J2000), DEC(J2000)) = (288.26452$^{\circ}$, 19.77350$^{\circ}$) with a 90\%-confidence error radius of about 0.61 arcsec \citep{GCNUVOTloc}. The event, soon classified as a gamma-ray burst \citep{GCNfermiGBMtrigger}, happened at a redshift of 0.151 as reported by X-shooter/VLT \citep{GCNredshift}, and is the brightest (at Earth) ever recorded by any gamma-ray burst monitor by a large margin. 
Furthermore, the detection of dust-scattered soft X-ray rings was reported \citep{ATELTiengoSwiftRings} through \textit{Swift}/XRT observations in the two days after the prompt emission. Such rings are produced by X-rays from the extremely bright prompt emission efficiently scattered at small angles by interstellar dust grains in our Galaxy \citep[e.g.,][]{MiraldaEscude1999}. The scattered X-rays are delayed with respect to direct ones, due to their longer path length from the source to the observer, with a delay that depends on the distance of the dust cloud traversed by the X-ray radiation. Thus, the rings are echoes of the prompt emission.

The Imaging X-ray Polarimetry Explorer (IXPE) is a space observatory with three identical telescopes designed to measure the polarization of astrophysical X-rays \citep{IXPE_calibration, IXPEOverviewII,IXPEOverviewIII}. Launched on 2021 December 9, IXPE is an international collaboration between NASA and the Italian Space Agency (ASI), and it has been operating since January of 2022. IXPE measures polarization using the photo-electric effect of X-rays absorbed in the gas gap of a Gas Pixel Detector (GPD) \citep{BellazziniGPD}. 
On 2022 October 11 at 23:35:35.184 UTC IXPE started the observation of GRB~221009A in response to a Target of Opportunity request \citep{GCNIXPEobsplan}. The location position was provided by a \textit{Swift}/UVOT observation \citep{GCNUVOTloc}.
The observation ended on 2022 October 14 at 00:46:44.184 UTC with an effective exposure of 94,122 s.

In this work we present the results of the IXPE observation carried out with the fully processed data and through a careful data analysis. This represents the first observation of X-ray polarization of a GRB afterglow, the first measurement of soft X-ray polarization of GRB prompt emission, and the first time we observe polarization properties in both prompt and afterglow phases of the same GRB.

After a brief general introduction on IXPE data analysis provided in Section~\ref{sec:analysisintro}, we devote Section~\ref{sec:core} to the data analysis and interpretation of the GRB afterglow emission, while Section~\ref{sec:rings} illustrates the data analysis and interpretation of the rings in association to the GRB prompt emission. Summary and conclusions are offered in Section~\ref{sec:concl}.

\section{IXPE Polarization Analysis}
\label{sec:analysisintro}
We analyze IXPE Level 2 processed data\footnote{IXPE data are publicly available on the HEASARC archive.}, combining the data collected by the three identical detector units (DUs). 
The time-integrated radial profile reveals inconsistency with the expectation from a point-like source, showing a profile that deviates from the instrument point spread function (PSF). In particular two excesses around the peak emission are visible (see Figure~\ref{fig:bkg_radial} and Figure~\ref{fig:elapse} in the Appendix). Such excess appears as rings around a bright core emission and are associated with dust-scattering halos \citep[e.g.,][]{10.1143/PTP.43.1224, MiraldaEscude1999}. To utilize the full potential of this observation an image- and time-resolved analysis has been carried out, as described in the next section.

Prior to the data analysis, we perform a first background rejection removing a fraction of background events, mostly composed of cosmic rays interacting in the sensitive area of the instrument. A residual cosmic-ray background component still remains and needs to be estimated and subtracted from the data as well. The correct modeling of such a component is particularly relevant to study the fainter extended emission of the dust-scattering rings. As a suitable background region cannot be extracted directly from this observation, due to the presence of the rings, we assess the expected X-ray background rate from previous IXPE observations. In particular, we consider three IXPE observations of low-rate point-like sources: the observation of 1ES~1959+650 carried out between 2022 June 9 and 2022 June 12 (BKG1);  the observation of BL Lacertae (BL Lac) which happened between 2022 July 7 and 2022 July 09 (BKG2); the observation of 3C~279 performed between 2022 June 12 and 2022 June 18 (BKG3). From each of these observations we extract the background spectrum and we simulate a long exposure (1 Ms) IXPE observation with the \textit{ixpeobssim} simulation tool \citep{ixpeobssim}. The three selected observations provide a good bracketing of the background emission. More details on the particle background rejection, the residual background simulation, scaling, and subtraction are reported in Appendix~\ref{app:A}.

Typically, for IXPE observations, the polarization information is extracted via two types of analyses: a polarimetric analysis and a spectropolarimetric analysis. For the former, we use the \textit{xpbin} routine of \textit{ixpeobssim} with the flag {\tt --algorithm PCUBE}. This routine computes the I-normalized Stokes parameters Q and U from the event-by-event Stokes parameters of the sample of selected events. The algorithm supports the calculation of the background-subtracted Stokes parameters, if a background template is provided. The polarization degree (PD) and polarization angle (PA) with associated errors are calculated from the Q/I and U/I parameters following the recipe of \cite{Stokes}. 

The spectropolarimetric analysis, as opposed to the simpler polarimetric analysis, accounts for the shape of the intensity spectrum. This analysis consists of the joint fit of the I, Q and U spectra and, for this work, we make use of the Multi-Mission Maximum Likelihood (\threeml) framework \footnote{\url{https://threeml.readthedocs.io/en/stable/index.html}} \citep{3ml}, which is publicly available and allows for both frequentist and Bayesian analysis approaches. Here we report the results of the frequentist analysis, but we verified that the Bayesian approach leads to the same results.  

Hereafter, we refer to the central region as the \core, while the inner and the outer rings are denoted \rone{} and \rtwo, respectively. In the next sections we will illustrate the data analyses and results for these different regions.

\section{The Core / Afterglow emission}
\label{sec:core}
\subsection{Data analysis}
\label{subsec:coreanal}

\begin{figure}[t]
    \centering    
    \includegraphics[height=7.5cm]{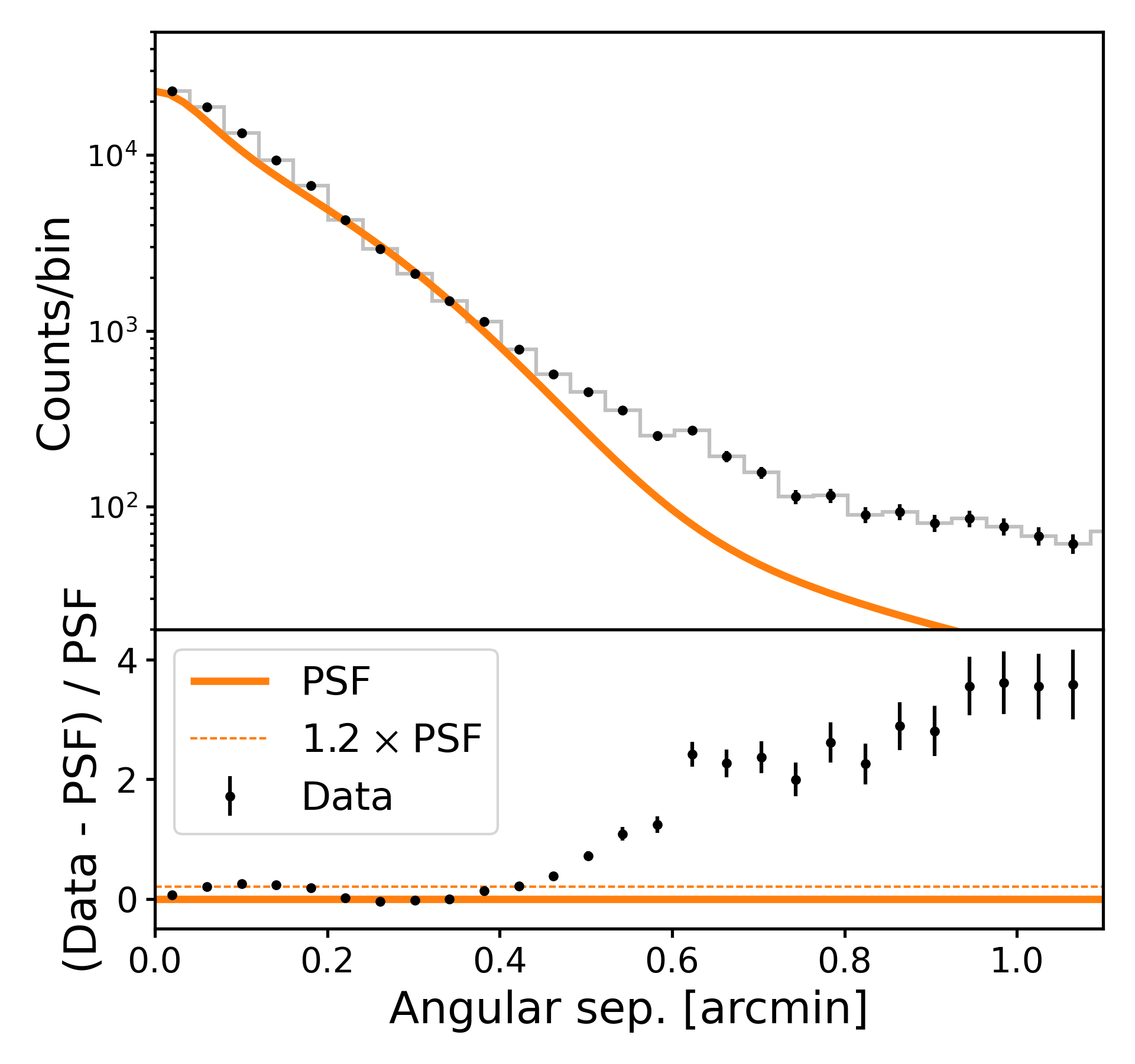}\quad\quad
    \includegraphics[height=7.5cm]{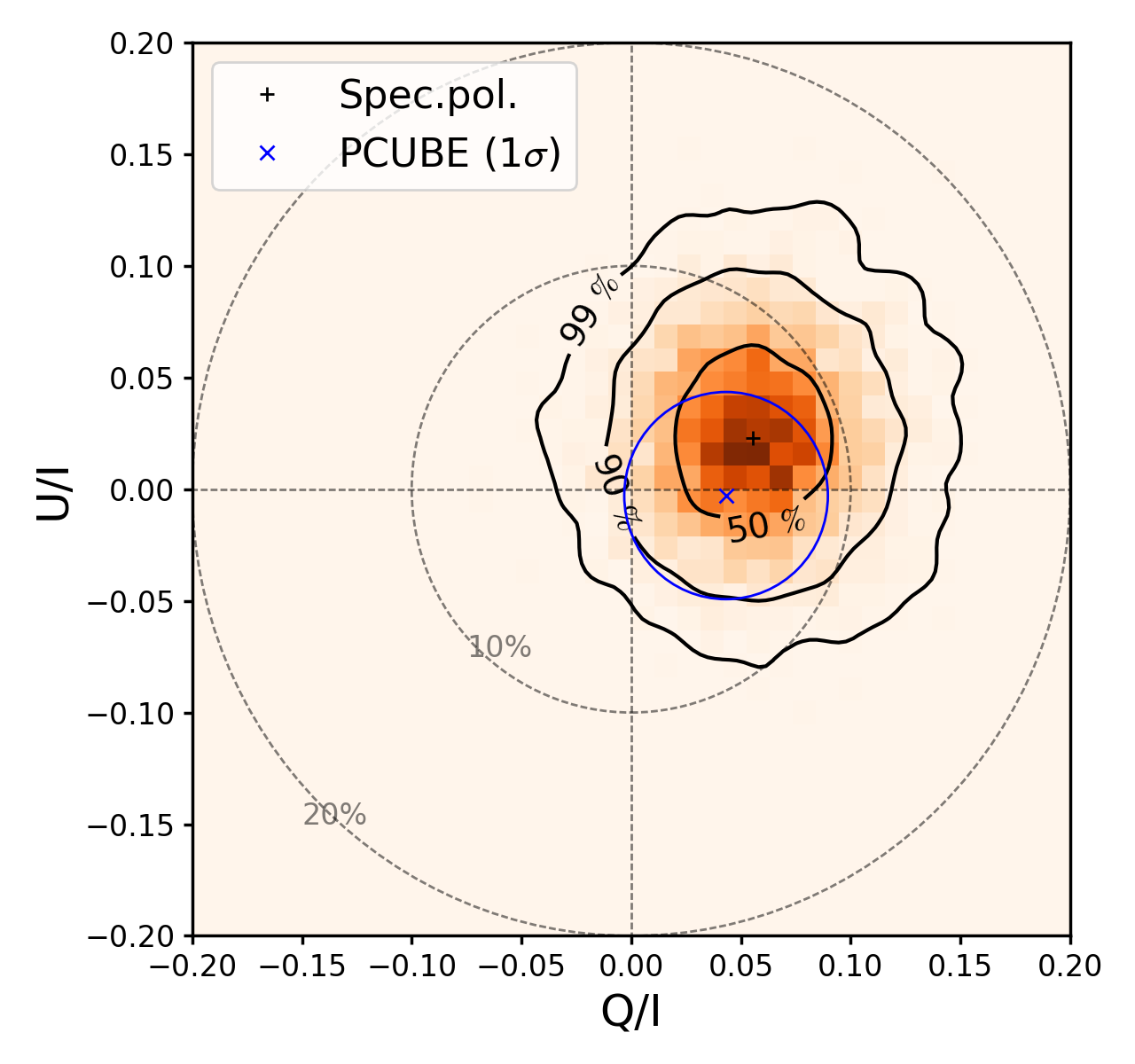}
    \caption{Left: Background-rejected radial profile around the core emission for DU1; as shown in the lower panel, the source profile starts deviating more than 20\% from the instrumental PSF at around 0.43 arcmin. The equivalent plots for DU2 and DU3 are not reported here as they carry the same information. Right: Q/I versus U/I plot; in orange we show the distribution resulting from the spectropolarimetric analysis and the 50\%, 90\% and 99\% C.L. contours in black. The blue cross and circle show the PCUBE analysis result and the related 1 sigma error.}
    \label{fig:core}
\end{figure}

We start with the analysis of the \core, which arises from the burst afterglow. We select the region as a disc centered on the brightest pixel\footnote{The image pixel size is $2.57\arcsec\times2.57\arcsec$.} of the IXPE image and radius of 26 arcsec (0.43 arcmin). Beyond this radius, the radial profile of the emission deviates from the PSF of the instrument by more than 20\%, as shown in Figure~\ref{fig:core} (left panel). Such a deviation informs us on possible contamination from the emission of dust-scattered X-rays (from the GRB prompt and/or afterglow emission) that we cannot fully resolve. We verified that the bright core emission from the central point-like source dominates the final result as we find consistent numbers within the one sigma uncertainty when varying slightly the selecting radius. According to the IXPE PSF, cutting at a radius of 26 arcsec eliminates less than $\sim$15\% of the total source emission.

Given the high photon statistics of the \core{} (signal-to-noise ratio $>100$ for any background), the results of the analysis are not affected by the choice of the background spectrum. Here, we report the results for the background template BKG2.
Through the PCUBE analysis, we find an unconstrained polarization in the 2--8 keV energy range and derive a 99\% C.L. upper limit of 16.1\%. No evolution with time or energy is observed.
For the spectropolarimetric analysis we model the observed spectrum with an absorbed power law decreasing in energy, with intrinsic parameters fixed to the values of the \textit{Swift}/XRT automated online analysis \citep{SwiftXRTauto}, which are consistent with other reported values \citep{SwiftRef1}. In particular, the intrinsic absorption parameters are fixed to $1.36\times10^{22}$cm$^{-2}$ \citep{SwiftXRTauto} at a redshift of 0.151 \citep{GCNredshift}, while the Galactic absorption value is fixed to $5.38\times10^{21}$cm$^{-2}$ \citep{Willingale2003}. To account for mismatches of inter-calibration among the different IXPE telescopes, a constant normalization is left free to vary for DU2 and DU3 with respect to DU1.

The best-fit values of the spectropolarimetric analysis are provided in Table~\ref{tab:summary}. We find a best-fit power-law index of $\Gamma_\core=1.98\pm0.03$, in agreement with expectations from a late afterglow emission (see Section~\ref{subsec:coreinter}). The polarization results are slightly more constraining than, but consistent with, the PCUBE analysis with a polarization degree of $6.1\pm3.0\%$. The right panel of Figure~\ref{fig:core} shows the Q/I versus U/I distribution of the \core\, emission. 

The Stokes parameters Q/I and U/I are expected to be normally distributed with respective means $\overline{Q/I}$ and $\overline{U/I}$, and equal standard deviations. An error contour in (Q/I, U/I) space is a circle of radius $\epsilon$ centered on ($\overline{Q/I}, \overline{U/I}$), where $(\epsilon/\sigma)^2$ is distributed as $\chi^2$(d.o.f=2). The probability that the observed polarization exceeds the measured value, under the null hypothesis of unpolarized emission, is 9.7\%. This is inconsistent with a zero degree of polarization at 90\% C.L. We therefore set a upper limit to the polarization degree (1D distribution) of 13.8\% at 99\% C.L. (12.1\% at 95\% C.L.).
For completeness, the I, Q, U spectra are reported in Figure~\ref{fig:spectra} (first row) in Appendix~\ref{app:C}.

\begin{deluxetable*}{ll}
\label{tab:summary}
\tabletypesize{\scriptsize}
\tablewidth{0pt} 
\tablecaption{Spectropolarimetric \textbf{\core} analysis}
\tablehead{
\colhead{Parameter} & \colhead{Value}}
    \startdata 
        ~~~~~ PD & ($6.1\pm3.0$)\% \\
        ~~~~~ PD U.L. 99\% C.L. ~~~~~~~~~~ & $<13.8\%$ ~~~~~~~ \\       
        ~~~~~ PD U.L. 95\% C.L. ~~~~~~~~~~ & $<12.1\%$ \\
        ~~~~~ $\Gamma_{\core{}}$ & $1.98\pm0.03$
    \enddata
    \tablecomments{Summary table of the spectropolarimetric analysis of the \core{}. 
    The spectropolarimetric fit is performed in the 2--8 keV energy band and it assumes Gaussian statistics. The PA is unconstrained. The best-fit Q/I and U/I constants are $(5.6 \pm 2.9)\times10^{-2}$ and $(2.8 \pm 2.9)\times10^{-2}$, respectively.}
\end{deluxetable*}


\subsection{Interpretation}
\label{subsec:coreinter}
According to current models, once beyond the early flaring stages, GRB afterglows arise via synchrotron processes from electrons accelerated through interactions of the GRB jet with the circumstellar material. This is consistent with observations from radio to high energies  of GRB afterglows \citep{kumar2015physics}. The physics is well understood and follows a set of closure relations \citep[e.g.,][]{Sari+00refresh} which, when observations fit a self-consistent picture, can be used to infer properties of the underlying emitting region through observables such as the spectral indices and rate of temporal decay. 
The synchrotron spectrum is described by a set of power laws with different spectral indices,  each with its own closure relation depending on the particle density distribution of the circumstellar environment.
We model the \core{} X-ray emission as observed by IXPE with this interpretation, in order to utilize the polarimetric observation to constrain intrinsic properties of the jet and our viewing angle.

We start by investigating the density of the interstellar matter around the GRB progenitor. 
The density profile in units of cm$^{-3}$,
$n(R)$, where $R$ is the distance from the central engine,  is parameterized by the index $k$, such that $n(R)\propto R^{-k}$. For example, $k=0$ corresponds to a constant density medium and $k=2$ describes a wind medium, and in-between values are also possible. A wind medium may be expected around long GRBs since they arise in the deaths of massive stars. The density profile affects the time evolution of the synchrotron break frequencies.
We assume that the IXPE energy range lies between the typical ($\nu_m$) and the cooling ($\nu_c$) synchrotron frequencies, and show that this assumption yields a consistent picture.

The time and energy evolution of GRB afterglow emission is described by \citep[see, e.g.,][]{Granot+02Dabreaks}
\begin{equation}
     F_\nu\propto t^{-\alpha} \nu^{-\beta}~.
\end{equation}
The \core{} spectrum is well fit by an absorbed power law with a photon index of $1.98\pm0.03$, which yields $\beta=\Gamma_{\core}-1=0.98\pm0.03$.
The temporal evolution of GRB afterglow usually shows a break, which causes the index $\alpha$ to increase. This steepening is proportional to $(\Gamma_j\theta_j)^2$, where $\Gamma_j$ is the jet Lorentz factor and $\theta_j$ the jet half opening angle. Taking into account the time evolution of the Lorentz factor, the increase in the temporal index will be $\Delta \alpha=(k-3)/(4-k)$. From the closure relations \citep[e.g.,][]{Sari+00refresh} between the temporal and spectral indices, we can express the index of the density profile $k$ as 
\begin{equation}
    k= \frac{2 (4 \alpha -6 \beta +3)}{2 \alpha -3 \beta +3}.
\end{equation}
For $\alpha=1.634\pm 0.015$, measured by \textit{Swift}-XRT\footnote{\textit{Swift}-XRT data were analyzed in the IXPE observation time window through the online tool: \url{https://www.swift.ac.uk/xrt_live_cat/01126853/}. IXPE's light curve shows a consistent time evolution, but we find a less precise estimation of the power index. Therefore, we adopt the \textit{Swift} value in our model.} and $\beta=0.98\pm0.03$, we get 
$k=2.20\pm0.05$. We note that using the Swift-XRT spectral index of $0.88\pm 0.15$, we get $k=2.35\pm0.21$. In our model we assume that the IXPE observation was preceded by an achromatic jet break at $\sim$1 day \citep{GCNjetbreak}. 

We will thus assume that the forward shock propagates in a wind medium with density $n(R)=A R^{-2}$, where $A=3.02\times 10^{35} A_{\star}~ {\rm cm}^{-1}$. To estimate $A_{\star}$ we introduce fiducial or base values for the energy density fraction in electrons and in magnetic fields: 
\begin{equation}
    \epsilon_e=10^{-1}\epsilon_{e,-1}~~,  ~~~~~~~~~~\epsilon_B=10^{-3}\epsilon_{B,-3}~~.
\end{equation}
Furthermore, we set the kinetic energy of the outflow to $E_{k,{\rm iso}}\approx 10^{55}$ erg and we use the $Q=10^xQ_x$ scaling convention for quantity $Q$ in cgs units. 
With the above choice of parameters, and neglecting the effect of inverse Compton scattering on the cooling, we have \citep[e.g.,][]{Granot+02Dabreaks}:
\begin{equation}
\centering
\begin{split}
\nu_m =~ & 3.6\times10^{12}~E_{k,55}^{1/2}~\epsilon_{e,-1}^{2}~\epsilon_{B,-3}^{1/2}~(t/3.5~\rm{d})^{-3/2} ~~Hz~,\\
\nu_c =~ & 2\times10^{18} ~E_{k,55}^{1/2}~ A_{\star,-1.5}^{-2}~  \epsilon_{B,-3}^{-3/2}~(t/3.5 ~\rm{d})^{1/2}~~  Hz~, 
\end{split}
\end{equation}
\noindent
confirming that indeed $\nu_m<\nu_{\rm IXPE}\lesssim\nu_{c}$ at the time of the IXPE observation, and this ordering persists at later times because $\nu_m\propto t^{-3/2}$ and $\nu_c\propto t^{1/2}$. In this spectral regime, the energy spectral index is given by $\beta=(p-1)/2$, where p is the power law index of the electron energy distribution ($dN_e/d\gamma_e\propto\gamma_e^{-p}$, where $\gamma_e$ is the electron's Lorentz factor). Using the $\beta$ derived from IXPE observation, we find $p=2.96\pm0.06$.

We can now estimate $A_{\star}$ by comparing the observed flux density at 10 keV, $F_{\nu,obs}\approx10^{-6}$ Jy, to the synchrotron model prediction \citep[e.g.,][]{Granot+02Dabreaks}, valid after the jet break: 
\begin{equation}
    A_\star\approx3\times10^{-1}~ E_{k,55}^{-(3+p)/2} \epsilon^{2-2p}_{e,-1} \epsilon_{B,-3}^{-(p+1)/2}~ {\rm cm}^{-1}. 
\end{equation}

\noindent
We note that $A_\star$ depends strongly on the $\epsilon_e$ parameter ($A_\star\propto\epsilon_e^{-4}$ for $p\approx 3$).

A separate constraint for our afterglow model comes from the measured jet break time, $t_{\rm jet}$, which scales as $t_{\rm jet}\propto E_k \theta_{j}^{4} A_\star^{-1}$ if the ratio between the jet opening angle $\theta_j$ and viewing angle $\theta_v$ is known. This parameter, and its position in time with respect to the time of the observation, is relevant for polarization, as it can be broadly associated with the time when the polarization degree lightcurve has a zero point and the polarization angle rotates by 90 degrees. 
In fact, for uniform (top-hat) jet structure with no sideways expansion, significant polarization arises from the break in symmetry of the visible surface. This surface is typically an annulus when projected to the plane of the sky. As the annulus grows, it encompasses a progressively larger fraction of the jet surface. Eventually, for an off-axis observer, the annulus will grow beyond the size of the jet on one side, while still collecting emission from the opposite side, resulting in net polarization. The polarization lightcurve exhibits the typical two-bump structure \citep{1999MNRAS.309L...7G, 1999ApJ...524L..43S}, where the jet break time approximately corresponds to the minimum between the bumps. Our model is constructed so that the PD zero point between the two bumps is at $\approx 1$~day, to match the estimated jet break time \citep{GCNjetbreak}.
We derive the expected polarization degree by integrating the intensity and polarization of the comoving volume elements of the jet over the equal arrival time surfaces \citep{1999ApJ...524L..43S, 2003ApJ...594L..83G,2021ApJ...913...58S,2022arXiv221012904P}.
For each comoving volume element, the maximum PD of a synchrotron-emitting, shock accelerated electron population with power-law distribution with index $p$ will be: ${\rm PD}=(p+1)/(p+7/3)\lesssim75\%$ \citep{rybicki79}. The observed polarization will be reduced from this value by integrating over all the parts of the jet that contribute to the flux at a given observer time \citep[see e.g.,][]{Lyutikov+03pol}.

\begin{figure}[t]
    \centering    
    \includegraphics[height=7.5cm]{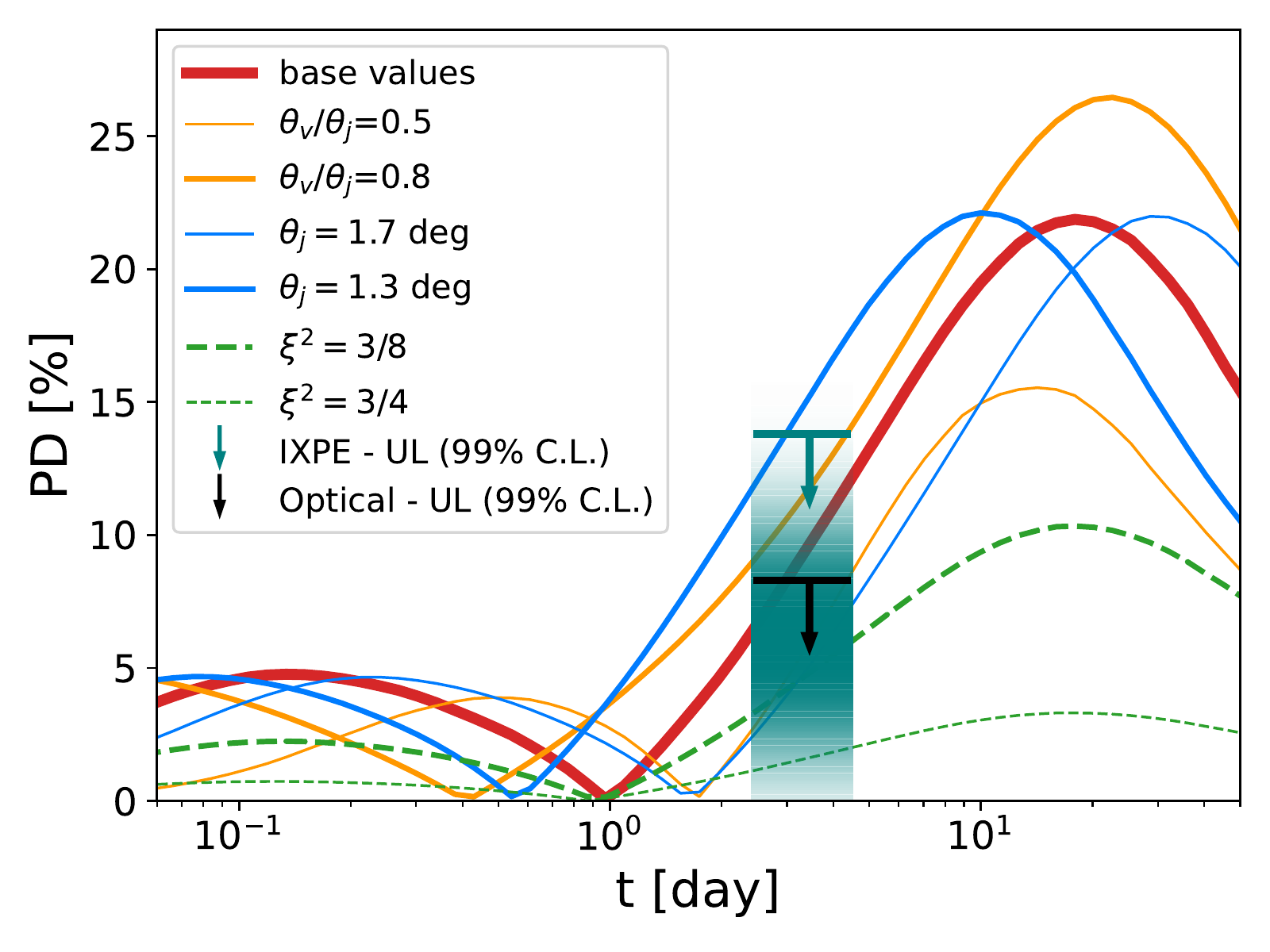} 
    \caption{
    Polarization lightcurves using a set of base parameter values ($\theta_j=1.5$~deg, $\theta_v=\frac{2}{3}\theta_j$, p=2.96, $E_{k}=10^{55}$ erg, $A_{\star}=3\times 10^{-1}$ cm$^{-1}$, $\xi=0$). We show the effect of changing the $\theta_v/\theta_j$ ratio, the jet opening angle, $\theta_j$ and the magnetic field ratio, $\xi$. The IXPE upper limits are shown in teal, while the black upper limit marks the upper limit of the contemporaneous optical observation. The shaded band shows a Gaussian modulation centered on the PD (darker shade) and width equal to the one sigma uncertainty on the PD (from the spectropolarimetric fit). We stress that we do not claim a measurement, which would require at least a 99\% C.L. significance.}
    \label{fig:xi}
\end{figure}

The evolution of the polarization as a function of time depends strongly on a variety of parameters. We take a set of parameters (base values) that give a polarization consistent with the IXPE spectropolarimetric measurement: jet opening angle
$\theta_j=1.5$~deg, viewing angle $\theta_v=\frac{2}{3}\theta_j$ \citep{1999MNRAS.309L...7G}, electron energy distribution index p=2.96, kinetic energy $E_{k}=10^{55}$ erg, density parameter $A_{\star}=3\times 10^{-1}$ cm$^{-1}$ and  $\xi=0$. 
The parameter $\xi$ is the ratio of the magnetic field strength in two directions defined as: $\xi^2= 2 \langle B^2_{||}\rangle/\langle B^2_{\perp}\rangle$. Here, $B_{||}$  and $B_{\perp}$ are the magnetic field parallel and perpendicular to the shock normal, respectively. The case $\xi=0$ yields the maximum attainable polarization for any given set of afterglow parameters. 
Our model with base values and several additional realizations is presented in Figure \ref{fig:xi}. 
In general terms, all realizations have zero points anchored at $\approx 1$ day and yield increasing PD at the time of the IXPE observations. For a given $\theta_v/\theta_j$ ratio, we can choose a set of parameters ($E_k$, $A_\star$ and $\theta_j$)  so that $t_{\rm jet}=1$ day is satisfied. A higher $\theta_v/\theta_j$ ratio results in a higher peak polarization and earlier jet break time, due to the higher level of asymmetry as we move away from the jet axis. 

All presented models in Figure \ref{fig:xi}, except the low jet opening angle, are consistent with the upper limit. Taking the PD=$6.1\pm3.0\%$ at face value, models with jet opening angle $\theta_j<1.5$ deg (while keeping all other base values fixed) are disfavored. Similarly, models with $\theta_v/\theta_j>2/3$ tend to overpredict the IXPE measurement. Assuming a magnetic field ratio, $\xi$, closer to 1 simply scales down the PD. In principle, any model that overpredicts the observations can be made consistent by appropriate choice of $\xi$. The IXPE measurement, considering the base values, favors cases where $\xi^2\lesssim1/2$.

Optical polarization observations occurred during the IXPE observation window at the Nordic Optical Telescope \citep{GCNoptpol}. The sky conditions allowed an estimation of an upper limit to the optical polarization degree of 8.3\% at 99\% C.L. \citep[5.1\% at 95\% C.L., ][]{GCNoptpol}. In Appendix \ref{app:B} we provide more details about the optical data reduction. 
The optical band falls in the same spectral regime as the X-rays ($\nu_m<\nu_{\rm optical}<\nu_c$) for most choices of parameters around the base values. Thus the optical upper limit can be used to constrain the models. The optical limit is slightly stronger, but gives qualitatively the same constraints as the X-ray limit.
As the above analysis assumes a homogeneous jet (top-hat profile), more elaborate profiles ---e.g., Gaussian jet or structured jet \citep{rossi2004polarization}--- could give different results.


\section{The Rings / Prompt emission}
\label{sec:rings}
\subsection{Data Analysis}
\label{subsec:ringsanal}
\begin{figure}[b]
    \centering
    \includegraphics[height=7.5cm]{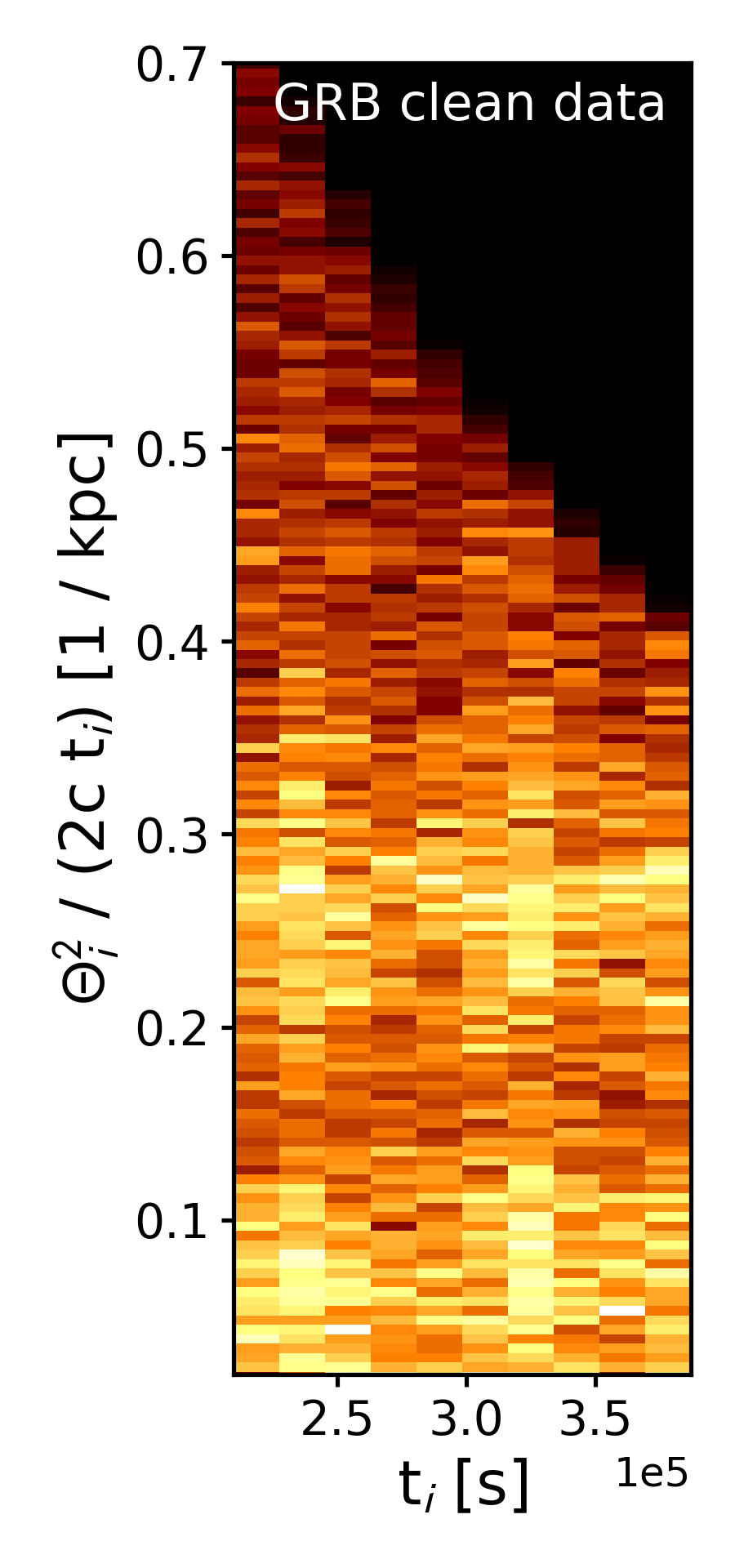}\quad
    \includegraphics[height=7.5cm]{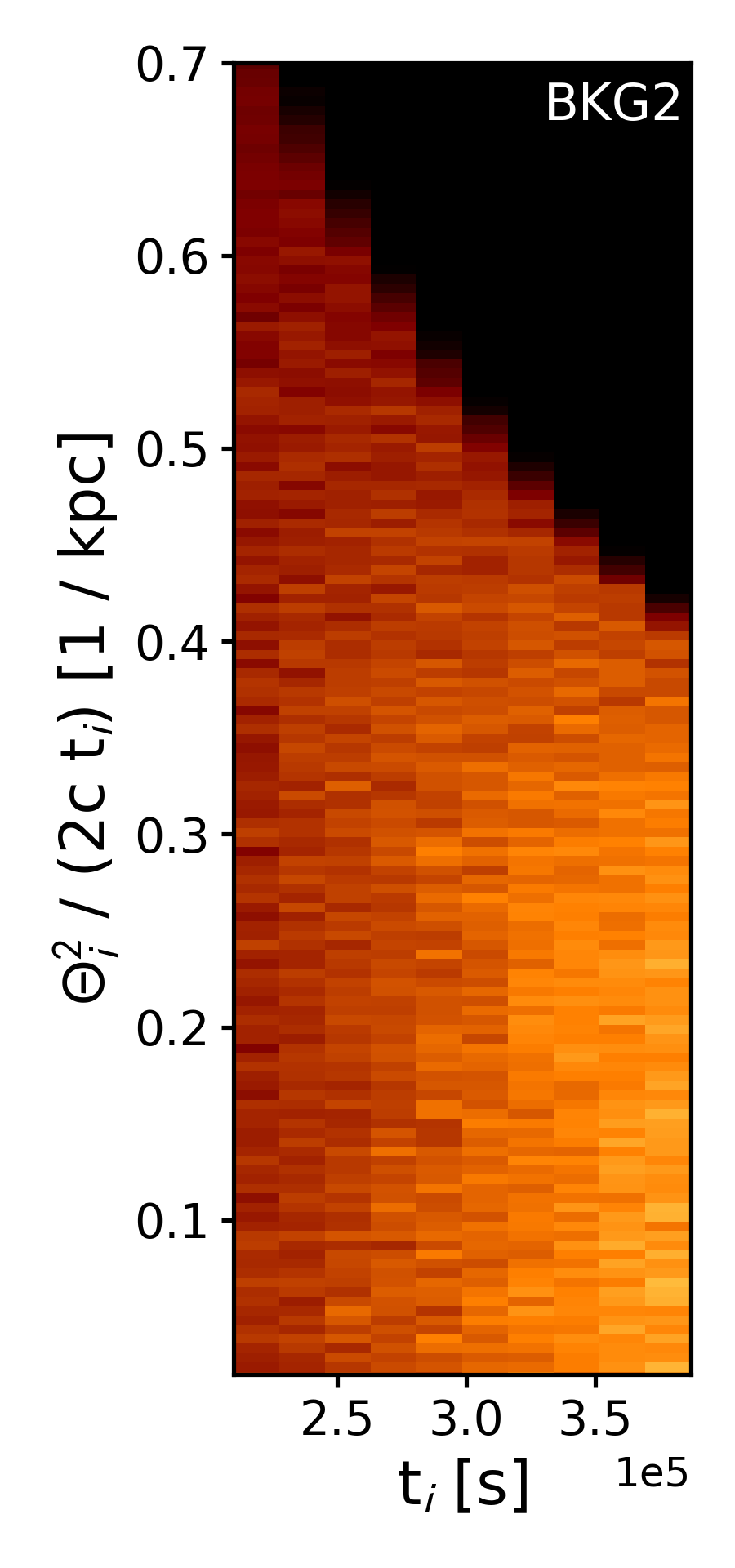}\quad
    \includegraphics[height=7.5cm]{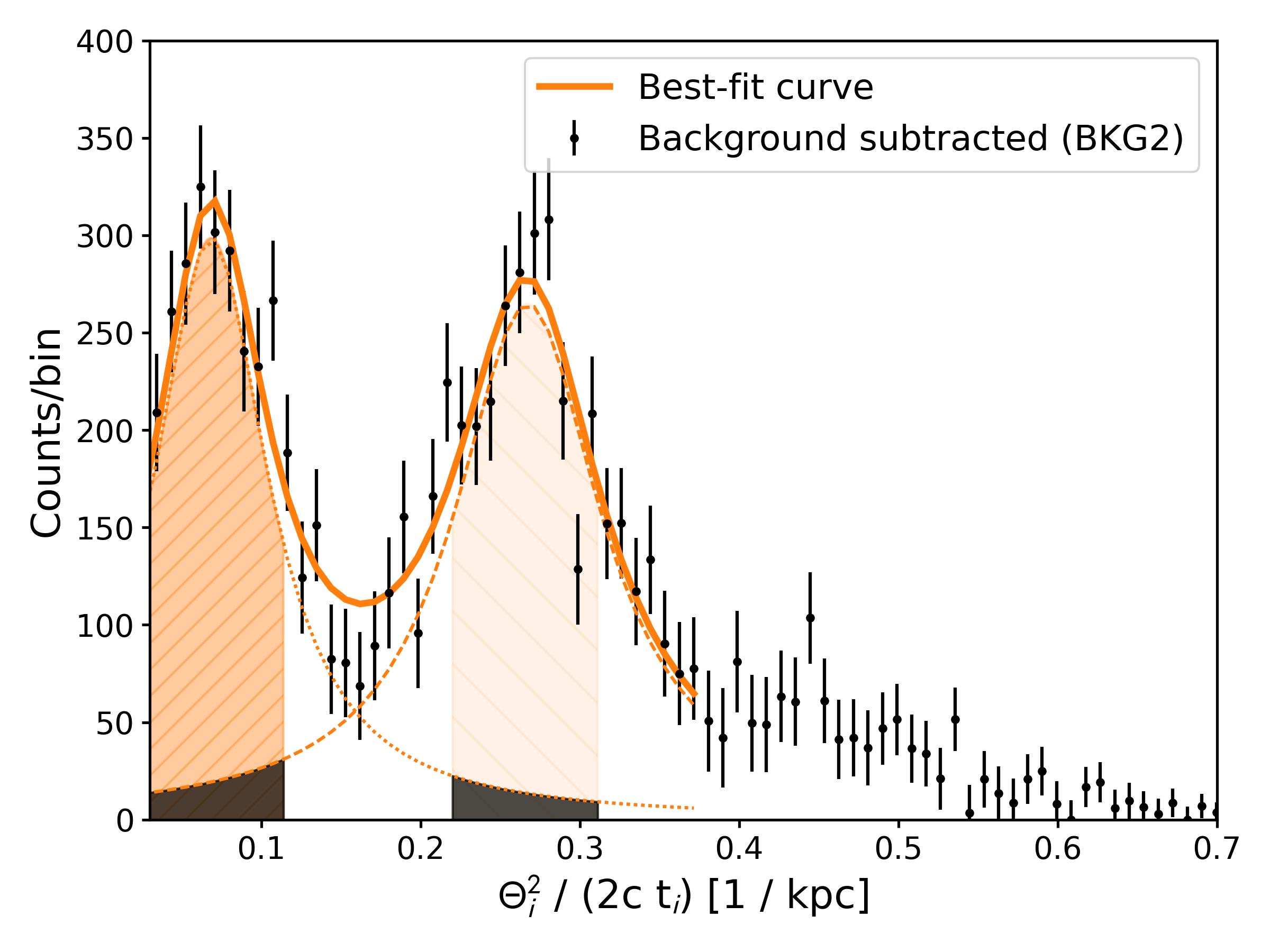}
    \caption{Study of the ring evolution in the IXPE observation. Left: distribution of the variable $K_i = 1/D_i$ in time bins (in terms of seconds after the trigger time). Middle: Equivalent to the figure on the left panel, but for the BKG2 template. Right: Background subtracted distribution of $1/D_i$ for the specific case of the BKG2 template (the equivalent distributions for BKG1 and BKG3 are provided in Appendix~\ref{app:C}).}
    \label{fig:ringevol}
\end{figure}

As mentioned in the introduction, the observed rings are the result of a known effect involving Galactic dust along the line-of-sight of a bright transient event. A fraction of the photons emitted in the prompt phase of the GRB are scattered by dust clouds in the Milky Way. Those scattered inwards towards the line of sight arrive at Earth after traveling a longer path length with respect to the unscattered ones. This results in a later arrival time of the scattered photons, with a delay that depends on the distance of the dust cloud to Earth and the scattering angle $\theta_s$. The angular size of the halos $\theta_h$ is related to the scattering angle, as $\theta_s(1-x)$, where $x$ is the ratio between the distance of the cloud and the distance of the source. Since we are dealing with a transient event at cosmological distance ($x \ll 1$), the approximation $\theta_h \sim \theta_s$ applies \citep{MiraldaEscude1999, Draine2003II}. 

Being produced by a short transient event, the rings expand radially in time. This arises as photons with different scattering angles travel different path lengths. In order to study the radial evolution of the rings and correctly select prompt, scattered photons as the rings expand, we devise a method inspired by the procedure described in \cite{TiengoMereghetti2006}. For each photon $i$ detected at a time $T_i$ and at a sky coordinate ($a_i$, $\delta_i$), we define the following variables

\begin{equation}
   \label{eq:ki}
   t_i = T_i - T_0 \quad\quad {\rm and}  \quad\quad
   K_i = [(a_i - a_B)^2 + (\delta_i - \delta_B)^2] / 2c t_i = \theta_i^2 / 2c t_i~,
\end{equation}

\noindent where ($a_B$, $\delta_B$) are the coordinates of the unscattered emission (the center of the \core{} in the IXPE image) and $\theta_i$ is the angular distance (in arcsec) of the photon $i$ from the point ($a_B$, $\delta_B$). The trigger time of the prompt emission is taken from the \textit{Fermi}/GBM\footnote{The time difference between the GBM trigger and the beginning on the IXPE observation is 209848 s. \textit{Fermi}-GBM triggered on a precursor event, about 210 s before the main brighter peak. We reasonably assume that the rings emission is an echo of the brightest part of the prompt phase. Hence we use the GBM trigger time plus 210 seconds. In any case, a difference of 210 seconds on the total time-distance between IXPE observation and the GBM trigger does not affect our results.} \citep{GCNfermiGBMtrigger}. The advantage of this approach is that in the plane $K_i$ vs $t_i$, shown in the left panel of Figure~\ref{fig:ringevol}, the expanding rings appear as horizontal bands, facilitating the event selection. We remove the dominant emission from the \core{} by removing events within 0.85 arcmin from the center to avoid contamination from the bright core (afterglow) emission. We estimate that the contamination from the \core{} emission at radial distances larger than 0.85 arcmin is less than 4\%. The distribution $n(K_i)$, after subtracting the simulated background events, is shown in the right panel of Figure~\ref{fig:ringevol}: the contribution of the two rings is prominent. We fit the distribution around the peaks with the sum of two Lorentzian functions, which approximate well the observed distribution. 

We define the event selection cut on the $K_i$ distribution as illustrated in Figure~\ref{fig:ringevol} (right panel). The area under each best-fit Lorentzian between R$^i_{{\rm min}}$ and R$^i_{{\rm max}}$ (orange areas in the plot), where $i = 1,2$ denotes \rone{} and \rtwo{} respectively, is at least a factor of twenty larger than the area under the other Lorentzian in the same range (gray areas in the plot). This ensures a negligible contamination from the emission of one ring onto the other. The edges of the selection for the wider ring are symmetric with respect to the peak, while the innermost edge of the smaller ring is naturally defined by our region cut off at 0.85 arcmin to exclude the \core{} emission. 

\begin{figure}[t]
    \centering    
    \includegraphics[height=7cm]{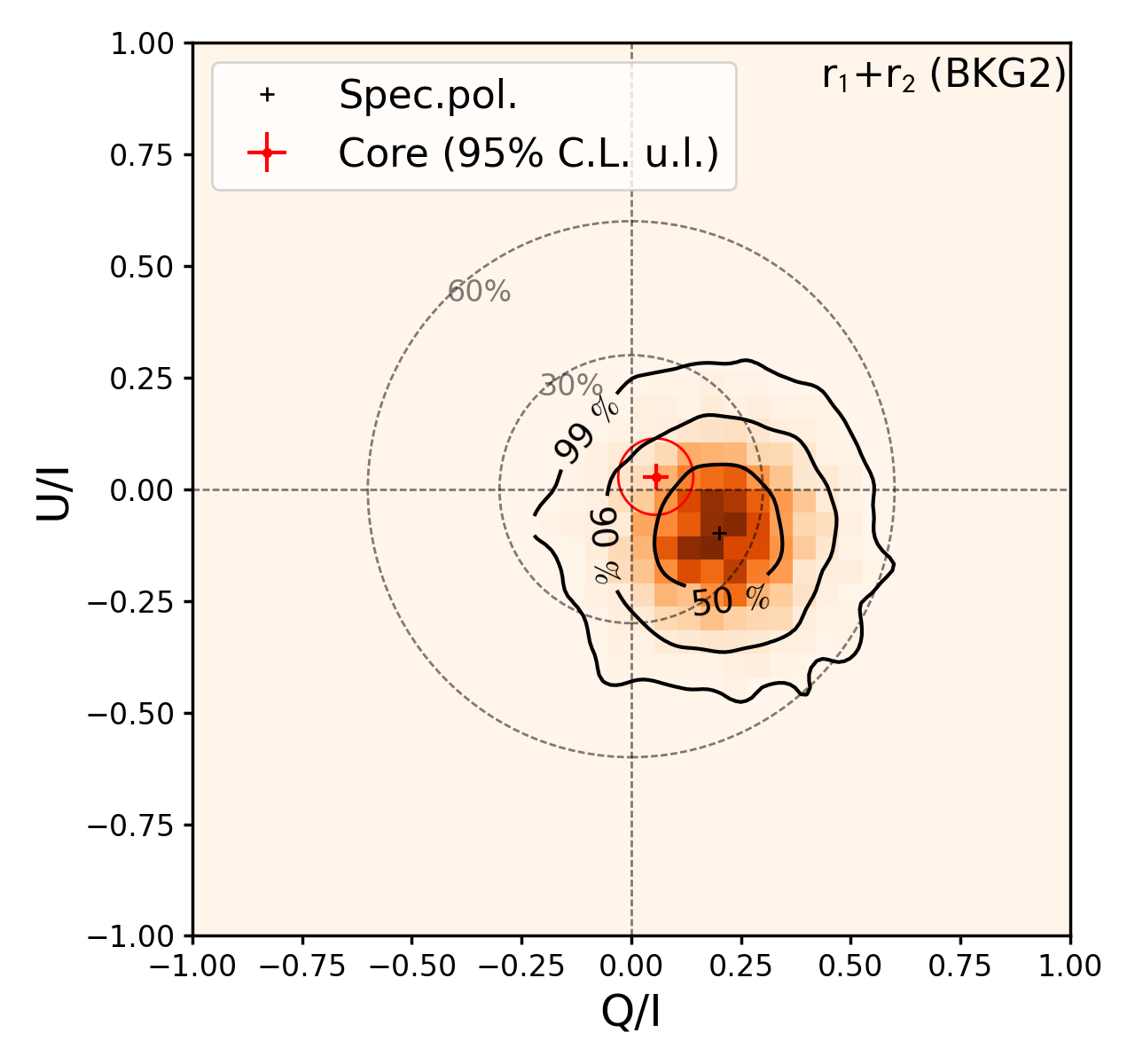}\quad\quad
    \includegraphics[height=7cm]{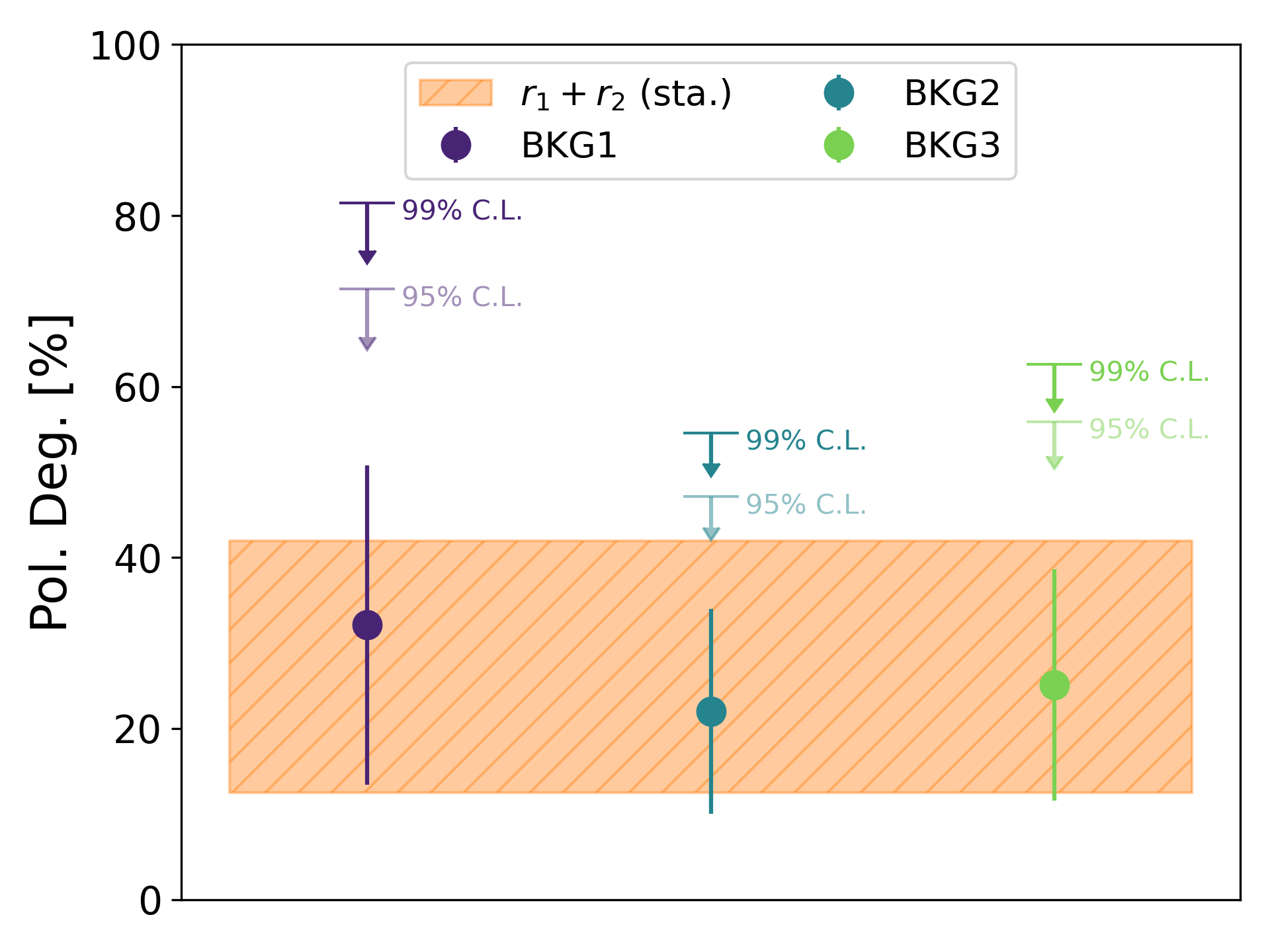}
    \caption{Left: Q/I versus U/I distribution of the polarization of the rings resulting from the spectropolarimetric analysis. The 50\%, 90\% and 99\% C.L. contours are shown in black. The red circle reports, for comparison, the 95\% upper limit to the PD of the \core/afterglow emission. Right: Results of the spectropolarimetric fit for the PD assuming different background templates. The colors violet, teal and green correspond to the different assumed backgrounds BKG1, BKG2 and BKG3, respectively. The orange band is centered on the average of the best-fit values weighted by their uncertainties, and has a width representative of the mean relative statistical error. }
    \label{fig:3mlfit_comp}
\end{figure}

Similar to the \core{} analysis, we proceed with the PCUBE polarization analysis in the 2--8 keV energy band. The observed spectra of the two rings are expected to be different because they are generated from the same prompt emission scattered at different angles. As discussed later on, given a scattering angle, the scattering efficiency of X-rays by dust grains is energy dependent \citep{Draine2003I}. This leads to the realization that combining the two ring selections into one single PCUBE analysis would be inaccurate. Furthermore, the low statistics of the individual ring selections prevents a proper background template subtraction for the PCUBE analysis. This implies that the estimated uncertainties are not accurate because they are computed on a boosted statistic that includes background events. The results of the PCUBE analysis for the individual rings are reported in Table~\ref{tab:summary3} of Appendix~\ref{app:C}. We find a PD$_{r_1} = 19.6\pm8.7$\% and PD$_{r_2} = 17.2\pm8.8$\%, in agreement with each other. These values indicate a higher polarization degree than what observed in the \core, though never exceeding the 99\% C.L. required to claim a detection.\footnote{The minimum detectable polarization at 99\% C.L. (for non background subtracted data) is $MDP_{99\%}=26.5\%$ and $MDP_{99\%} = 26.8\%$ for \rone{} and \rtwo, respectively}

The spectropolarimetric fit, allowing a proper combination of the rings selections, can give a more accurate estimation of the underlying polarization. 
The phenomenological model we define to describe the rings emission allows for the spectral parameters of the rings to be different while sharing common polarization parameters. The spectra of both rings are modeled as absorbed simple power laws, while we assume constant polarization parameters. The intrinsic and Galactic absorption parameters are kept fixed to the same values adopted for the \core{} analysis.
As opposed to the PCUBE analysis, we perform the subtraction of the background spectrum. We test different background assumptions, subtracting the spectra derived from BKG1, BKG2, and BKG3 templates, to which we applied the analogous event selection as for \rone{} and \rtwo{}.

The results are summarized in Table~\ref{tab:summary2}. We find that \rone{} has a best-fit photon index, averaged over the values obtained assuming different backgrounds, $\Gamma_\rone=2.89$, with a relative statistical error of about 7\% and negligible systematic uncertainty due to the choice of the background spectrum. \rtwo{} has a steeper spectrum, with photon index $\Gamma_\rtwo=3.98$ with a relative statistical error of about 6\% as well as a 7\% relative systematic error associated with different background assumptions. This is also illustrated in Figure~\ref{fig:ring_intrinsic_corrections} in Appendix~\ref{app:C}. As we will discuss in the next section, such a difference in spectral index between the two rings is expected owing to the energy dependence of the X-ray scattering cross section. We note that, for the case of \rtwo, the estimated photon index found assuming BKG1 shows a larger statistical error: the softer spectrum of the emission from this ring with respect to \rone{} makes the measurement more sensitive to the spectral characteristics of the subtracted background (Figure~\ref{fig:bkg_spec} in Appendix~\ref{app:C} shows that BKG1 has the hardest spectrum).

As for the polarization, we find that the PD value and uncertainty depend upon the assumed background. The statistical-error-weighted average is ($27.2\pm15.1~{\rm (sta.)}\pm 4.0~ {\rm (sys.)}$)\%, where the statistical error is the average among the statistical uncertainties obtained assuming different backgrounds, and the systematic uncertainty is given by the variation of the best-fit value assuming different backgrounds. Figure~\ref{fig:3mlfit_comp} shows the results for the different background subtractions. The significance of this result, tested against the null hypothesis of unpolarized emission, is about 81\% C.L.. The 1D 99\% C.L. upper limit on the PD varies between 54.6\% and 81.5\%, depending on the assumed background. Such a difference is due to the low-statistic regime we have for the rings data selection, which causes the statistical uncertainty to be strongly affected by small changes of the subtracted background.

The comparison of the best-fit PDs found assuming different backgrounds is illustrated in the right plot of Figure~\ref{fig:3mlfit_comp}. For completeness, the Q/I versus U/I distributions obtained for the different assumed background are provided in Figure~\ref{fig:polarplots} in Appendix~\ref{app:C}.
Note that, given the different approaches and handling of the background, the PCUBE and spectropolarimetric analyses are not directly comparable in this case.

\begin{table}[ht]
    \centering
    \begin{tabular}{r|l}
    \toprule 
    \multicolumn{2}{c}{{\large Summary table of the \textbf{rings} analysis}}\\
    \hline \rule{0pt}{4ex} 
        PD (BKG1) & $(32.1\pm19.2$)\% \\
        PD (BKG2) & $(22.0\pm12.6$)\% \\
        PD (BKG3)  & $(25.1\pm13.5$)\% \\
        PD$_{{\rm ave}}$ & $27.2^{\pm 15.1~ {\rm (sta.)}}_{\pm 4.0~ {\rm (sys.)}}$ \% \\
        PD U.L. 99\% C.L. & $<[54.6\% - 81.5\%]$  \\       
        PD U.L. 95\% C.L. & $<[47.1\% - 71.4\%]$  \\
        $\Gamma_{\rone{}}$ & $2.89^{\pm 0.20~ {\rm (sta.)}}_{\pm 0.07~ {\rm (sys.)}}$ \\
        $\Gamma_{\rtwo{}}$ & $3.98^{\pm 0.25~ {\rm (sta.)}}_{\pm 0.30~ {\rm (sys.)}}$ \\
    \toprule             
    \end{tabular}
    \caption{Summary table of the spectropolarimetric analysis for the rings emission. 
    The spectropolarimetric fit is performed assuming Poissonian statistics of the background-rejected data with background-subtraction applied. We report the best-fit PD value for each of the three assumed background models, as well as the weighted average with associated statistical and systematic errors. The upper limits are strongly dependent on the assumed subtracted background: we report the range of values defined by the minimum and maximum value obtained. The PA is unconstrained.}
    \label{tab:summary2}
\end{table}

Figure~\ref{fig:polarplots} in Appendix~\ref{app:C} reports the Q/I versus U/I plots for the different background assumptions. Additionally, we show in the same figure the equivalent plots for the spectropolarimetric fit of the individual rings. Furthermore, Figure~\ref{fig:pcubes_sings} and Table~\ref{tab:summary3} report the results of the PCUBE analysis of the individual rings resolved in two logarithmic energy bins between 2 and 8 keV. We refer the reader to the Appendix for further discussion on this matter.

\subsection{Interpretation}
\label{subsec:ringsinter}
As discussed in \cite{Draine2003I}, the effect of the dust scattering at such a small angles is not expected to alter the intrinsic polarization of the incoming radiation (see their Figure 5). However, an explicit demonstration of this statement in the X-ray band is not directly discussed in the literature. Therefore we investigated the effect on polarization from reflection, scattering and transmission considering the common dust compounds. All of the above processes lead to a negligible effect on the polarization of the X-ray radiation, as discussed in Appendix~\ref{app:dustpoleff}. Therefore, we can reasonably assume that any polarization observed from the X-ray scattering halos is attributable to the original emission.

A high polarization degree (PD$\gtrsim 20\%$) in the prompt phase, when viewing the jet at angles smaller than the opening angle, $\theta_v<\theta_j$, can be achieved by synchrotron emission in an ordered, toroidal magnetic field configuration \citep{Toma+09pol}. Alternatively, high polarization can be achieved by random magnetic fields or Compton drag models, in a geometry where we are viewing the jet close to its edge, $\theta_j  \lesssim \theta_v<\theta_j + 1/\Gamma_j$ \citep{Granot+03highpol021206}. This scenario will result in a very early jet break and potentially high PD in the afterglow, which is disfavored by the observations. In what follows, therefore, we will focus on the ordered synchrotron scenario. 

We estimate the polarization degree integrated over the duration of the prompt emission. The PD mainly depends on the photon index, the viewing angle, and the product of the jet opening angle and the Lorentz factor, $y_j=(\theta_j \Gamma_j)^2$. For the IXPE observation, only the low-energy photon index is relevant. We consider the two extreme values 0.62 and 1.25, which correspond to the minimum and maximum best-fit values of the prompt GRB intrinsic spectral index given by the assumed background bracketing (see Section \ref{subsec:dust}). For $\Gamma_j=700$ \citep{2022arXiv221114200L}, $\theta_j=1.5$ deg, and $\theta_v=\frac{2}{3}\theta_j$ we obtain 16\% and 36\%, respectively. This range is consistent with the measured upper limits.

\begin{figure}[t]
    \centering    
    \includegraphics[height=6cm]{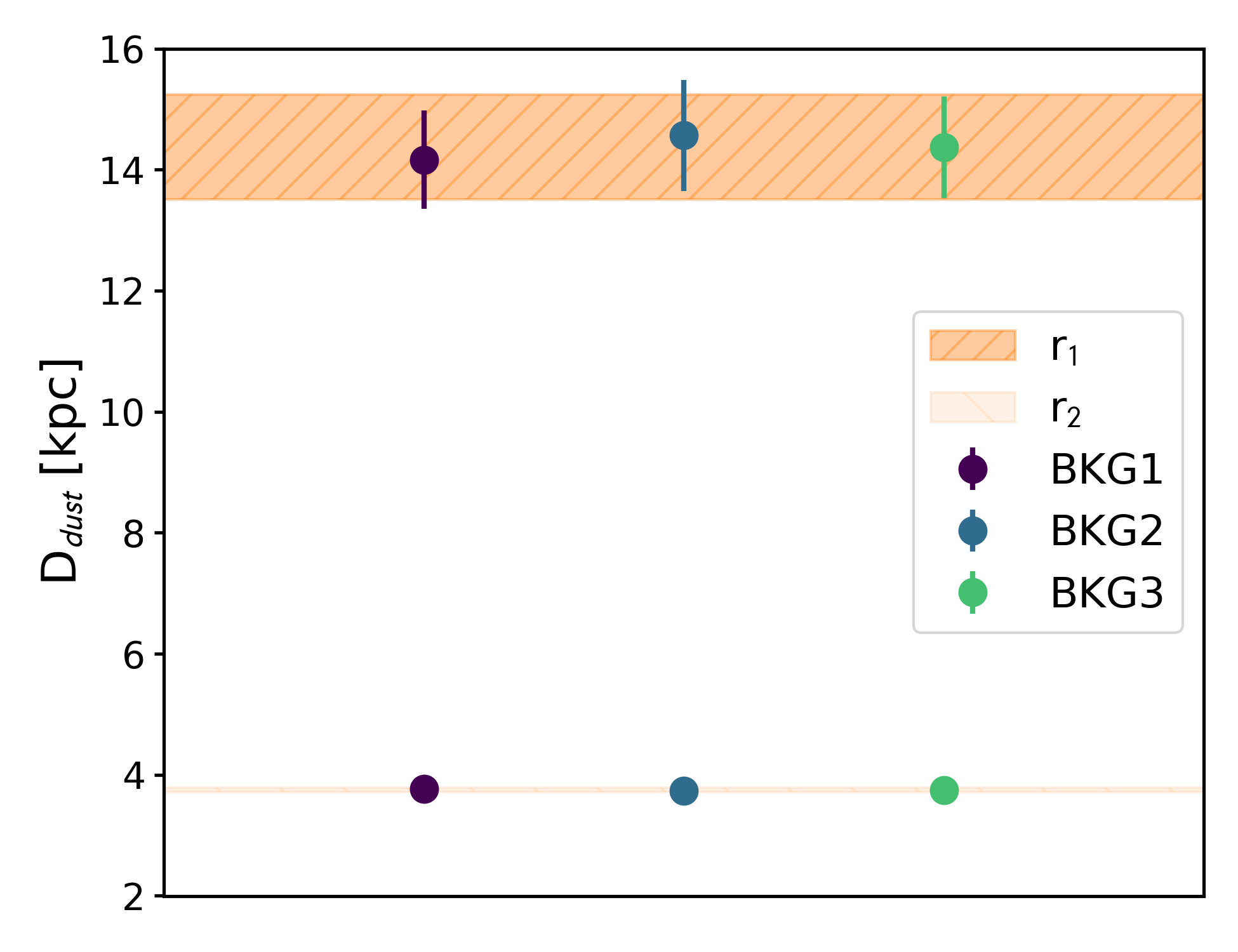}\quad
    \includegraphics[height=6cm]{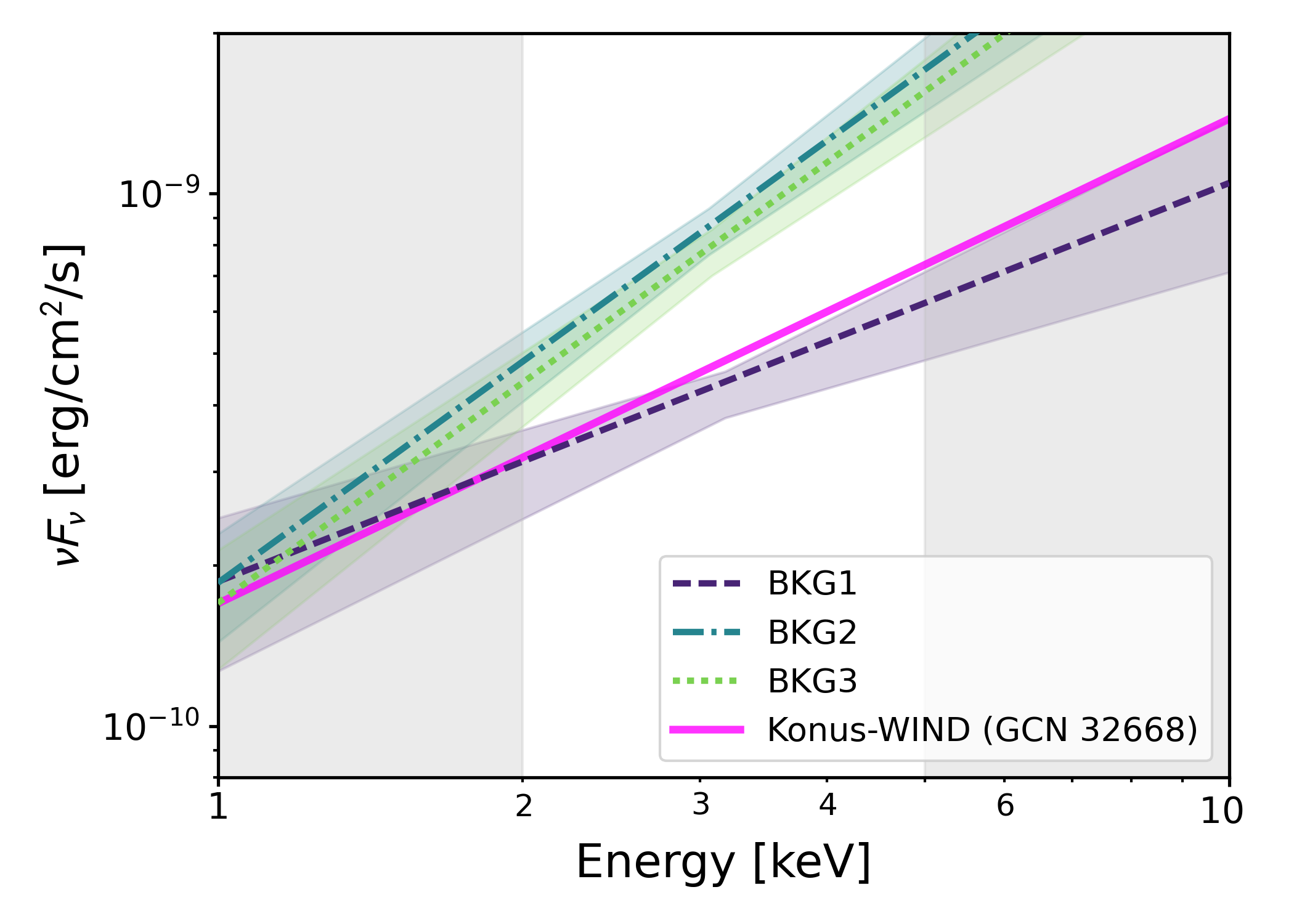}
    \caption{Left: Best-fit average distance of the dust clouds. The orange bands are centered on the average of the best-fit values weighted by their uncertainties, and have a width representative of the mean relative statistical error for the two dust clouds responsible for \rone{} and \rtwo{} emissions. Right: Derived intrinsic GRB prompt spectrum. The light-grey regions cover the energy ranges excluded in the fitting procedure.
    In both plots, the colors violet, teal and green correspond to the different assumed backgrounds BKG1, BKG2 and BKG3, respectively.}
    \label{fig:intrinsic}
\end{figure}

\subsection{Dust clouds and intrinsic GRB prompt emission} 
\label{subsec:dust}
In this section we derive some constraints on the dust clouds' distance and optical depth. We attempt to derive the intrinsic spectrum of the GRB prompt emission from the observed rings spectra. However, such considerations are limited by the imaging capabilities of our instrument with respect to other missions that were observing the burst at the same time. We therefore anticipate that the higher angular resolution and wider field of view of XMM-Newton and \textit{Swift}/XRT data, possibly resolving the presence of more than 2 rings, will better determine the characteristics of the dust clouds visible at the time of the IXPE observation. Constraining the spectral parameters of the rings from independent observations might help constrain the IXPE polarization parameters. This will be explored in a follow-up paper.

The dust cloud distance $D_{{\rm dust}}$ is given by
\begin{equation}
   D_{{\rm dust}} = 2c \, \frac{\Delta t}{\theta_h^2}, 
   \label{eq:ddust}
\end{equation}
\noindent where $c$ is the speed of light, $\Delta t$ is the difference between the time of the burst $T_0$ and the time of the observation of the rings. 
Hence, from the fit of the distribution $K_i$ defined in Eq.~\ref{eq:ki}, we can easily derive the distance of the clouds. In fact, the best-fit center of each Lorentzian is the inverse of the distance in kpc of the related dust clouds. We find the dust cloud associated with \rone{} to be at a distance of 14.41 kpc with a relative statistical error of 6\%. Considering the Galactic latitude of the GRB, a cloud at such distance would be located at about 1.1 kpc above the Galactic plane. The second cloud, responsible for \rtwo{} emission, is estimated to be at a distance of 3.75 kpc with a relative statistical error of 1\%.\footnote{Note that the farthest dust cloud is different from the (closer) ones producing the rings observed in the earlier \textit{Swift}-XRT observations reported in \cite{ATELTiengoSwiftRings}. The closer cloud in IXPE observation is consistent with the farthest one in \textit{Swift} observation.} The variance of this measurement, given by the different assumed backgrounds, is negligible with respect to the statistical error, as shown in Figure~\ref{fig:intrinsic} (left panel).
The half-width at half-maximum of the two Lorentzian curves correspond to $\sim (9.5\pm1.1)$ kpc and $\sim(0.7\pm0.1)$ kpc for \rone{} and \rtwo{}, respectively. Such values are larger than the values expected from the effect of the IXPE PSF\footnote{DU-averaged half-power diameter of $\sim$26'', \citep{IXPE_calibration}) would correspond to a full-width at half-maximum computed at the median time of the observation of $\sim$5.7 kpc and $\sim$0.8 kpc for \rone{} and \rtwo{}, respectively.} by a factor of $\sim$3 (for \rone{}) and $\sim$2 (for \rtwo{}). 
This could be symptomatic of a non-negligible thickness of the dust clouds, or, more likely, the presence of several rings within \rone{} and \rtwo{}, which we do not resolve.

As mentioned in the previous section, the X-ray emission of the ring is the echo of the prompt GRB emission scattered by Galactic dust clouds. Therefore, assuming the characteristics of the dust to be known, it is possible to infer the intrinsic spectrum of the prompt soft X-ray emission of the GRB from the dust-scattered spectra. 
The relation between the scattered spectra and the intrinsic one reads \citep{TiengoMereghetti2006}: 
\begin{equation}
\label{eq:intrinsic}
    \Phi_{r}(E, \theta_{1}, \theta_{2}) = \Phi(E)\tau_d(E)(g(\theta_{1}, E)-g(\theta_{2},E))~,
\end{equation}
where the intrinsic spectrum is modeled with an absorbed power law decreasing in energy, $\theta_{1}$ and $\theta_{2}$ are the rings extents at the beginning and at the end of the observations, and the function 
\begin{equation}
    g(\theta, E) = \frac{(\theta / \theta_s(E))^2}{1 + (\theta / \theta_s(E))^2} ,  ~~~~~~~~~ \theta_s(E) = 360''\left(\frac{E}{{\rm keV}}\right)^{-1} 
\end{equation}
accounts for both the fraction of halo we do not observe (because it lies outside the IXPE observing window) and the dependence of the scattering angle on the energy\footnote{$\theta_s(E)$ is the median scattering angle for photons of energy E, and the equation is a good approximation for photons of energy $>0.5$ keV \citep{Draine2003II}.}. In fact, larger (smaller) scattering angles correspond to a higher probability of scattering lower (higher) energy X-ray photons, which results in a steeper (harder) spectrum \citep{Draine2003I}.

According to \cite{Draine_2004}, the total scattering optical depth of the dust for photons between 0.8 and 10 keV can be estimated as
\begin{equation}
    \tau_d = \tau_{d_1} + \tau_{d_2} + \tau_{{\rm foreground}} = 0.15 A_v (E/{\rm keV})^{-1.8}
\end{equation} 
\noindent with $\tau_{d_1}$ and $\tau_{d_2}$ being the optical depths associated with the two dust clouds that produce the echo rings \rone{} and \rtwo{}, respectively, and $\tau_{{\rm foreground}}$ is the total optical depth between us and the first dust cloud we detect. 
$A_v$ is the V-band total Galactic extinction in magnitudes in the direction of the GRB. We assume $A_v=4.2$ mag as reported in the circulars by the VLT and JWST groups \citep{VLTGCN, WebbGCN}.

According to the measurements of \cite{NeckelKlare1980}, the total $A_v$ up to 3 kpc in the direction of GRB~221009A is about 3.3 mag. As the first cloud is at 3.75 kpc, this sum of three terms should be reasonably valid. Note that variations of the assumed value of $A_v$ affect the normalization of the intrinsic GRB prompt spectrum, but do not affect the slope of the power law.

We perform a maximum likelihood analysis by simultaneously fitting both rings spectra with the model in Eq.~\ref{eq:intrinsic}. As for the spectropolarimetric analysis, we have repeated the analysis three times for the different background models. Figure~\ref{fig:intrinsic} (right panel) illustrates the results of this fit. 

The fit is performed between 2 and 5 keV to avoid the low count statistics part of the ring spectra at high energy. The best-fit estimate of the fraction of the total optical depth associated with the farther and closer dust clouds is $n\sim0.36$ and $(1-n)\sim0.64$ respectively, and is not affected by the choice of the background. This translates into extinction values of $A_{v,d_1}\sim0.33$ and $A_{v,d_2}\sim0.57$. 

As for the intrinsic parameters of the GRB, we find that the prompt GRB power law has a photon index between 0.62 and 1.25 with a relative statistical uncertainty of the order of 10\%. This spectral index could be directly compared to the index inferred by the STIX observation \citep{GCNstix} and to the extrapolation of the lower-end of the energy spectrum measured by \textit{Fermi}/GBM. \textit{Konus}-Wind, which detects photons down to 20 keV in energy, has released a preliminary analysis of the prompt emission of this burst \citep{GCNkonuswind}. These authors find a time-averaged spectrum at the onset of the brightest phase of the event with a low-energy photon index of $1.09\pm0.01$. Such value lies within range defined by the best-fit values we find assuming different background models (see Figure~\ref{fig:intrinsic}, right panel).
Once confirmed, the direct observations of the prompt emission spectrum can provide a potential way to determine the best IXPE background model to use. In fact, the most representative background model could be selected based on agreement of these IXPE-inferred values with an externally measured index value. At the time of this work, however, such information is not publicly available, so we leave these considerations to a future work. Considering the bracketing given by the intrinsic spectra derived assuming different backgrounds, we infer a total fluence in the 1--10 keV band of $F=[1.6-6.1]\times10^{-4}$ erg/cm$^{2}$. The fluence is obtained from the integrated flux between 1 and 10 keV and multiplied by the IXPE total time of the observation, to account for the missing fluence due to Earth occultation time. The range of values we report for the fluence is based on the best-fit parameters of the intrinsic power-law model, ignoring statistical uncertainties which are of the order of 10--15\%.

\section{Conclusion}\label{sec:concl}

IXPE observed GRB~221009A from October 11 at 23:35:35 UTC to October 14 at 00:46:44 UTC for an effective exposure to the target of 94,122 s. The imaging capability of the instrument revealed the presence of a bright core emission, associated with the GRB afterglow, and the extended emission of two expanding dust-scattering halo rings. Such emission is an echo of the GRB prompt emission and therefore carries information about the latter. 

We studied the linear polarization properties of the \core/afterglow emission, and 
derived an upper limit on the polarization degree of 13.8\% at the 99\% C.L. The temporal and spectral parameters of the afterglow at the time of the IXPE observation are consistent with a forward shock propagating in a wind-like medium, with X-ray emission arising from synchrotron processes. The observed upper limit on the polarization degree favors a jet opening angle to be wider than 1.5 degrees, and a viewing angle wider than 2/3 of the jet opening angle (with some underlying assumptions). Also, scenarios with an equal magnetic field strength in the two directions parallel and perpendicular to the shock normal seem to be disfavored.

The polarization analysis of the combined dust-scattering rings revealed a non-significant polarization degree around ($27.2\pm 15.1{\rm (sta.)}\pm4.0{\rm (sys.)}$)\% with 99\% C.L. upper limit ranging between 54.6\% and 81.5\% depending on the assumed background. This result is in some tension with extremely high polarization measurements of other GRBs in the prompt phase \citep{2003Natur.423..415C, 2019ApJ...884..123C}. We also derive a photon index for the intrinsic GRB prompt spectrum between 0.62 and 1.25, depending on the background model considered. We note that this range includes the \textit{Konus}-WIND low-energy spectral index derived at energies above 20 keV as reported in \cite{GCNkonuswind}.
Considering the best-fit spectra, a scenario involving toroidal, ordered magnetic fields when the viewing angle is smaller than the jet opening angle, predicts high polarization degree up to 36\%, compatible with the observed upper limits. The upper limits on polarization from the IXPE observation exclude the case where we are observing close to the edge of a sharp transition in the jet.

Aside from the polarization properties of GRB~221009A, the main focus of this work, we could derive some constraints on the Galactic dust clouds distance. Through the time evolution of the emission from the two dust-scattering halos that we resolve, we estimated an average distance of the clouds to be about 14.41 and 3.75 kpc for the inner and outer ring, respectively. The width of the halos compared to the width expected from the effect of PSF suggests the presence of several unresolved halos within the two halos observed by IXPE. Contemporaneous observations by instruments with better angular resolution can inform us whether or not this is true. 

Future joint analyses exploiting contemporaneous observations from different instruments could be beneficial to constrain the spectral parameters and, therefore, better single out the polarization signature of the rings/prompt emission. Furthermore, independent polarization measurements from other instruments assessing a different energy regime, for either afterglow or prompt emission, will help to understand the full phenomenology behind this exceptional event. Works along these lines are already ongoing and will be the subject of upcoming publications. 

On a final note, we remark that the IXPE observation of GRB~221009A is, on its own account, exceptional and unique. We assessed, for the first time, the observation of soft X-ray linear polarization from the late afterglow emission of a GRB. Also for the first time, thanks to the peculiar location of GRB~221009A in the sky -- so close to the Galactic plane -- we were able to assess the polarization properties of the prompt emission in the same observation through the radiation scattered off the Galactic dust. Aside from providing valuable information about this peculiar event, this IXPE observation is a proof of observational feasibility for future nearby bright transient events. This, several years from now, could inspire new directions for the IXPE mission and widen IXPE's science portfolio to include fast-transient events, opening a new door for time-domain high-energy astrophysics.\\

\section*{Acknowledgements}
We thank Hintz Amenitsch for fruitful discussions on X-ray scattering at small angles. 
We also thank Joe Bright for pointing out a miscalculation of the $k$ parameter in an earlier version of this paper. A special acknowledgement goes to developers of the Slack team-work platform, which played a crucial role in enabling fast and efficient communication among several different teams.
We thank I. Negueruela for the careful optical polarization observations at the Nordic Optical Telescope. Based on observations made with the Nordic Optical Telescope, owned in collaboration by the University of Turku and Aarhus University, and operated jointly by Aarhus University, the University of Turku and the University of Oslo, representing Denmark, Finland and Norway, the University of Iceland and Stockholm University at the Observatorio del Roque de los Muchachos, La Palma, Spain, of the Instituto de Astrof\'isica de Canarias. The data presented here were obtained with ALFOSC, which is provided by the Instituto de Astrof\'isica de Andaluc\'ia (IAA) under a joint agreement with the University of Copenhagen and NOT.
MN acknowledges the support by NASA under award number 80GSFC21M0002. PV acknowledges support from NASA grant NNM11AA01A. IXPE-related research at Boston University is supported in part by U.S. National Science Foundation grant AST-2108622. SM and AT acknowledge financial support from the Italian MUR through grant PRIN 2017LJ39LM. \\
The Imaging X ray Polarimetry Explorer (IXPE) is a joint US and Italian mission.  The US contribution is supported by the National Aeronautics and Space Administration (NASA) and led and managed by its Marshall Space Flight Center (MSFC), with industry partner Ball Aerospace (contract NNM15AA18C).  The Italian contribution is supported by the Italian Space Agency (Agenzia Spaziale Italiana, ASI) through contract ASI-OHBI-2017-12-I.0, agreements ASI-INAF-2017-12-H0 and ASI-INFN-2017.13-H0, and its Space Science Data Center (SSDC) with agreements ASI-INAF-2022-14-HH.0 and ASI-INFN 2021-43-HH.0, and by the Istituto Nazionale di Astrofisica (INAF) and the Istituto Nazionale di Fisica Nucleare (INFN) in Italy.  This research used data products provided by the IXPE Team (MSFC, SSDC, INAF, and INFN) and distributed with additional software tools by the High-Energy Astrophysics Science Archive Research Center (HEASARC), at NASA Goddard Space Flight Center (GSFC).

\bibliography{_grb221009A}{}

\begin{thebibliography}{}
\expandafter\ifx\csname natexlab\endcsname\relax\def\natexlab#1{#1}\fi
\providecommand{\url}[1]{\href{#1}{#1}}
\providecommand{\dodoi}[1]{doi:~\href{http://doi.org/#1}{\nolinkurl{#1}}}
\providecommand{\doeprint}[1]{\href{http://ascl.net/#1}{\nolinkurl{http://ascl.net/#1}}}
\providecommand{\doarXiv}[1]{\href{https://arxiv.org/abs/#1}{\nolinkurl{https://arxiv.org/abs/#1}}}

\bibitem[{{Abbasi} {et~al.}(2022){Abbasi}, {Ackermann}, {Adams}, {Aguilar},
  {Ahlers}, {Ahrens}, {Alameddine}, {Alves}, {Amin}, {Andeen}, {Anderson},
  {Anton}, {Arg{\"u}elles}, {Ashida}, {Athanasiadou}, {Axani}, {Bai},
  {Balagopal}, {Barwick}, {Basu}, {Baur}, {Bay}, {Beatty}, {Becker}, {Becker
  Tjus}, {Beise}, {Bellenghi}, {Benda}, {BenZvi}, {Berley}, {Bernardini},
  {Besson}, {Binder}, {Bindig}, {Blaufuss}, {Blot}, {Boddenberg}, {Bontempo},
  {Book}, {Borowka}, {B{\"o}ser}, {Botner}, {B{\"o}ttcher}, {Bourbeau},
  {Bradascio}, {Braun}, {Brinson}, {Bron}, {Brostean-Kaiser}, {Burley},
  {Busse}, {Campana}, {Carnie-Bronca}, {Chen}, {Chen}, {Chirkin}, {Choi},
  {Clark}, {Clark}, {Classen}, {Coleman}, {Collin}, {Connolly}, {Conrad},
  {Coppin}, {Correa}, {Cowen}, {Cross}, {Dappen}, {Dave}, {Clercq}, {DeLaunay},
  {L{\'o}pez}, {Dembinski}, {Deoskar}, {Desai}, {Desiati}, {de Vries}, {de
  Wasseige}, {DeYoung}, {Diaz}, {D{\'\i}az-V{\'e}lez}, {Dittmer}, {Dujmovic},
  {DuVernois}, {Ehrhardt}, {Eller}, {Engel}, {Erpenbeck}, {Evans}, {Evenson},
  {Fan}, {Fazely}, {Fedynitch}, {Feigl}, {Fiedlschuster}, {Fienberg}, {Finley},
  {Fischer}, {Fox}, {Franckowiak}, {Friedman}, {Fritz}, {F{\"u}rst}, {Gaisser},
  {Gallagher}, {Ganster}, {Garcia}, {Garrappa}, {Gerhardt}, {Ghadimi},
  {Glaser}, {Glauch}, {Gl{\"u}senkamp}, {Goehlke}, {Gonzalez}, {Goswami},
  {Grant}, {Gr{\'e}goire}, {Griswold}, {G{\"u}nther}, {Gutjahr}, {Haack},
  {Hallgren}, {Halliday}, {Halve}, {Halzen}, {Ha Minh}, {Hanson}, {Hardin},
  {Harnisch}, {Haungs}, {Helbing}, {Henningsen}, {Hettinger}, {Hickford},
  {Hignight}, {Hill}, {Hill}, {Hoffman}, {Hoshina}, {Hou}, {Huber}, {Huber},
  {Hultqvist}, {H{\"u}nnefeld}, {Hussain}, {Hymon}, {In}, {Iovine}, {Ishihara},
  {Jansson}, {Japaridze}, {Jeong}, {Jin}, {Jones}, {Kang}, {Kang}, {Kang},
  {Kappes}, {Kappesser}, {Kardum}, {Karg}, {Karl}, {Karle}, {Katz}, {Kauer},
  {Kellermann}, {Kelley}, {Kheirandish}, {Kin}, {Kiryluk}, {Klein}, {Kochocki},
  {Koirala}, {Kolanoski}, {Kontrimas}, {K{\"o}pke}, {Kopper}, {Kopper},
  {Koskinen}, {Koundal}, {Kovacevich}, {Kowalski}, {Kozynets}, {Krupczak},
  {Kun}, {Kurahashi}, {Lad}, {Lagunas Gualda}, {Larson}, {Lauber}, {Lazar},
  {Lee}, {Leonard}, {Leszczy{\'n}ska}, {Li}, {Lincetto}, {Liu}, {Liubarska},
  {Lohfink}, {Mariscal}, {Lu}, {Lucarelli}, {Ludwig}, {Luszczak}, {Lyu}, {Ma},
  {Madsen}, {Mahn}, {Makino}, {Mancina}, {Mari{\c{s}}}, {Martinez-Soler},
  {Maruyama}, {McCarthy}, {McElroy}, {McNally}, {Mead}, {Meagher}, {Mechbal},
  {Medina}, {Meier}, {Meighen-Berger}, {Merckx}, {Micallef}, {Mockler},
  {Montaruli}, {Moore}, {Morse}, {Moulai}, {Mukherjee}, {Naab}, {Nagai},
  {Naumann}, {Necker}, {Nguyễn}, {Niederhausen}, {Nisa}, {Nowicki},
  {Obertacke Pollmann}, {Oehler}, {Oeyen}, {Olivas}, {O'Sullivan}, {Pandya},
  {Pankova}, {Park}, {Parker}, {Paudel}, {Paul}, {P{\'e}rez de los Heros},
  {Peters}, {Peterson}, {Philippen}, {Pieper}, {Pizzuto}, {Plum}, {Popovych},
  {Porcelli}, {Prado Rodriguez}, {Pries}, {Przybylski}, {Raab}, {Rack-Helleis},
  {Raissi}, {Rameez}, {Rawlins}, {Rea}, {Rechav}, {Rehman}, {Reichherzer},
  {Reimann}, {Renzi}, {Resconi}, {Reusch}, {Rhode}, {Richman}, {Riedel},
  {Roberts}, {Robertson}, {Roellinghoff}, {Rongen}, {Rott}, {Ruhe},
  {Ryckbosch}, {Rysewyk Cantu}, {Safa}, {Saffer}, {Salazar-Gallegos},
  {Sampathkumar}, {Sanchez Herrera}, {Sandrock}, {Santander}, {Sarkar},
  {Sarkar}, {Satalecka}, {Schaufel}, {Schieler}, {Schindler}, {Schmidt},
  {Schneider}, {Schneider}, {Schr{\"o}der}, {Schumacher}, {Schwefer},
  {Sclafani}, {Seckel}, {Seunarine}, {Sharma}, {Shefali}, {Shimizu}, {Silva},
  {Skrzypek}, {Smithers}, {Snihur}, {Soedingrekso}, {Sogaard}, {Soldin},
  {Spannfellner}, {Spiczak}, {Spiering}, {Stamatikos}, {Stanev}, {Stein},
  {Stettner}, {Stezelberger}, {St{\"u}rwald}, {Stuttard}, {Sullivan},
  {Taboada}, {Ter-Antonyan}, {Thwaites}, {Tilav}, {Tischbein}, {Tollefson},
  {T{\"o}nnis}, {Toscano}, {Tosi}, {Trettin}, {Tselengidou}, {Tung}, {Turcati},
  {Turcotte}, {Twagirayezu}, {Ty}, {Unland Elorrieta}, {Valtonen-Mattila},
  {Vandenbroucke}, {van Eijndhoven}, {Vannerom}, {van Santen},
  {Veitch-Michaelis}, {Verpoest}, {Walck}, {Wang}, {Watson}, {Weaver},
  {Weigel}, {Weindl}, {Weldert}, {Wendt}, {Werthebach}, {Weyrauch},
  {Whitehorn}, {Wiebusch}, {Willey}, {Williams}, {Wolf}, {Wrede}, {Wulff},
  {Xu}, {Yanez}, {Yildizci}, {Yoshida}, {Yu}, {Yuan}, {Zhang}, {Zhelnin},
  {Goldstein}, {Wood}, \& {Fermi Gamma-ray Burst Monitor}}]{ICgrbsearch}
{Abbasi}, R., {Ackermann}, M., {Adams}, J., {et~al.} 2022, \apj, 939, 116,
  \dodoi{10.3847/1538-4357/ac9785}

\bibitem[{{Abbott} {et~al.}(2017)}]{GW170817-GRB170817A}
{Abbott}, B.~P., {et~al.} 2017, \apjl, 828, \dodoi{10.3847/2041-8213}

\bibitem[{{Baldini} {et~al.}(2022){Baldini}, {Bucciantini}, {Lalla}, {Ehlert},
  {Manfreda}, {Negro}, {Omodei}, {Pesce-Rollins}, {Sgr{\`o}}, \&
  {Silvestri}}]{ixpeobssim}
{Baldini}, L., {Bucciantini}, N., {Lalla}, N.~D., {et~al.} 2022, SoftwareX, 19,
  101194, \dodoi{10.1016/j.softx.2022.101194}

\bibitem[{{Bellazzini} {et~al.}(2006){Bellazzini}, {Angelini}, {Baldini},
  {Bitti}, {Brez}, {Cavalca}, {Del Prete}, {Kuss}, {Latronico}, \&
  {Omodei}}]{BellazziniGPD}
{Bellazzini}, R., {Angelini}, F., {Baldini}, L., {et~al.} 2006, Nuclear
  Instruments and Methods in Physics Research A, 560, 425,
  \dodoi{10.1016/j.nima.2006.01.046}

\bibitem[{{Birenbaum} \& {Bromberg}(2021)}]{2021MNRAS.506.4275B}
{Birenbaum}, G., \& {Bromberg}, O. 2021, \mnras, 506, 4275,
  \dodoi{10.1093/mnras/stab193610.48550/arXiv.2105.04574}

\bibitem[{{Burns} {et~al.}(2021){Burns}, {Svinkin}, {Hurley}, {Wadiasingh},
  {Negro}, {Younes}, {Hamburg}, {Ridnaia}, {Cook}, {Cenko}, {Aloisi}, {Ashton},
  {Baring}, {Briggs}, {Christensen}, {Frederiks}, {Goldstein}, {Hui}, {Kaplan},
  {Kasliwal}, {Kocevski}, {Roberts}, {Savchenko}, {Tohuvavohu}, {Veres}, \&
  {Wilson-Hodge}}]{Burns2021}
{Burns}, E., {Svinkin}, D., {Hurley}, K., {et~al.} 2021, \apjl, 907, L28,
  \dodoi{10.3847/2041-8213/abd8c8}

\bibitem[{{Burrows} {et~al.}(2005){Burrows}, {Hill}, {Nousek}, {Kennea},
  {Wells}, {Osborne}, {Abbey}, {Beardmore}, {Mukerjee}, {Short}, {Chincarini},
  {Campana}, {Citterio}, {Moretti}, {Pagani}, {Tagliaferri}, {Giommi},
  {Capalbi}, {Tamburelli}, {Angelini}, {Cusumano}, {Br{\"a}uninger}, {Burkert},
  \& {Hartner}}]{Swift-XRT}
{Burrows}, D.~N., {Hill}, J.~E., {Nousek}, J.~A., {et~al.} 2005, \ssr, 120,
  165, \dodoi{10.1007/s11214-005-5097-2}

\bibitem[{{Chattopadhyay} {et~al.}(2019){Chattopadhyay}, {Vadawale}, {Aarthy},
  {Mithun}, {Chand}, {Ratheesh}, {Basak}, {Rao}, {Bhalerao}, {Mate}, {Arvind},
  {Sharma}, \& {Bhattacharya}}]{2019ApJ...884..123C}
{Chattopadhyay}, T., {Vadawale}, S.~V., {Aarthy}, E., {et~al.} 2019, \apj, 884,
  123, \dodoi{10.3847/1538-4357/ab40b7}

\bibitem[{{Coburn} \& {Boggs}(2003)}]{2003Natur.423..415C}
{Coburn}, W., \& {Boggs}, S.~E. 2003, \nat, 423, 415,
  \dodoi{10.1038/nature01612}

\bibitem[{{Costa} \& {Frontera}(2011)}]{2011NCimR..34..585C}
{Costa}, E., \& {Frontera}, F. 2011, Nuovo Cimento Rivista Serie, 34, 585,
  \dodoi{10.1393/ncr/i2011-10069-0}

\bibitem[{{Costantini} \& {Corrales}(2022)}]{CostantiniCorrales}
{Costantini}, E., \& {Corrales}, L. 2022, arXiv e-prints, arXiv:2209.05261.
\newblock \doarXiv{2209.05261}

\bibitem[{{Covino} \& {Gotz}(2016)}]{2016A&AT...29..205C}
{Covino}, S., \& {Gotz}, D. 2016, Astronomical and Astrophysical Transactions,
  29, 205, \dodoi{10.48550/arXiv.1605.03588}

\bibitem[{{Dichiara} {et~al.}(2022){Dichiara}, {Gropp}, {Kennea}, {Kuin},
  {Lien}, {Marshall}, {Tohuvavohu}, \& {Williams}}]{GCNUVOTloc}
{Dichiara}, S., {Gropp}, J.~D., {Kennea}, J.~A., {et~al.} 2022, GCN, 32632, 1

\bibitem[{{Draine}(2003{\natexlab{a}})}]{Draine2003II}
{Draine}, B.~T. 2003{\natexlab{a}}, \apj, 598, 1026, \dodoi{10.1086/379123}

\bibitem[{{Draine}(2003{\natexlab{b}})}]{Draine2003I}
---. 2003{\natexlab{b}}, \apj, 598, 1017, \dodoi{10.1086/379118}

\bibitem[{Draine \& Bond(2004)}]{Draine_2004}
Draine, B.~T., \& Bond, N.~A. 2004, \apj, 617, 987, \dodoi{10.1086/425609}

\bibitem[{{Draine} \& {Lee}(1984)}]{DraineLee1984}
{Draine}, B.~T., \& {Lee}, H.~M. 1984, \apj, 285, 89, \dodoi{10.1086/162480}

\bibitem[{{D’Avanzo} {et~al.}(2022){D’Avanzo}, {Ferro}, {Brivio},
  {Bernardini}, {Fugazza}, {Campana}, \& other}]{GCNjetbreak}
{D’Avanzo}, P.~d., {Ferro}, M., {Brivio}, R., {et~al.} 2022, GCN, 32755, 1

\bibitem[{{Evans} {et~al.}(2007){Evans}, {Beardmore}, {Page}, {Tyler},
  {Osborne}, {Goad}, {O'Brien}, {Vetere}, {Racusin}, {Morris}, {Burrows},
  {Capalbi}, {Perri}, {Gehrels}, \& {Romano}}]{SwiftXRTauto}
{Evans}, P.~A., {Beardmore}, A.~P., {Page}, K.~L., {et~al.} 2007, \aap, 469,
  379, \dodoi{10.1051/0004-6361:20077530}

\bibitem[{{Fong} {et~al.}(2015){Fong}, {Berger}, {Margutti}, \&
  {Zauderer}}]{2015ApJ...815..102F}
{Fong}, W., {Berger}, E., {Margutti}, R., \& {Zauderer}, B.~A. 2015, \apj, 815,
  102, \dodoi{10.1088/0004-637X/815/2/102}

\bibitem[{{Frederiks} {et~al.}(2022){Frederiks}, {Lysenko}, {Ridnaia},
  {Svinkin}, {Tsvetkova}, {Ulanov}, \& {Cline}}]{GCNkonuswind}
{Frederiks}, D., {Lysenko}, A., {Ridnaia}, A., {et~al.} 2022, GCN, 32668, 1

\bibitem[{{Frontera}(2019)}]{2019RLSFN..30S.171F}
{Frontera}, F. 2019, Rendiconti Lincei. Scienze Fisiche e Naturali, 30, 171,
  \dodoi{10.1007/s12210-019-00766-z}

\bibitem[{{Galama} {et~al.}(1998){Galama}, {Vreeswijk}, {van Paradijs},
  {Kouveliotou}, {Augusteijn}, {B{\"o}hnhardt}, {Brewer}, {Doublier},
  {Gonzalez}, {Leibundgut}, {Lidman}, {Hainaut}, {Patat}, {Heise}, {in't Zand},
  {Hurley}, {Groot}, {Strom}, {Mazzali}, {Iwamoto}, {Nomoto}, {Umeda},
  {Nakamura}, {Young}, {Suzuki}, {Shigeyama}, {Koshut}, {Kippen}, {Robinson},
  {de Wildt}, {Wijers}, {Tanvir}, {Greiner}, {Pian}, {Palazzi}, {Frontera},
  {Masetti}, {Nicastro}, {Feroci}, {Costa}, {Piro}, {Peterson}, {Tinney},
  {Boyle}, {Cannon}, {Stathakis}, {Sadler}, {Begam}, \&
  {Ianna}}]{1998Natur.395..670G}
{Galama}, T.~J., {Vreeswijk}, P.~M., {van Paradijs}, J., {et~al.} 1998, \nat,
  395, 670, \dodoi{10.1038/27150}

\bibitem[{{Ghisellini} \& {Lazzati}(1999)}]{1999MNRAS.309L...7G}
{Ghisellini}, G., \& {Lazzati}, D. 1999, \mnras, 309, L7,
  \dodoi{10.1046/j.1365-8711.1999.03025.x}

\bibitem[{{Gill} {et~al.}(2021){Gill}, {Kole}, \& {Granot}}]{Gill2021}
{Gill}, R., {Kole}, M., \& {Granot}, J. 2021, Galaxies, 9, 82,
  \dodoi{10.3390/galaxies9040082}

\bibitem[{{Granot}(2003)}]{Granot+03highpol021206}
{Granot}, J. 2003, \apjl, 596, L17, \dodoi{10.1086/379110}

\bibitem[{{Granot} \& {K{\"o}nigl}(2003)}]{2003ApJ...594L..83G}
{Granot}, J., \& {K{\"o}nigl}, A. 2003, \apjl, 594, L83, \dodoi{10.1086/378733}

\bibitem[{{Granot} \& {Sari}(2002)}]{Granot+02Dabreaks}
{Granot}, J., \& {Sari}, R. 2002, \apj, 568, 820, \dodoi{10.1086/338966}

\bibitem[{Hayakawa(1970)}]{10.1143/PTP.43.1224}
Hayakawa, S. 1970, Progress of Theoretical Physics, 43, 1224,
  \dodoi{10.1143/PTP.43.1224}

\bibitem[{{Hovatta} {et~al.}(2016){Hovatta}, {Lindfors}, {Blinov}, {Pavlidou},
  {Nilsson}, {Kiehlmann}, {Angelakis}, {Fallah Ramazani}, {Liodakis},
  {Myserlis}, {Panopoulou}, \& {Pursimo}}]{Hovatta2016}
{Hovatta}, T., {Lindfors}, E., {Blinov}, D., {et~al.} 2016, \aap, 596, A78,
  \dodoi{10.1051/0004-6361/201628974}

\bibitem[{{Huang} {et~al.}(2022){Huang}, {Hu}, {Chen}, {Zha}, {Liu}, {Yao},
  {Cao}, \& {Experiment}}]{2022GCN.32677....1H}
{Huang}, Y., {Hu}, S., {Chen}, S., {et~al.} 2022, GCN, 32677, 1

\bibitem[{{Hulsman}(2020)}]{hulsman2020polar}
{Hulsman}, J. 2020, in Society of Photo-Optical Instrumentation Engineers
  (SPIE) Conference Series, Vol. 11444, Society of Photo-Optical
  Instrumentation Engineers (SPIE) Conference Series, 114442V,
  \dodoi{10.1117/12.2559374}

\bibitem[{{Izzo} {et~al.}(2022){Izzo}, {Saccardi}, {Fynbo}, {Palmerio},
  {Malesani}, {Agui Fernandez}, {et~al.}}]{VLTGCN}
{Izzo}, L., {Saccardi}, A., {Fynbo}, J.~P.~U., {et~al.} 2022, GCN, 32765, 1

\bibitem[{{Kennea} {et~al.}(2022){Kennea}, {Tohuvavohu}, {Osborne}, {Page},
  {Beardmore}, {Melandri}, {Sbarrato}, {D'Avanzo}, {Burrows}, \&
  {Bissaldi}}]{SwiftRef1}
{Kennea}, J.~A., {Tohuvavohu}, A., {Osborne}, J.~P., {et~al.} 2022, GCN, 32651,
  1

\bibitem[{Kislat {et~al.}(2015)Kislat, Clark, Beilicke, \&
  Krawczynski}]{Stokes}
Kislat, F., Clark, B., Beilicke, M., \& Krawczynski, H. 2015, Astroparticle
  Physics, 68, 45, \dodoi{https://doi.org/10.1016/j.astropartphys.2015.02.007}

\bibitem[{Kole {et~al.}(2020)Kole, De~Angelis, Berlato, Burgess, Gauvin,
  Greiner, Hajdas, Li, Li, Pollo, {et~al.}}]{kole2020polar}
Kole, M., De~Angelis, N., Berlato, F., {et~al.} 2020, \aap, 644, A124

\bibitem[{Kumar \& Zhang(2015)}]{kumar2015physics}
Kumar, P., \& Zhang, B. 2015, Physics Reports, 561, 1

\bibitem[{Kuwata {et~al.}(2022)Kuwata, Toma, Kimura, Tomita, \&
  Shimoda}]{kuwata2022synchrotron}
Kuwata, A., Toma, K., Kimura, S.~S., Tomita, S., \& Shimoda, J. 2022, arXiv
  preprint arXiv:2208.09242

\bibitem[{{Laor} \& {Draine}(1993)}]{LaorDraine1993}
{Laor}, A., \& {Draine}, B.~T. 1993, \apj, 402, 441, \dodoi{10.1086/172149}

\bibitem[{{Lesage} {et~al.}(2022){Lesage}, {Veres}, {Roberts}, {Burns}, \&
  {Bissaldi}}]{GCNLesage}
{Lesage}, S., {Veres}, P., {Roberts}, O., {Burns}, E., \& {Bissaldi}, E. 2022,
  GCN, 32642, 1

\bibitem[{{Levan} {et~al.}(2022){Levan}, {Barclay}, {Burns}, {Cenko},
  {Chrimes}, {D’Avanzo}, {et~al.}}]{WebbGCN}
{Levan}, A., {Barclay}, T., {Burns}, E., {et~al.} 2022, GCN, 32821, 1

\bibitem[{{Lindfors} {et~al.}(2022){Lindfors}, {Nilsson}, {Liodakis}, \&
  {Negueruela}}]{GCNoptpol}
{Lindfors}, E., {Nilsson}, K., {Liodakis}, I., \& {Negueruela}, I. 2022, GCN,
  32995, 1

\bibitem[{{Liu} {et~al.}(2022){Liu}, {Zhang}, \& {Wang}}]{2022arXiv221114200L}
{Liu}, R.-Y., {Zhang}, H.-M., \& {Wang}, X.-Y. 2022, arXiv e-prints,
  arXiv:2211.14200.
\newblock \doarXiv{2211.14200}

\bibitem[{Lumb {et~al.}(2002)Lumb, Warwick, Page, \& Luca}]{Lumb_2002}
Lumb, D.~H., Warwick, R.~S., Page, M., \& Luca, A.~D. 2002, Astronomy {\&}
  Astrophysics, 389, 93, \dodoi{10.1051/0004-6361:20020531}

\bibitem[{{Lyutikov} {et~al.}(2003){Lyutikov}, {Pariev}, \&
  {Blandford}}]{Lyutikov+03pol}
{Lyutikov}, M., {Pariev}, V.~I., \& {Blandford}, R.~D. 2003, \apj, 597, 998,
  \dodoi{10.1086/378497}

\bibitem[{{McConnell}(2017)}]{McConnell}
{McConnell}, M.~L. 2017, \nar, 76, 1, \dodoi{10.1016/j.newar.2016.11.001}

\bibitem[{{McConnell} {et~al.}(2021){McConnell}, {Baring}, {Bloser}, {Briggs},
  {Ertley}, {Fletcher}, {Gaskin}, {Gelmis}, {Goldstein}, {Grove}, {Hartmann},
  {Hui}, {Jenke}, {Kippen}, {Kislat}, {Kocevski}, {Kole}, {Krizmanic},
  {Legere}, {Littenberg}, {Martin}, {McBreen}, {McQueen}, {Meegan}, {O{\~n}ate
  Melecio}, {Pearce}, {Preece}, {Produit}, {Ryan}, {Sturner}, {Veres},
  {Vestrand}, {Wilson-Hodge}, \& {Zhang}}]{mcconnell2021large}
{McConnell}, M.~L., {Baring}, M., {Bloser}, P., {et~al.} 2021, in Society of
  Photo-Optical Instrumentation Engineers (SPIE) Conference Series, Vol. 11821,
  UV, X-Ray, and Gamma-Ray Space Instrumentation for Astronomy XXII, ed. O.~H.
  {Siegmund}, 118210P, \dodoi{10.1117/12.2594737}

\bibitem[{{Miralda-Escud{\'e}}(1999)}]{MiraldaEscude1999}
{Miralda-Escud{\'e}}, J. 1999, \apj, 512, 21, \dodoi{10.1086/306767}

\bibitem[{Mundell {et~al.}(2013)Mundell, Kopa{\v{c}}, Arnold, Steele, Gomboc,
  Kobayashi, Harrison, Smith, Guidorzi, Virgili, {et~al.}}]{mundell2013highly}
Mundell, C., Kopa{\v{c}}, D., Arnold, D., {et~al.} 2013, Nature, 504, 119

\bibitem[{{Neckel} \& {Klare}(1980)}]{NeckelKlare1980}
{Neckel}, T., \& {Klare}, G. 1980, \aaps, 42, 251

\bibitem[{{Negro} {et~al.}(2022){Negro}, {Manfreda}, \&
  {Omodei}}]{GCNIXPEobsplan}
{Negro}, M., {Manfreda}, A., \& {Omodei}, N. 2022, GCN, 32690, 1

\bibitem[{{Nilsson} {et~al.}(2018){Nilsson}, {Lindfors}, {Takalo}, {Reinthal},
  {Berdyugin}, {Sillanp{\"a}{\"a}}, {Ciprini}, {Halkola}, {Hein{\"a}m{\"a}ki},
  {Hovatta}, {Kadenius}, {Nurmi}, {Ostorero}, {Pasanen}, {Rekola}, {Saarinen},
  {Sainio}, {Tuominen}, {Villforth}, {Vornanen}, \& {Zaprudin}}]{Nilsson2018}
{Nilsson}, K., {Lindfors}, E., {Takalo}, L.~O., {et~al.} 2018, \aap, 620, A185,
  \dodoi{10.1051/0004-6361/201833621}

\bibitem[{{O'Dell} {et~al.}(2019){O'Dell}, {Attin{\`a}}, {Baldini},
  {Barbanera}, {Baumgartner}, {Bellazzini}, {Bladt}, {Bongiorno}, {Brez},
  {Cavazzuti}, {Citraro}, {Costa}, {Deininger}, {Del Monte}, {Dietz}, {Di
  Lalla}, {Donnarumma}, {Elsner}, {Fabiani}, {Ferrazzoli}, {Guy}, {Kalinowski},
  {Kaspi}, {Kelley}, {Kolodziejczak}, {Latronico}, {Lefevre}, {Lucchesi},
  {Manfreda}, {Marshall}, {Masciarelli}, {Matt}, {Minuti}, {Muleri}, {Nasimi},
  {Nuti}, {Orsini}, {Osborne}, {Perri}, {Pesce-Rollins}, {Peterson},
  {Pinchera}, {Puccetti}, {Ramsey}, {Ratheesh}, {Romani}, {Santoli},
  {Sciortino}, {Sgr{\`o}}, {Smith}, {Spandre}, {Soffitta}, {Tennant}, {Tobia},
  {Trois}, {Wedmore}, {Weisskopf}, {Xie}, {Zanetti}, {Alexander}, {Allen},
  {Amici}, {Antoniak}, {Bonino}, {Borotto}, {Breeding}, {Brienza}, {Bygott},
  {Caporale}, {Cardelli}, {Ceccanti}, {Centrone}, {Di Persio}, {Evangelista},
  {Ferrie}, {Footdale}, {Forsyth}, {Foster}, {Gunji}, {Gurnee}, {Hibbard},
  {Johnson}, {Kelly}, {Kilaru}, {La Monaca}, {Le Roy}, {Loffredo}, {Magazzu},
  {Marengo}, {Marrocchesi}, {Massaro}, {McCracken}, {McEachen}, {Mereu},
  {Mitchell}, {Mitsuishi}, {Morbidini}, {Mosti}, {Negro}, {Oppedisano},
  {Pacheco}, {Paggi}, {Pavelitz}, {Pentz}, {Piazzola}, {Porter}, {Profeti},
  {Ranganathan}, {Rankin}, {Root}, {Rubini}, {Ruswick}, {Sanchez}, {Scalise},
  {Schindhelm}, {Speegle}, {Tamagawa}, {Tardiola}, {Walden}, {Weddendorf}, \&
  {Welch}}]{IXPEOverviewII}
{O'Dell}, S.~L., {Attin{\`a}}, P., {Baldini}, L., {et~al.} 2019, in Society of
  Photo-Optical Instrumentation Engineers (SPIE) Conference Series, Vol. 11118,
  UV, X-Ray, and Gamma-Ray Space Instrumentation for Astronomy XXI, ed. O.~H.
  {Siegmund}, 111180V, \dodoi{10.1117/12.2530646}

\bibitem[{{Pedreira} {et~al.}(2022){Pedreira}, {Fraija}, {Dichiara}, {Veres},
  {Dainotti}, {Galvan-Gamez}, {Becerra}, \& {Betancourt
  Kamenetskaia}}]{2022arXiv221012904P}
{Pedreira}, A.~C. C. d. E.~S., {Fraija}, N., {Dichiara}, S., {et~al.} 2022,
  arXiv e-prints, arXiv:2210.12904.
\newblock \doarXiv{2210.12904}

\bibitem[{{Pillera} {et~al.}(2022){Pillera}, {Bissaldi}, {Omodei}, {La Mura},
  \& {Longo}}]{ATelLATrefined}
{Pillera}, R., {Bissaldi}, E., {Omodei}, N., {La Mura}, G., \& {Longo}, F.
  2022, The Astronomer's Telegram, 15656, 1

\bibitem[{Rossi {et~al.}(2004)Rossi, Lazzati, Salmonson, \&
  Ghisellini}]{rossi2004polarization}
Rossi, E.~M., Lazzati, D., Salmonson, J.~D., \& Ghisellini, G. 2004, \mnras,
  354, 86

\bibitem[{{Rybicki} \& {Lightman}(1979)}]{rybicki79}
{Rybicki}, G.~B., \& {Lightman}, A.~P. 1979, {Radiative processes in
  astrophysics} (New York, Wiley-Interscience, 1979.~393 p.)

\bibitem[{{Sari}(1999)}]{1999ApJ...524L..43S}
{Sari}, R. 1999, \apjl, 524, L43,
  \dodoi{10.1086/31229410.48550/arXiv.astro-ph/9906503}

\bibitem[{{Sari} \& {M{\'e}sz{\'a}ros}(2000)}]{Sari+00refresh}
{Sari}, R., \& {M{\'e}sz{\'a}ros}, P. 2000, \apjl, 535, L33,
  \dodoi{10.1086/312689}

\bibitem[{{Shimoda} \& {Toma}(2021)}]{2021ApJ...913...58S}
{Shimoda}, J., \& {Toma}, K. 2021, \apj, 913, 58,
  \dodoi{10.3847/1538-4357/abf2c2}

\bibitem[{{Soffitta} {et~al.}(2020){Soffitta}, {Attin{\`a}}, {Baldini},
  {Barbanera}, {Baumgartner}, {Bellazzini}, {Bladt}, {Bongiorno}, {Brez},
  {Castellano}, {Carpentiero}, {Castronuovo}, {Cavalli}, {Cavazzuti},
  {D'Amico}, {Citraro}, {Costa}, {Deininger}, {D'Alba}, {Del Monte}, {Diets},
  {Di Lalla}, {Di Marco}, {Di Persio}, {Donnarumma}, {Elsner}, {Fabiani},
  {Ferrazzoli}, {Guy}, {Kalinowski}, {Kolodziejczak}, {Latronico}, {Lefevre},
  {Lorenzi}, {Lucchesi}, {Maldera}, {Manfreda}, {Mangraviti}, {Marshall},
  {Masciarelli}, {Matt}, {Minuti}, {Muleri}, {Nasimi}, {Negri}, {Nuti},
  {Orsini}, {Osborne}, {Pilia}, {Perri}, {Pesce-Rollins}, {Peterson},
  {Pinchera}, {Puccetti}, {Ramsey}, {Ratheesh}, {Romani}, {Sarra}, {Santoli},
  {Sciortino}, {Sgr{\`o}}, {Smith}, {Spandre}, {Tennant}, {Tobia}, {Trois},
  {Vimercati}, {Wedmnore}, {Weisskopf}, {Xie}, {Zanetti}, {Alexander}, {Allen},
  {Amici}, {Antonelli}, {Antoniak}, {Bachetti}, {Bonino}, {Borotto},
  {Breeding}, {Brienza}, {Bygott}, {Cardelli}, {Ceccanti}, {Centrone},
  {Evangelista}, {Ferrie}, {Forsyth}, {Foster}, {Gurnee}, {Hibbard}, {Johnson},
  {Kelly}, {Kilaru}, {La Monaca}, {Le Roy}, {Loffredo}, {Magazzu'}, {Marengo},
  {Marrocchesi}, {Massaro}, {Morbidini}, {McCracken}, {McEachen}, {Mereu},
  {Mitchell}, {Mitsuishi}, {Mosti}, {Nigro}, {Nuti}, {Oppedisano}, {Pacheco},
  {Paggi}, {Pavelitz}, {Pentz}, {Piazzolla}, {Porter}, {Profeti},
  {Ranganathan}, {Rankin}, {Root}, {Rubini}, {Ruswick}, {Sanchez}, {Scalise},
  {Schindhelm}, {Speegle}, {Tamagawa}, {Tardiola}, {Walden}, {Weddendorf},
  {Welch}, {Head}, {Gray}, {Mize}, {O'Dell}, {Schroeder}, {Thomas}, {Bagget},
  {Dolan}, {Ferrant}, {Footdale}, {Garelick}, {Johnson}, \&
  {Seek}}]{IXPEOverviewIII}
{Soffitta}, P., {Attin{\`a}}, P., {Baldini}, L., {et~al.} 2020, in Society of
  Photo-Optical Instrumentation Engineers (SPIE) Conference Series, Vol. 11444,
  Society of Photo-Optical Instrumentation Engineers (SPIE) Conference Series,
  1144462, \dodoi{10.1117/12.2567001}

\bibitem[{{Stringer} \& {Lazzati}(2020)}]{2020ApJ...892..131S}
{Stringer}, E., \& {Lazzati}, D. 2020, \apj, 892, 131,
  \dodoi{10.3847/1538-4357/ab76d2}

\bibitem[{{Tamagawa} {et~al.}(2003){Tamagawa}, {Kawai}, {Yoshida}, {Shirasaki},
  {Torii}, {Sakamoto}, {Suzuki}, {Urata}, {Nakagawa}, {Takahashi}, {Sato},
  {Matsuoka}, {Vanderspek}, {Crew}, {Doty}, {Villasenor}, {Butler}, {Ricker},
  {Lamb}, {Graziani}, {Donaghy}, {Fenimore}, {Galassi}, {Yamauchi},
  {Takagishi}, \& {Hatsukade}}]{2003ICRC....5.2741T}
{Tamagawa}, T., {Kawai}, N., {Yoshida}, A., {et~al.} 2003, in International
  Cosmic Ray Conference, Vol.~5, International Cosmic Ray Conference, 2741

\bibitem[{{Tiengo} \& {Mereghetti}(2006)}]{TiengoMereghetti2006}
{Tiengo}, A., \& {Mereghetti}, S. 2006, \aap, 449, 203,
  \dodoi{10.1051/0004-6361:20054162}

\bibitem[{{Tiengo} {et~al.}(2022){Tiengo}, {Pintore}, {Mereghetti}, \&
  {Salvaterra}}]{ATELTiengoSwiftRings}
{Tiengo}, A., {Pintore}, F., {Mereghetti}, S., \& {Salvaterra}, R. 2022, The
  Astronomer's Telegram, 15661, 1

\bibitem[{{Toma} {et~al.}(2009){Toma}, {Sakamoto}, {Zhang}, {Hill},
  {McConnell}, {Bloser}, {Yamazaki}, {Ioka}, \& {Nakamura}}]{Toma+09pol}
{Toma}, K., {Sakamoto}, T., {Zhang}, B., {et~al.} 2009, \apj, 698, 1042,
  \dodoi{10.1088/0004-637X/698/2/1042}

\bibitem[{Tomsick {et~al.}(2021)Tomsick, Boggs, Zoglauer, Wulf, Mitchell,
  Phlips, Sleator, Brandt, Shih, Roberts, {et~al.}}]{tomsick2021compton}
Tomsick, J.~A., Boggs, S.~E., Zoglauer, A., {et~al.} 2021, arXiv preprint
  arXiv:2109.10403

\bibitem[{{Ugarte Postigo} {et~al.}(2022){Ugarte Postigo}, {Izzo}, {Pugliese},
  {Xu}, {Schneider}, {Fynbo}, {Tanvir}, {Malesani}, {Saccardi}, {Kann},
  {Wiersema}, {Gompertz}, {Thoene}, \& {Levan}}]{GCNredshift}
{Ugarte Postigo}, A.~d., {Izzo}, L., {Pugliese}, G., {et~al.} 2022, GCN, 32648,
  1

\bibitem[{Urata {et~al.}(2019)Urata, Toma, Huang, Asada, Nagai, Takahashi,
  Petitpas, Tashiro, \& Yamaoka}]{urata2019first}
Urata, Y., Toma, K., Huang, K., {et~al.} 2019, The Astrophysical Journal
  Letters, 884, L58

\bibitem[{{Urata} {et~al.}(2022){Urata}, {Toma}, {Covino}, {Wiersema}, {Huang},
  {Shimoda}, {Kuwata}, {Nagao}, {Asada}, {Nagai}, {Takahashi}, {Chung},
  {Petitpas}, {Yamaoka}, {Izzo}, {Fynbo}, {de Ugarte Postigo}, {Arabsalmani},
  \& {Tashiro}}]{Urata2022radiopol}
{Urata}, Y., {Toma}, K., {Covino}, S., {et~al.} 2022, arXiv e-prints,
  arXiv:2212.05085.
\newblock \doarXiv{2212.05085}

\bibitem[{{Veres} {et~al.}(2022){Veres}, {Burns}, {Bissaldi}, {Lesage}, \&
  {Roberts}}]{GCNfermiGBMtrigger}
{Veres}, P., {Burns}, E., {Bissaldi}, E., {Lesage}, S., \& {Roberts}, O. 2022,
  GCN, 32636, 1

\bibitem[{{Vianello} {et~al.}(2015){Vianello}, {Lauer}, {Younk}, {Tibaldo},
  {Burgess}, {Ayala}, {Harding}, {Hui}, {Omodei}, \& {Zhou}}]{3ml}
{Vianello}, G., {Lauer}, R.~J., {Younk}, P., {et~al.} 2015, arXiv e-prints,
  arXiv:1507.08343.
\newblock \doarXiv{1507.08343}

\bibitem[{{Weisskopf} {et~al.}(2022){Weisskopf}, {Soffitta}, {Baldini},
  {Ramsey}, {O'Dell}, {Romani}, {Matt}, {Deininger}, {Baumgartner},
  {Bellazzini}, {Costa}, {Kolodziejczak}, {Latronico}, {Marshall}, {Muleri},
  {Bongiorno}, {Tennant}, {Bucciantini}, {Dovciak}, {Marin}, {Marscher},
  {Poutanen}, {Slane}, {Turolla}, {Kalinowski}, {Di Marco}, {Fabiani},
  {Minuti}, {La Monaca}, {Pinchera}, {Rankin}, {Sgro'}, {Trois}, {Xie},
  {Alexander}, {Allen}, {Amici}, {Andersen}, {Antonelli}, {Antoniak},
  {Attin{\`a}}, {Barbanera}, {Bachetti}, {Baggett}, {Bladt}, {Brez}, {Bonino},
  {Boree}, {Borotto}, {Breeding}, {Brienza}, {Bygott}, {Caporale}, {Cardelli},
  {Carpentiero}, {Castellano}, {Castronuovo}, {Cavalli}, {Cavazzuti},
  {Ceccanti}, {Centrone}, {Citraro}, {D'Amico}, {D'Alba}, {Di Gesu}, {Del
  Monte}, {Dietz}, {Di Lalla}, {Persio}, {Dolan}, {Donnarumma}, {Evangelista},
  {Ferrant}, {Ferrazzoli}, {Ferrie}, {Footdale}, {Forsyth}, {Foster},
  {Garelick}, {Gunji}, {Gurnee}, {Head}, {Hibbard}, {Johnson}, {Kelly},
  {Kilaru}, {Lefevre}, {Roy}, {Loffredo}, {Lorenzi}, {Lucchesi}, {Maddox},
  {Magazzu}, {Maldera}, {Manfreda}, {Mangraviti}, {Marengo}, {Marrocchesi},
  {Massaro}, {Mauger}, {McCracken}, {McEachen}, {Mize}, {Mereu}, {Mitchell},
  {Mitsuishi}, {Morbidini}, {Mosti}, {Nasimi}, {Negri}, {Negro}, {Nguyen},
  {Nitschke}, {Nuti}, {Onizuka}, {Oppedisano}, {Orsini}, {Osborne}, {Pacheco},
  {Paggi}, {Painter}, {Pavelitz}, {Pentz}, {Piazzolla}, {Perri},
  {Pesce-Rollins}, {Peterson}, {Pilia}, {Profeti}, {Puccetti}, {Ranganathan},
  {Ratheesh}, {Reedy}, {Root}, {Rubini}, {Ruswick}, {Sanchez}, {Sarra},
  {Santoli}, {Scalise}, {Sciortino}, {Schroeder}, {Seek}, {Sosdian}, {Spandre},
  {Speegle}, {Tamagawa}, {Tardiola}, {Tobia}, {Thomas}, {Valerie}, {Vimercati},
  {Walden}, {Weddendorf}, {Wedmore}, {Welch}, {Zanetti}, \&
  {Zanetti}}]{IXPE_calibration}
{Weisskopf}, M.~C., {Soffitta}, P., {Baldini}, L., {et~al.} 2022, Journal of
  Astronomical Telescopes, Instruments, and Systems, 8, 026002,
  \dodoi{10.1117/1.JATIS.8.2.026002}

\bibitem[{{Willingale} {et~al.}(2013){Willingale}, {Starling}, {Beardmore},
  {Tanvir}, \& {O'Brien}}]{Willingale2003}
{Willingale}, R., {Starling}, R.~L.~C., {Beardmore}, A.~P., {Tanvir}, N.~R., \&
  {O'Brien}, P.~T. 2013, \mnras, 431, 394, \dodoi{10.1093/mnras/stt175}

\bibitem[{Woosley \& Bloom(2006)}]{woosley2006supernova}
Woosley, S., \& Bloom, J. 2006, Annu. Rev. Astron. Astrophys., 44, 507

\bibitem[{{Xiao} {et~al.}(2022){Xiao}, {Krucker}, \& R.}]{GCNstix}
{Xiao}, H., {Krucker}, S., \& R., D. 2022, GCN, 32661, 1

\bibitem[{Zhang(2018)}]{zhang2018physics}
Zhang, B. 2018, The physics of gamma-ray bursts (Cambridge University Press)

\end{thebibliography}
\bibliographystyle{aasjournal}

\appendix

\section{Background Handling} \label{app:A}
The vast majority of the background events for the IXPE telescope are instrumental in origin, e.g., cosmic rays that trigger the detector and are reconstructed as photons by the reconstruction algorithm. On top of those events, a weak X-ray background is also expected \citep{Lumb_2002}. A fraction of the background events can be identified and rejected by looking at the track morphology. The remaining fraction is indistinguishable from genuine X-ray-triggered events and constitutes a residual background that must be treated statistically.

We adopted a two-step strategy to remove the background events: first we apply a background rejection and then a background subtraction, as detailed here below.

\begin{figure}[ht]
    \centering
    \includegraphics[width=\textwidth]{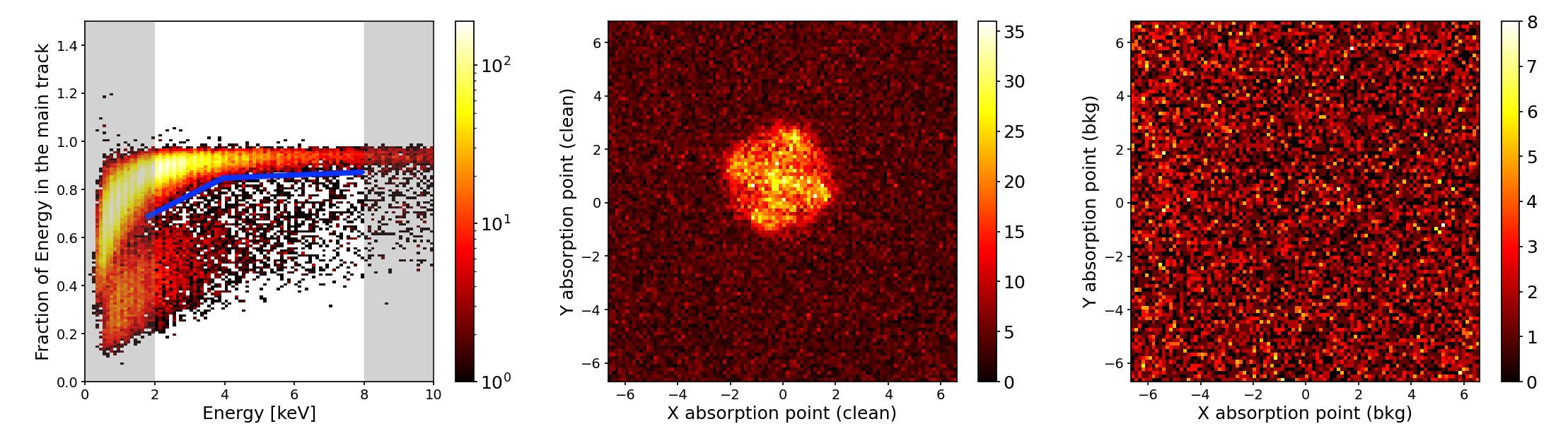}
    \caption{Left: background rejection cut based on the energy fraction contained in the main cluster as a function of the reconstructed energy. The events that are below the blue line are removed. The gray shaded areas mark the energy ranges outside the fiducial range for IXPE data analyses. Center and right: diagnostic maps produced to check the efficiency of the cut: total events map (middle) and rejected events map (right). The maps are in detector coordinates, so the central source appears blurred following a pattern caused by deliberate dithering of the satellite \citep{IXPE_calibration}. No apparent residuals of the central source are visible in the background map on the right. }
    \label{fig:bkg_rej}
\end{figure}

\paragraph{Background rejection}
Typical X-ray events, compared to charge cosmic-ray events, display a higher fraction of energy deposit associated to the main track\footnote{The first step of IXPE reconstruction algorithm is a clustering stage meant to identify a group of adjacent pixels that recorded a charge value above a noise-rejection threshold. For typical X-ray-induced events, the charge deposit associated with the photo-electron produces a single main cluster, while additional, spurious clusters are caused by noise fluctuations. The reconstruction algorithm assumes the cluster with the higher charge deposit to correspond to the main track. On the other side, charged cosmic-ray-induced events may produce several, disconnected clusters of charge inside the detector with similar energy deposit. It follows that cosmic rays display a lower fraction of energy deposit associated to the main track with respect to the total energy of the event.} over the total energy of the event.
Based on such a difference, a rejection cut can be devised to remove the portion of events that are of clear cosmic-ray nature. The left panel of Figure~\ref{fig:bkg_rej} illustrates the energy fraction deposited in the main track of the event as a function of the reconstructed energy: the blue line marks the rejection event cut we apply for events between 2 and 8 keV. 

\begin{figure}[t]
    \centering
    \includegraphics[width=0.32\textwidth]{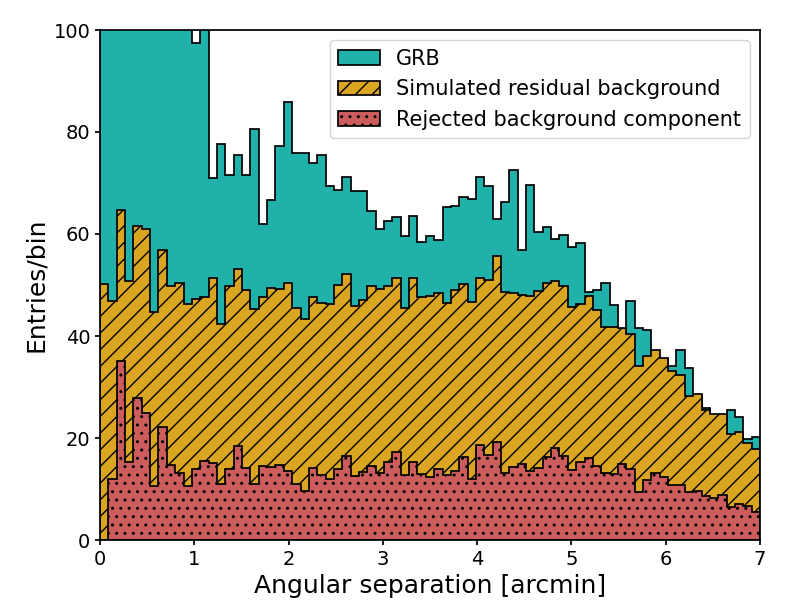}
    \includegraphics[width=0.32\textwidth]{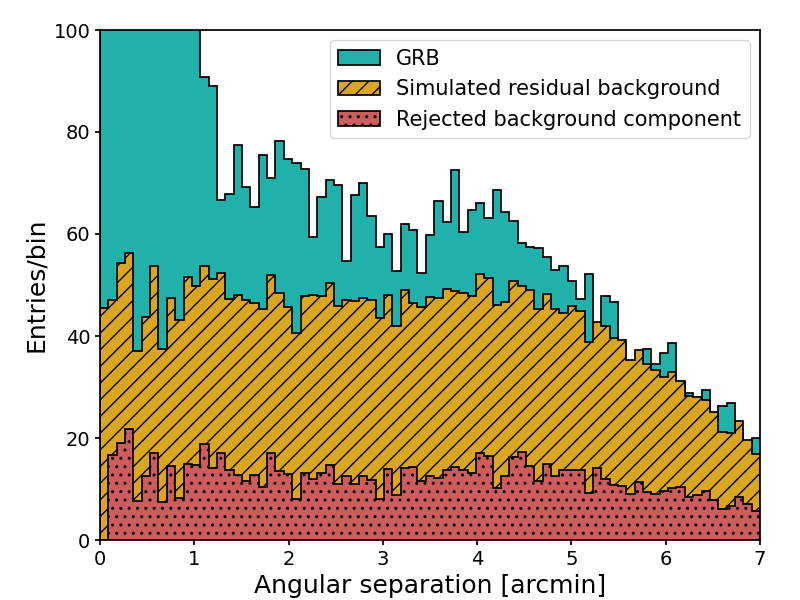}
    \includegraphics[width=0.32\textwidth]{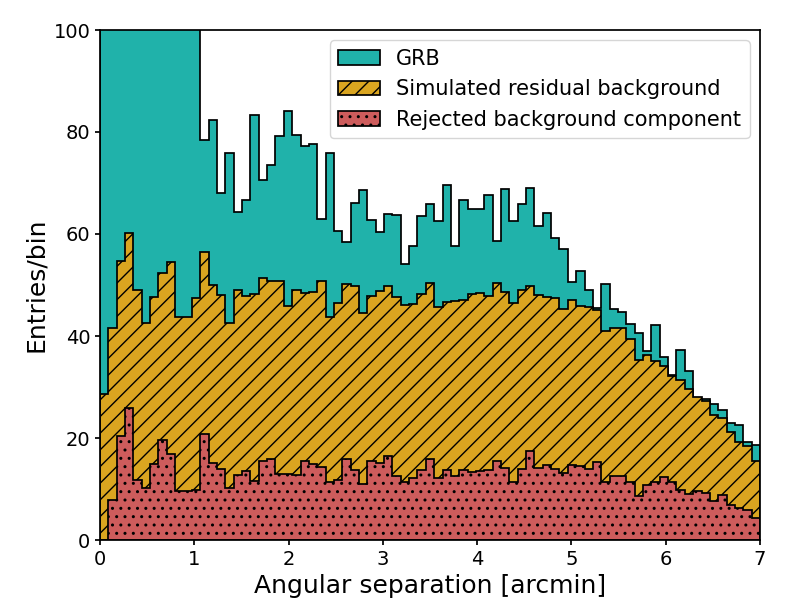}
    \caption{Comparison of the radial profiles of the GRB observation (water green) with the sum of the rejected background (dotted red) and the simulated residual background from BL Lac (hatched yellow) for the three IXPE detector units, zoomed on the vertical scale to exclude the large central peak in correspondence of the \core{} and better show the region of the two rings.}
    \label{fig:bkg_radial}
\end{figure}

We verify that the rejected events do not manifest any trace of the observed target (see the comparison between the middle and right panels of Figure~\ref{fig:bkg_rej}) and that they do not carry any significant polarization. In the region of the point source the fraction of the rejected background events reach at most 0.6\% in the case of DU2 (see also Fig. ~\ref{fig:bkg_radial}). 

\paragraph{Background subtraction}
The residual background needs to be estimated, simulated, and subtracted. The standard approach consists of selecting a region of the image in the field of view far from the point source, avoiding the edges where the sensitivity of the instrument degrades. However, in the case of extended sources (e.g. the dust-scattering rings we detect in this observation), this method cannot be applied. To address the issue, we estimate the residual background from a previous IXPE observation of a relatively faint source. For this work we 
considered: 1) the observation of 1ES~1959+650 carried out between 2022 June 9 and 2022 June 12; 2) the observation of 3C~279 performed between 2022 June 12 and 2022 June 18 3) the observation of BL Lacertae (BL Lac) which happened between 2022 July 7 and 2022 July 09. Due to changes in IXPE operations, the observations prior to June 09 would not provide background estimations suitable for this data analysis, and therefore have not been considered. These particular sources are point-like and have a small count rate ($<$ 0.2 Hz), which gives us a wide region of high noise to signal ratio to characterize the background. The first two observations are close in time and show a similar background spectrum, while the background obtained from the BL Lac observation shows a lower background rate: this provides a good bracketing for our analysis. 

The same background rejection procedure is applied to the data of all the observations considered. The residual background spectrum is derived by selecting the events in an annulus centered on the source with inner and outer radius of 1.2' and 5.5' respectively. For each background spectrum we simulate an IXPE observation using the \textit{ixpeobssim} simulation tool. Events are generated uniformly on the surface of the detectors and then projected in the sky using a realistic model for the pointing history that accounts for satellite dithering \citep{IXPE_calibration}. In order to reduce the statistical uncertainty, background templates are simulated with a longer exposure (1 Ms) compared to the GRB observation, then re-weighted appropriately to the respective livetime ratio before the subtraction. 

Figure~\ref{fig:bkg_radial} shows the radial profiles for the three detector units in celestial coordinates: the data of the observation of GRB~221009A are compared to the rejected background and the simulated background (here we show the case for the background extracted from the BL Lac observation, as an example). 

\begin{figure}[t]
    \centering
    \includegraphics[width=5.8cm]{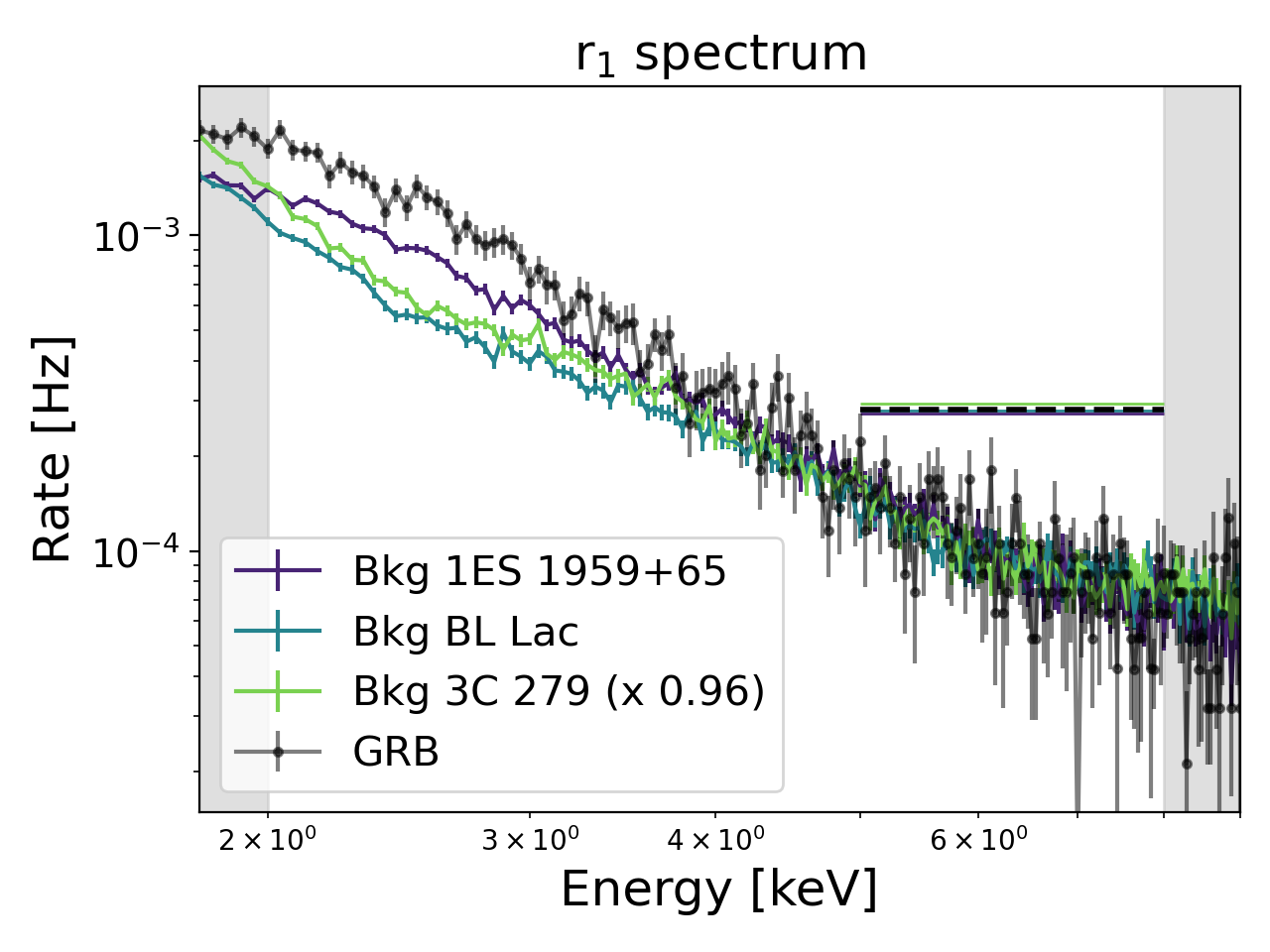}
    \includegraphics[width=5.8cm]{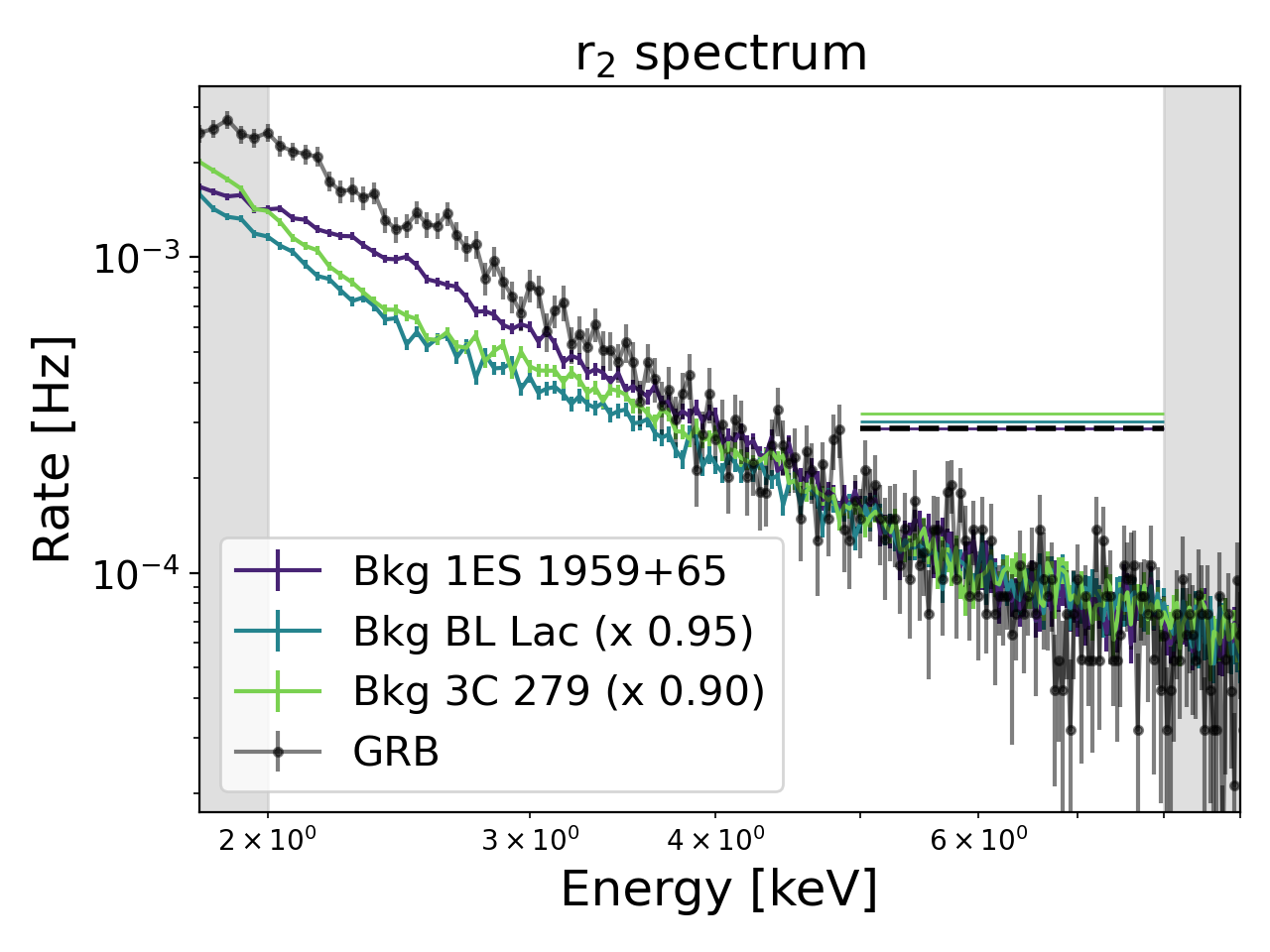}
    \includegraphics[width=5.5cm]{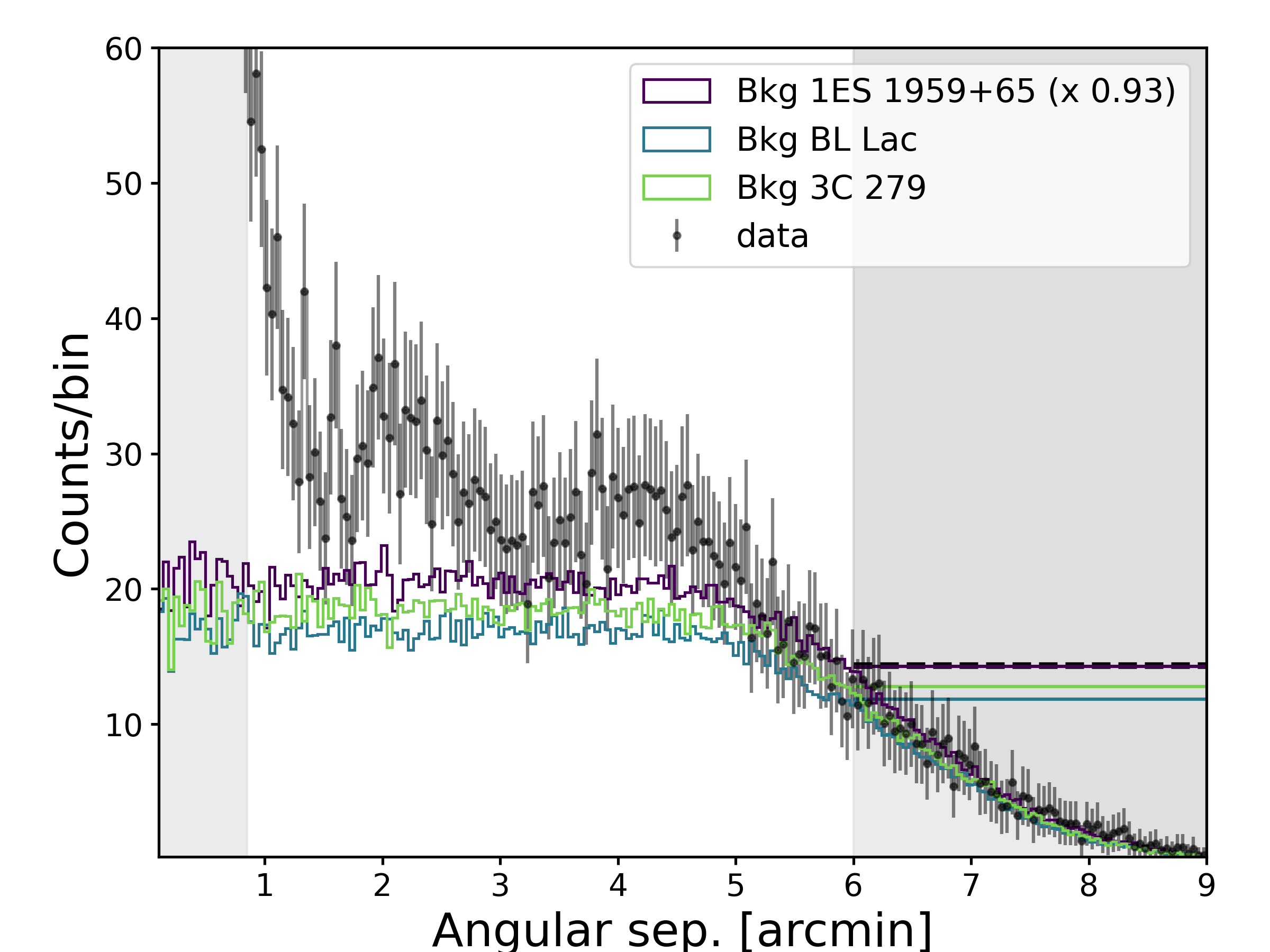}
    \caption{ Left and Middle: simulated background spectra of \rone{} and \rtwo{} event selections compared to the GRB observed ring spectra. The horizontal lines show the values of the integral of the spectra between 5 and 8 keV. Right: the radial profile of the simulated background templates compared to the GRB data profile for DU1. The horizontal lines show the values of the integral of the profiles above 6 arcmin. In all plots the gray shaded areas mark the regions of the parameters space excluded from the analysis.}
    \label{fig:bkg_spec}
\end{figure}

\paragraph{Background scaling} Due to statistical fluctuations in the low-count regime of the GRB rings data, the simulated backgrounds need to be scaled in order to never overshoot the data at high energies and at the edges of the field of view, namely where the background is expected to dominate. To define the scaling factor for each background, we estimate 1) the integral of the background spectra for both \rone{} and \rtwo{} selections between 5 and 8 keV and 2) the integral of the radial profile above 6', then we derive their ratio with the corresponding values of the GRB data. The ratios are reported in the label of Figure~\ref{fig:bkg_spec} and the horizontal lines show visually how the value of the integrals compare to each other.  The right panel of Figure~\ref{fig:bkg_spec} shows the radial profile of all the three simulated background templates, appropriately scaled to the livetime of the GRB observation. The scaling factor for each background is defined by the most extreme values among the ratios of \rone{} spectra, \rtwo{} spectra and radial profile. The background derived from the 1ES~1959+65 observation is hence scaled down by a factor of 1.07, the one derived from the BL Lac observation by a factor of 1.05, and the one from 3C~279 by a factor of 1.10. Table~\ref{tab:cnts} reports the number of counts for the GRB and for simulated background templates (re-weighted to account for the different live times) in the 2--8 keV band for the three region selections of our analysis.

\begin{deluxetable*}{lccc}
 \label{tab:cnts}
\tabletypesize{\scriptsize}
\tablewidth{0pt} 
\tablecaption{Count Statisitcs}
\tablehead{
\colhead{Tot. counts for} & \colhead{\rone} & \colhead{\core} & \colhead{\rtwo}}
    \startdata 
        ~~~GRB data & 16121 & 5450 & 5502~~~\\
        ~~~Bkg 1ES~1959+65  & 135  & 4099 & 4313 ~~~\\
        ~~~Bkg BL Lac & 105 & 3263 & 3497 ~~~\\
        ~~~Bkg 3C~279 & 124 & 3776 & 4017 ~~~\\
    \enddata
    \tablecomments{Total and background counts in the 2--8 keV band for the \core, \rone, and \rtwo{} selections. These numbers refer to the background rejected data. The background counts are computed by multiplying the background rate by the GRB observation live time. }
\end{deluxetable*}

\section{Additional Considerations} 
\label{app:CC}

\subsection{Optical polarization data analysis} 
\label{app:B}
During the IXPE pointing we also performed optical polarization observations in the R-band at the Nordic Optical Telescope \citep{GCNoptpol}. The observations were obtained using the Alhambra Faint Object Spectrograph and Camera (ALFOSC) in the standard linear polarimetric mode that includes a $\lambda$/2 retarder followed by calcite.  At the time of the observations (2022 October 12 at 20:15UT) the sky conditions were clear with 1.2 arcsecond seeing. However, GRB221009A is located in a crowded Galactic field. This resulted in the extraordinary beam of a nearby bright star to overlap with the ordinary beam of the source. As such, the standard polarimetric analysis was not possible \citep[see e.g.]{Hovatta2016,Nilsson2018}. Instead, we performed careful modelling of the point spread function. We used the second brightest star within the ALFOSC field of view to create a model of the PSF, which was then subtracted from each image separately. This process, however, can result in background artifacts. To mitigate the effect of any artifact we used a small aperture of 1.5 arcsec radius to perform the measurements using standard formulas. 

\subsection{Effect of dust scattering on X-rays polarization}
\label{app:dustpoleff}
We investigated the effect on polarization from reflection, scattering and transmission considering the dominant dust compounds, Carbon and silicates \citep[see e.g.][ for a recent discussion of the topic]{CostantiniCorrales}.
At the small angles that we observe, even assuming a coherent reflection angle, any polarization induced by reflection of X-rays would result in a negligible modulation of less than $10^{-5}$, or a PD$\sim$0.001\%. These values were obtained using the Center for X-Ray Optics database and online tools\footnote{\url{https://henke.lbl.gov/optical\_constants/}}. 
Polarization from transmission is expected to be negligible as well for X-rays of energies at the peak of IXPE sensitivity, given that the common dust compounds do not show K or L shell edges there. A fraction of the scattered light might have a polarization status affected by big spheroidal dust grains via Mie scattering. We checked this by using the python package \textit{Miepython}\footnote{\url{https://miepython.readthedocs.io/}}, which calculates light scattering according to the Mie theory and Rayleigh–Gans approximation, and adopting the X-ray refraction index for silicates provided by \cite{DraineLee1984} and \cite{LaorDraine1993}. We find that at the scattering angles we are considering, the PD due to refraction is less than $5.5\times10^{-5}$ at 2 keV for a binary population of grains (e.g.: perfectly aligned, elongated and not aligned, spherical grains) with a power-law size distribution with an index of -3.5 \citep{CostantiniCorrales}. Therefore, we can reasonably assume that any polarization observed from the X-ray scattering halos is attributable to the original emission. 

\subsection{Additional plots}
\label{app:C}
\begin{figure}[ht]
    \centering    
    \includegraphics[height=10cm]{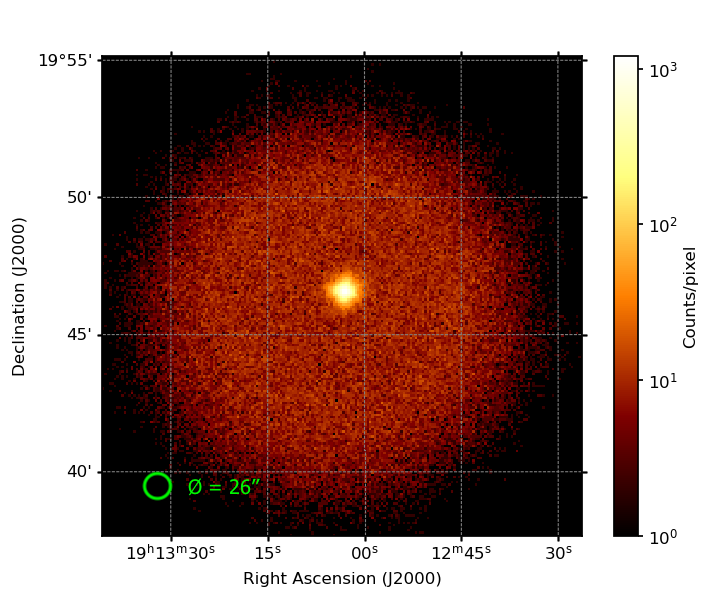}
    \caption{IXPE counts map combining the 3 DU observations obtained with the \textit{xpbin} routine of \textit{ixpeobssim} with the flag {\tt --algorithm CMAP}. The \core/afterglow emission dominates the image, however the fainter halos are already visible to the attentive eye. The green circle has a diameter that approximates the IXPE DU-averaged PSF.}
    \label{fig:cmap}
\end{figure}

\begin{figure}[ht]
    \centering    
    \includegraphics[width=\textwidth]{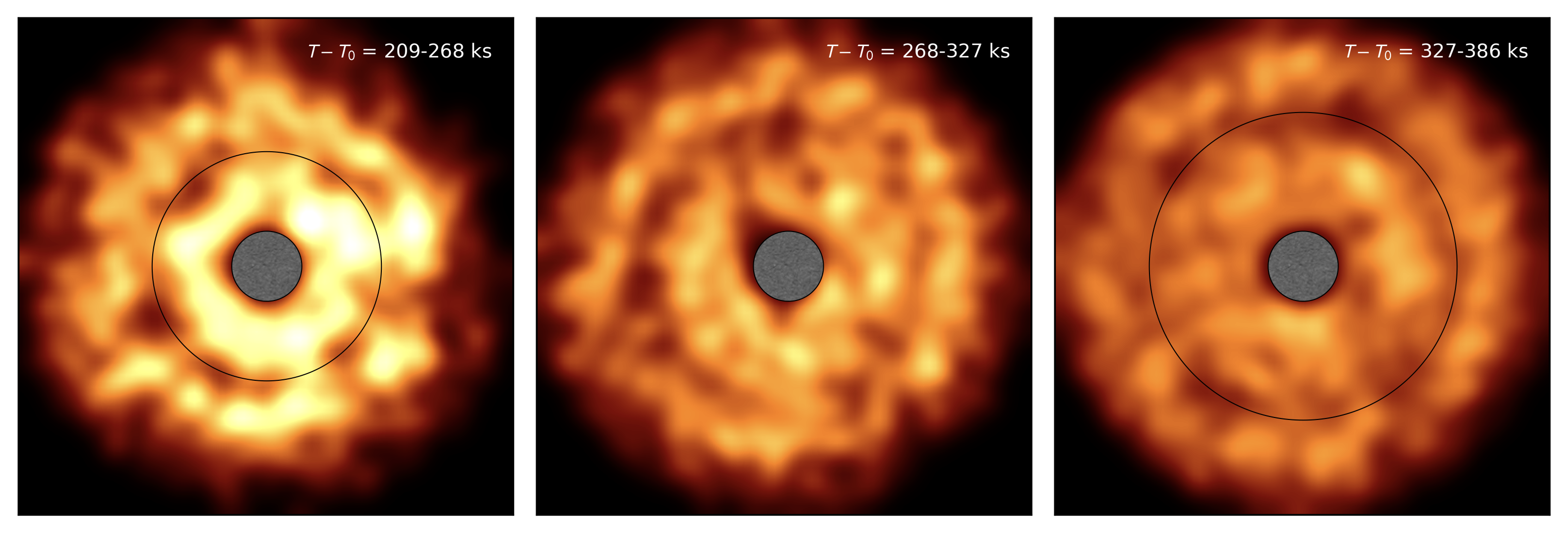} 
    \caption{Time evolution of the dust-scattering halos. The counts maps are generated in three time bins combining the data of the 3 DUs. The \core{} region has been removed to better show the rings. The images have been smoothed with a Gaussian beam for visualization purposes. To guide the eye, we added a thin circle in the first and last images of the sequence to mark the minimum of the counts gap between the two evolving rings.}
    \label{fig:elapse}
\end{figure}

\begin{deluxetable*}{ccc|cc}
 \label{tab:summary3}
\tabletypesize{\scriptsize}
\tablewidth{0pt} 
\tablecaption{Summary table of the PCUBE rings analysis}
\tablehead{
 \colhead{~~~} & \multicolumn2c{\rone} & \multicolumn2c{\rtwo} \\
         $\Delta$E  & PD  & ~~~PD u.l.(99\%)~~~ & PD  & ~~~PD u.l.(99\%)~~~\\[-0.1cm]
         {\tiny keV} & {\tiny[\%]} & {\tiny [\%]} & {\tiny [\%]}  & {\tiny [\%]} }
    \startdata 
         ~~~ 2--8 & $19.6\pm8.7$  & $<$42.0 & $17.2\pm8.8$  & $<$39.9 \\[0.2cm]
         ~~~ 2--4 & $27.2\pm8.3$  & $<$48.6 & $5.5\pm8.3$   & $<$26.9 \\
         ~~~ 4--8 & ~~~$15.5\pm11.5$~~~ & $<$45.1 & $25.1\pm11.6$ & $<$55.0 \\[0.2cm]
    \enddata
    \tablecomments{Results of the PCUBE analysis between 2 and 8 keV and resolved in 2 logarithmic energy bins. This analysis is performed on the background-rejected (not background subtracted) data. This implies that 1) the estimated uncertainties are not accurate because they are computed on a boosted statistic that includes background events (a big fraction of the total, see Table~\ref{tab:cnts}); and 2) the results of the PCUBE analysis are not directly comparable to those resulting form the spectropolarimetric analysis. The latter represents a more accurate analysis. For the 2$-$8 keV PCUBE analysis the minimum detectable polarization at 99\% C.L. is $MDP_{99\%}=26.5\%$ and $MDP_{99\%} = 26.8\%$ for \rone{} and \rtwo, respectively. \\
    Note that in the 2--4 keV bin for \rone{} the PD might seem to exceed the 99\% C.L.. However, in this case, the test-statistic follows a $\chi^2$ distribution with 4 d.o.f, accounting for the two energy bins considered and 2D Q-U space. This gives a 3\% probability of finding a $\chi^2$ value equal to or exceeding the observed one in case unpolarized emission, which means that we are compatible with the null hypothesis within the 97\% C.L.. Such significance is even lower if we account for the trials due to both rings selections: in this case we should derive the significance from a $\chi^2$ distribution with 8 d.o.f..}
\end{deluxetable*}

\begin{figure}[ht]
    \centering
    \includegraphics[height=6.5cm]{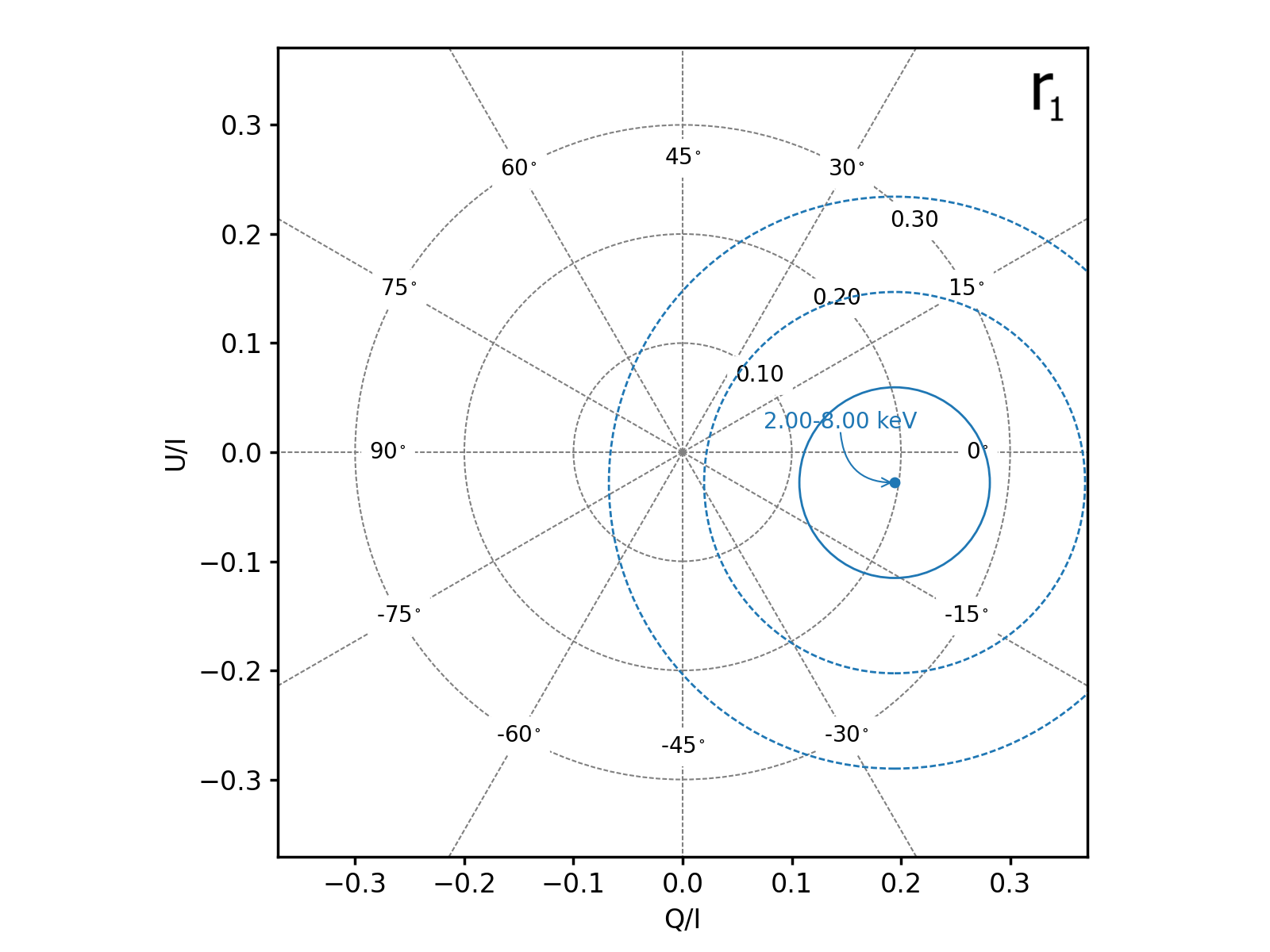}
    \includegraphics[height=6.5cm]{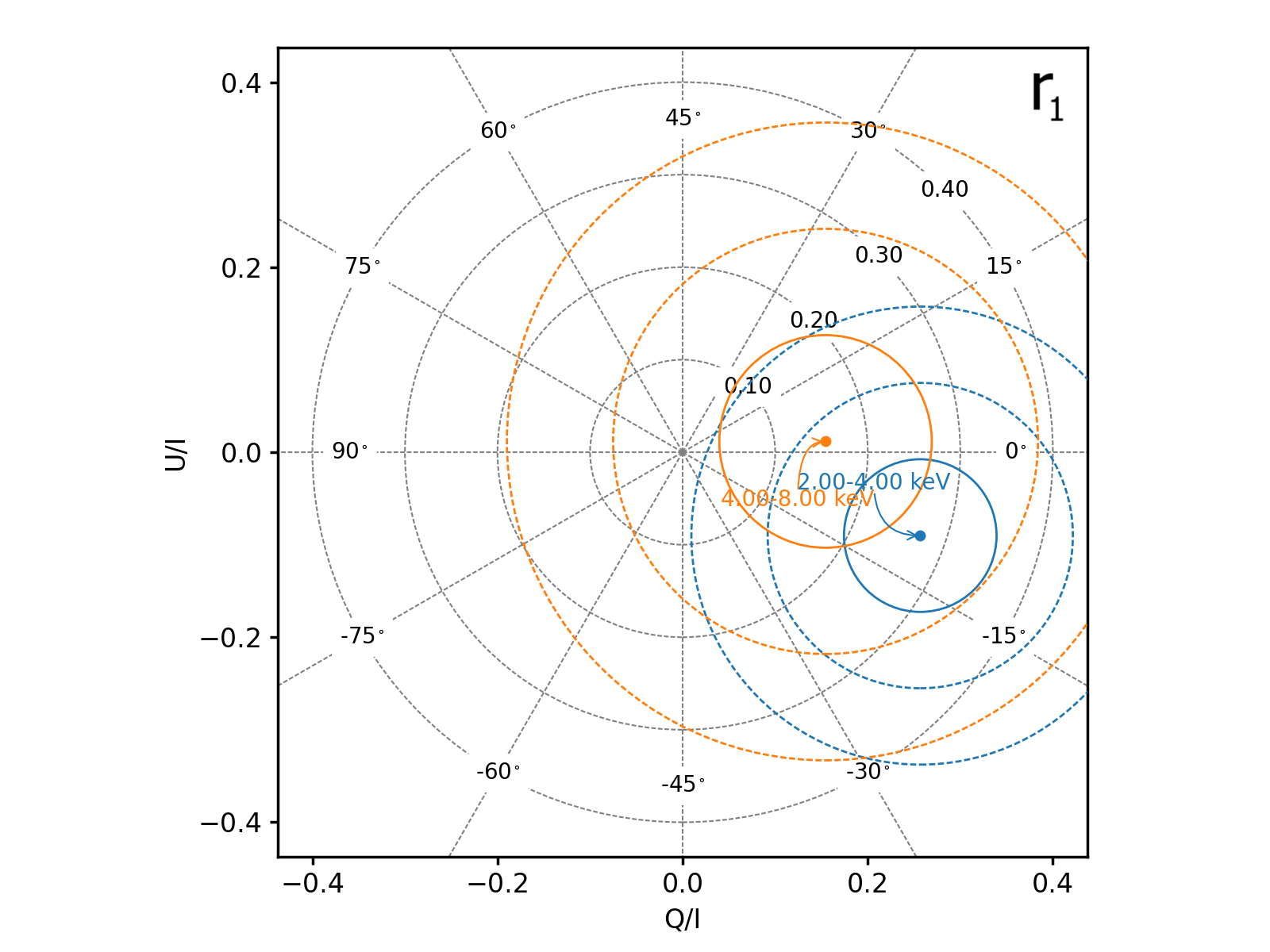}\\
    \includegraphics[height=6.5cm]{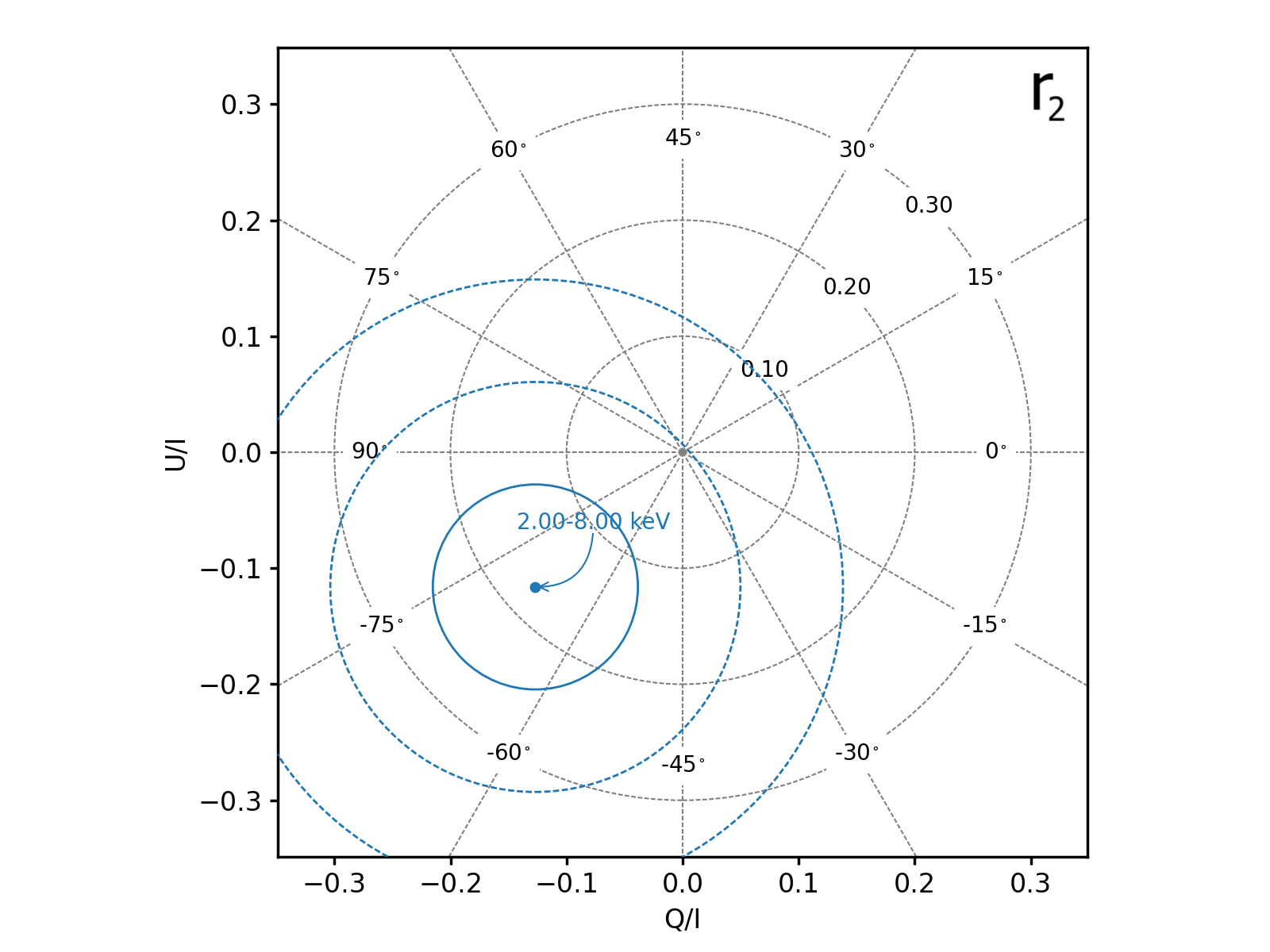}
    \includegraphics[height=6.5cm]{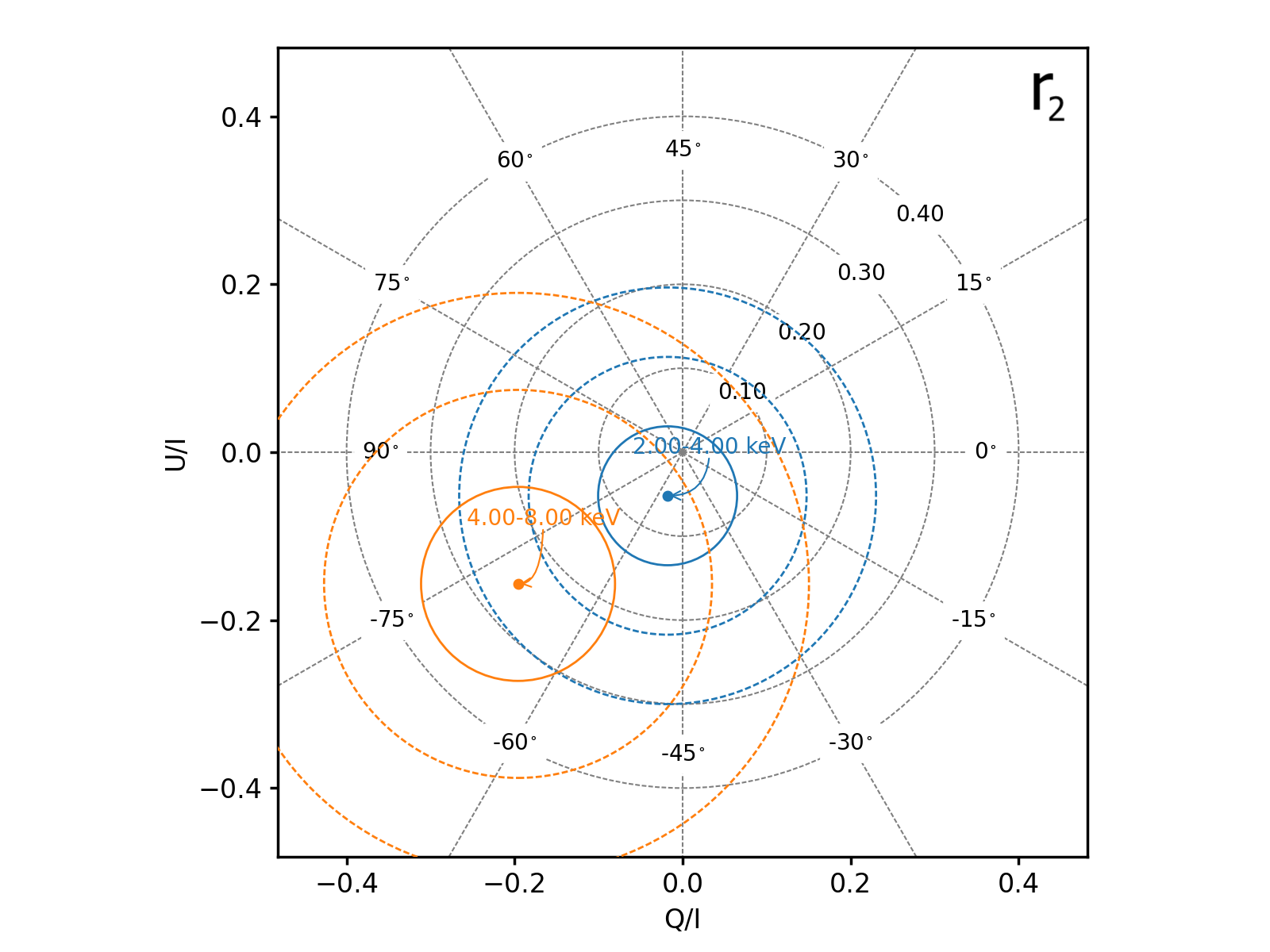}\\
    \caption{Polarization PCUBE analysis results for \rone{} (top row) and \rtwo{} (middle row), for one energy bin 2--8 keV (left column) and two logarithmic energy bins 2--4 keV and 4--8 keV (right column). Tab.\ref{tab:summary3} provides the values and 1$\sigma$ errors on the PD and PA for the energy-resolved PCUBE analysis. Note that the PCUBE analysis of the rings emission is performed without a proper background subtraction (we refer to the caption of Tab.\ref{tab:summary3} for further discussion on the caveats of this point).\\
    The PAs of the two rings seem to be significantly different, however: as mentioned in the main text, the uncertainties are underestimated. Polarization measurements with significance below the 99\% C.L. are not considered as a detectios and the PAs are to be considered unconstrained. Furthermore, under the assumption that the two rings originate from the same prompt emission, there is no (known) reason to expect the polarization of the two rings to be intrinsically rotated. This fact is symptomatic of fluctuations due to the low photon statistics of the signal, and further justifies our approach of combining the two rings for a more accurate spectropolarimetric analysis.\\
    }
    \label{fig:pcubes_sings}
\end{figure}

\begin{figure}[ht]
    \centering    
    \includegraphics[width=0.31\textwidth,height=0.23\textwidth]{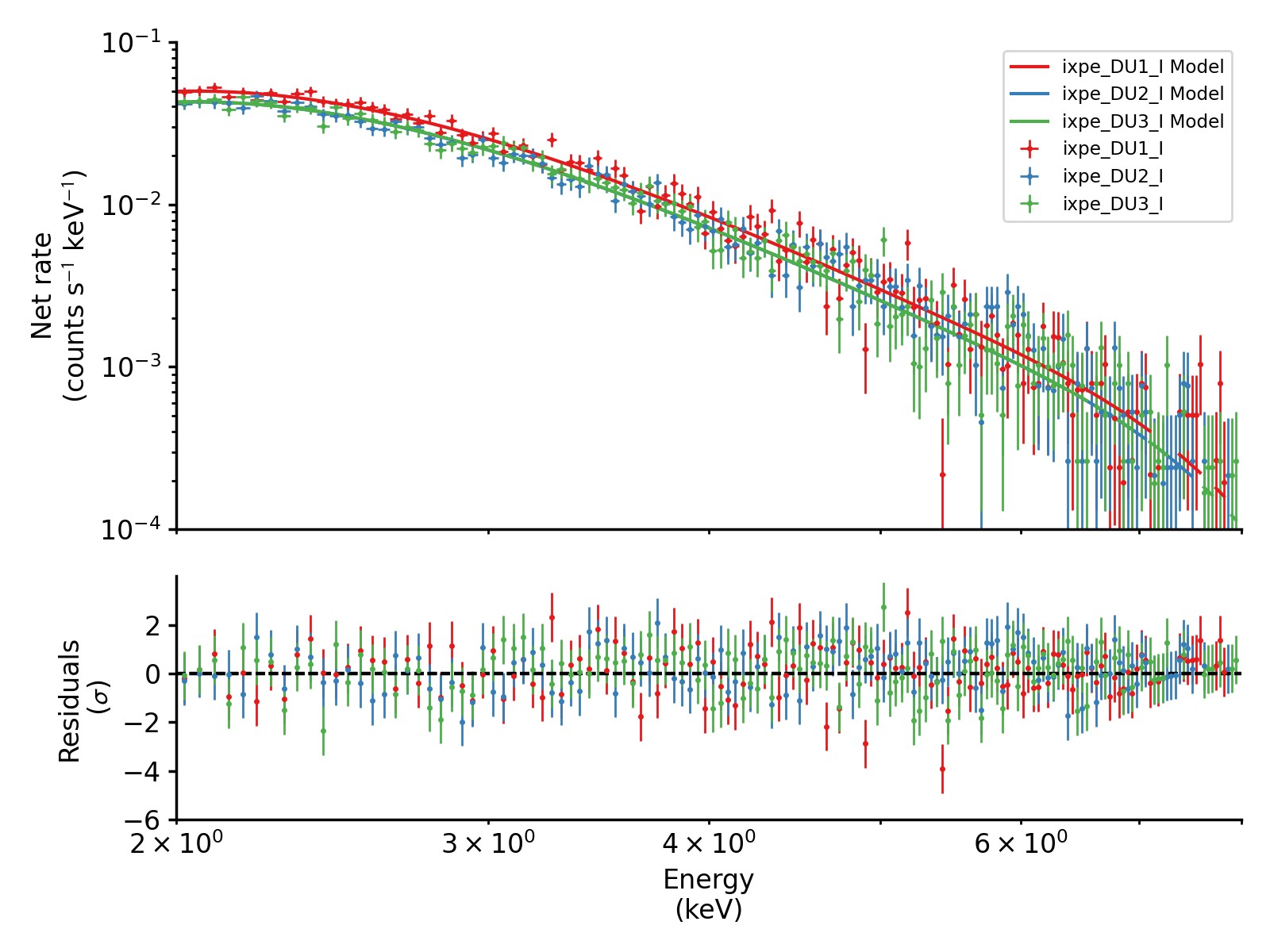} 
    \includegraphics[width=0.31\textwidth,height=0.23\textwidth]{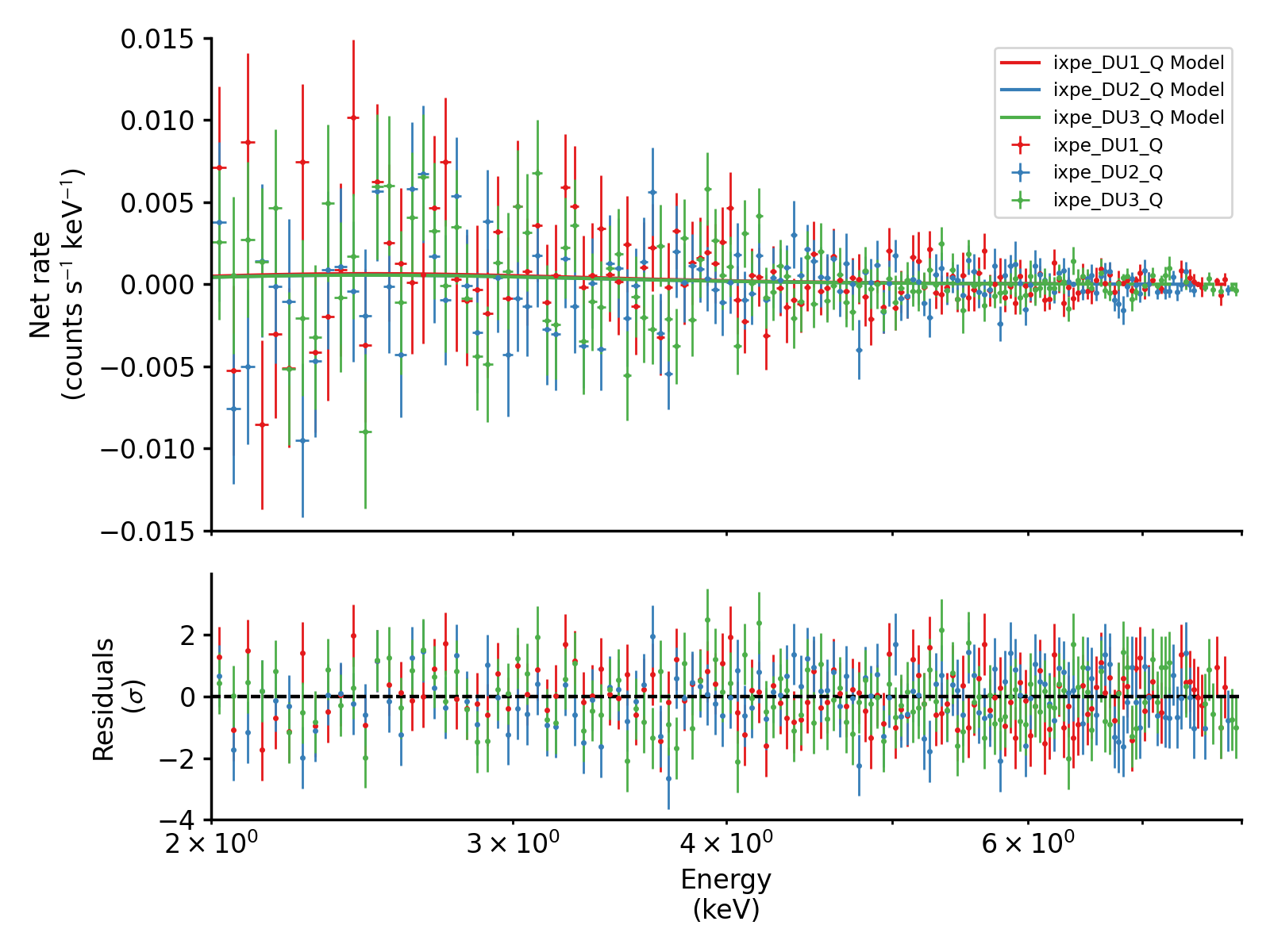}
    \includegraphics[width=0.31\textwidth,height=0.23\textwidth]{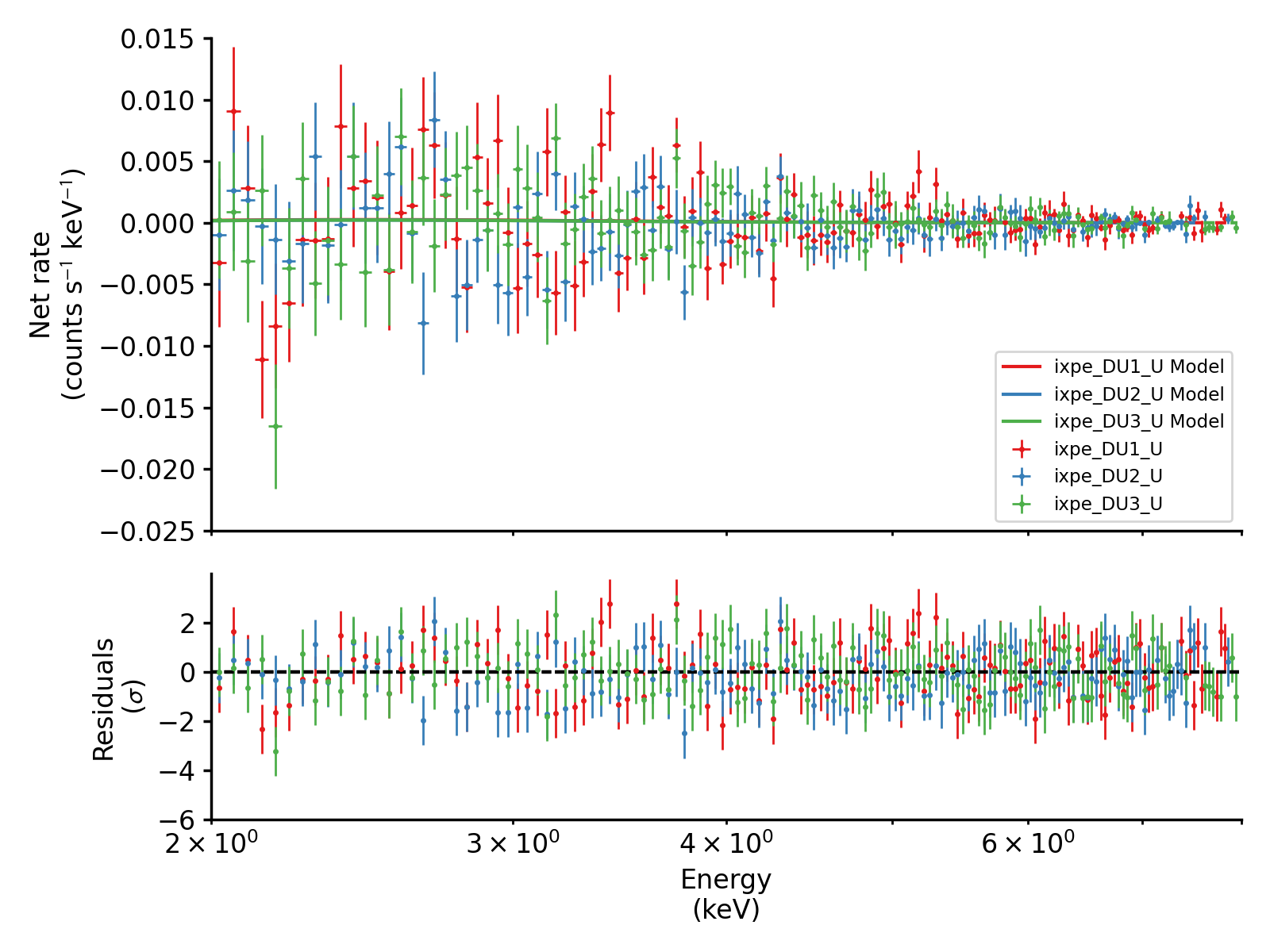}
    \includegraphics[width=0.31\textwidth,height=0.23\textwidth]{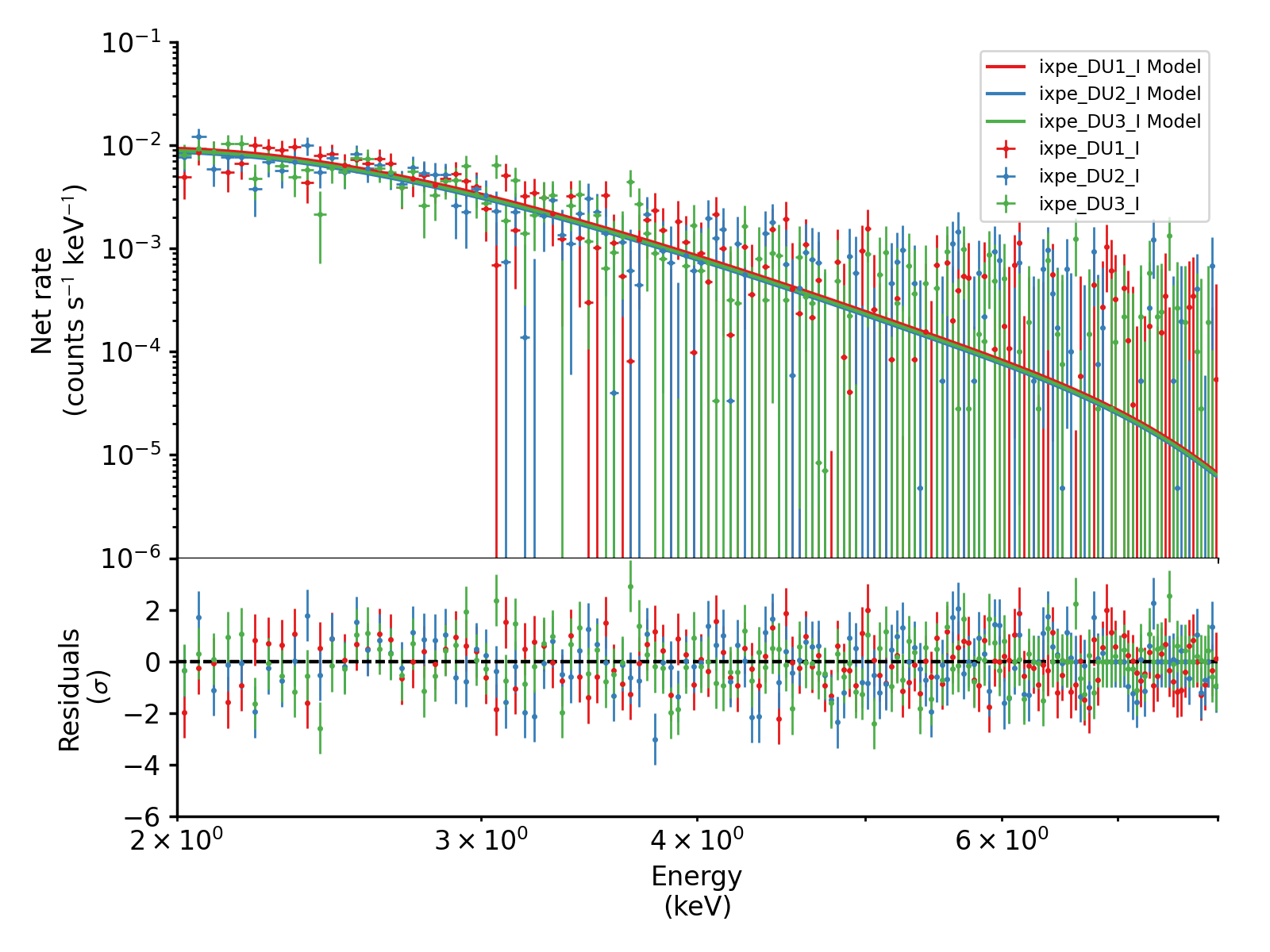}
    \includegraphics[width=0.31\textwidth,height=0.23\textwidth]{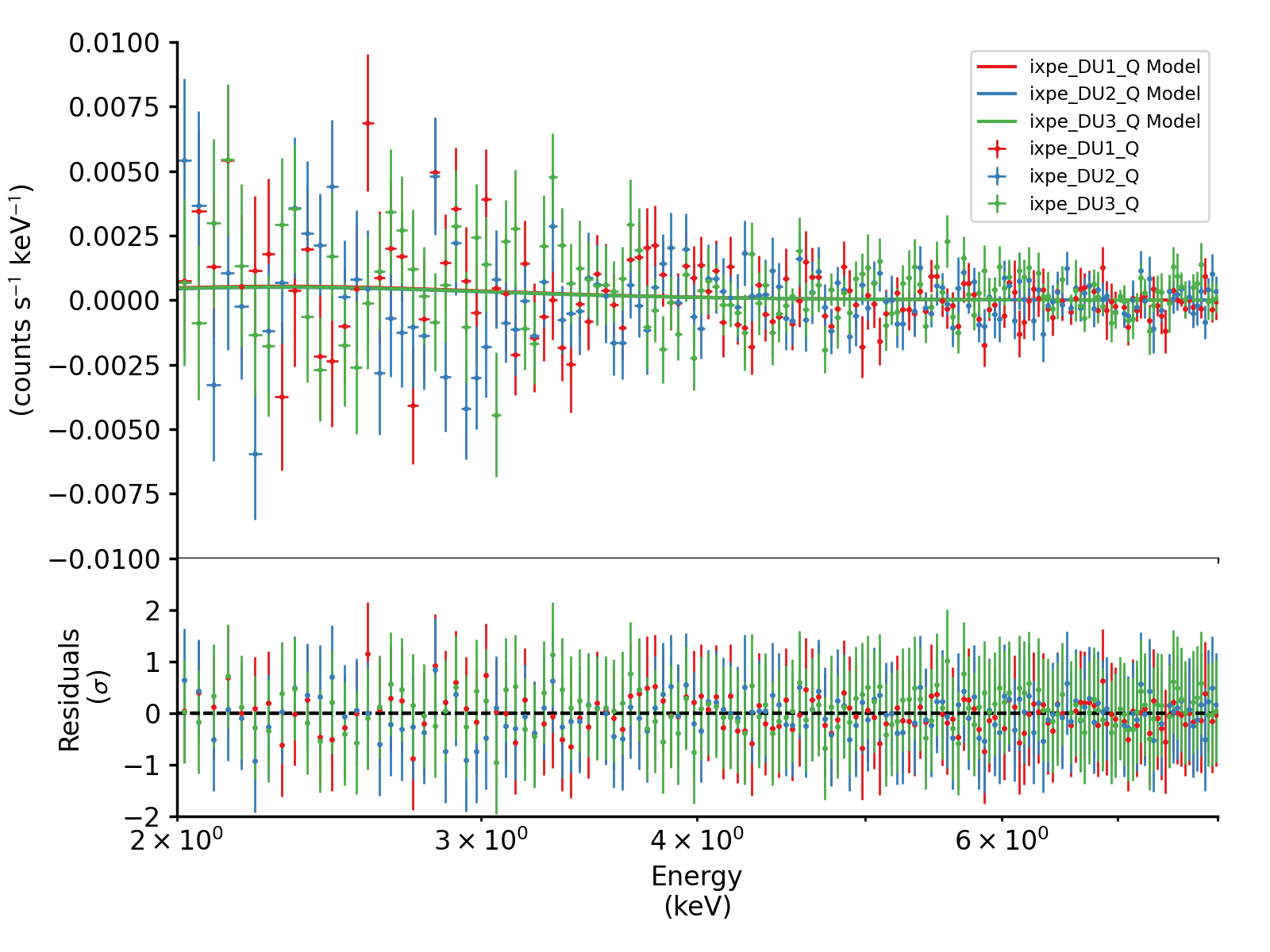}
    \includegraphics[width=0.31\textwidth,height=0.23\textwidth]{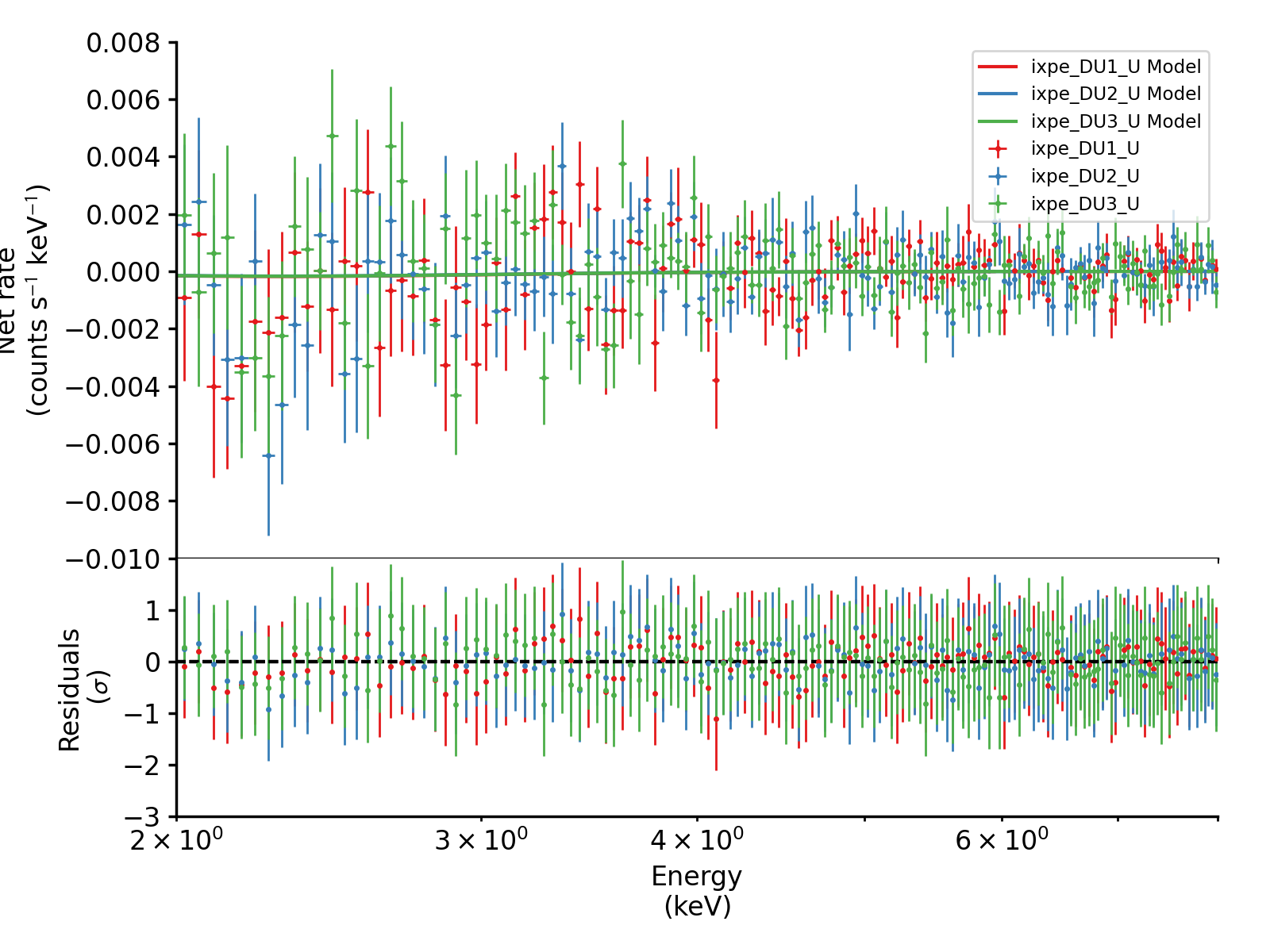}\\
    \includegraphics[width=0.31\textwidth,height=0.23\textwidth]{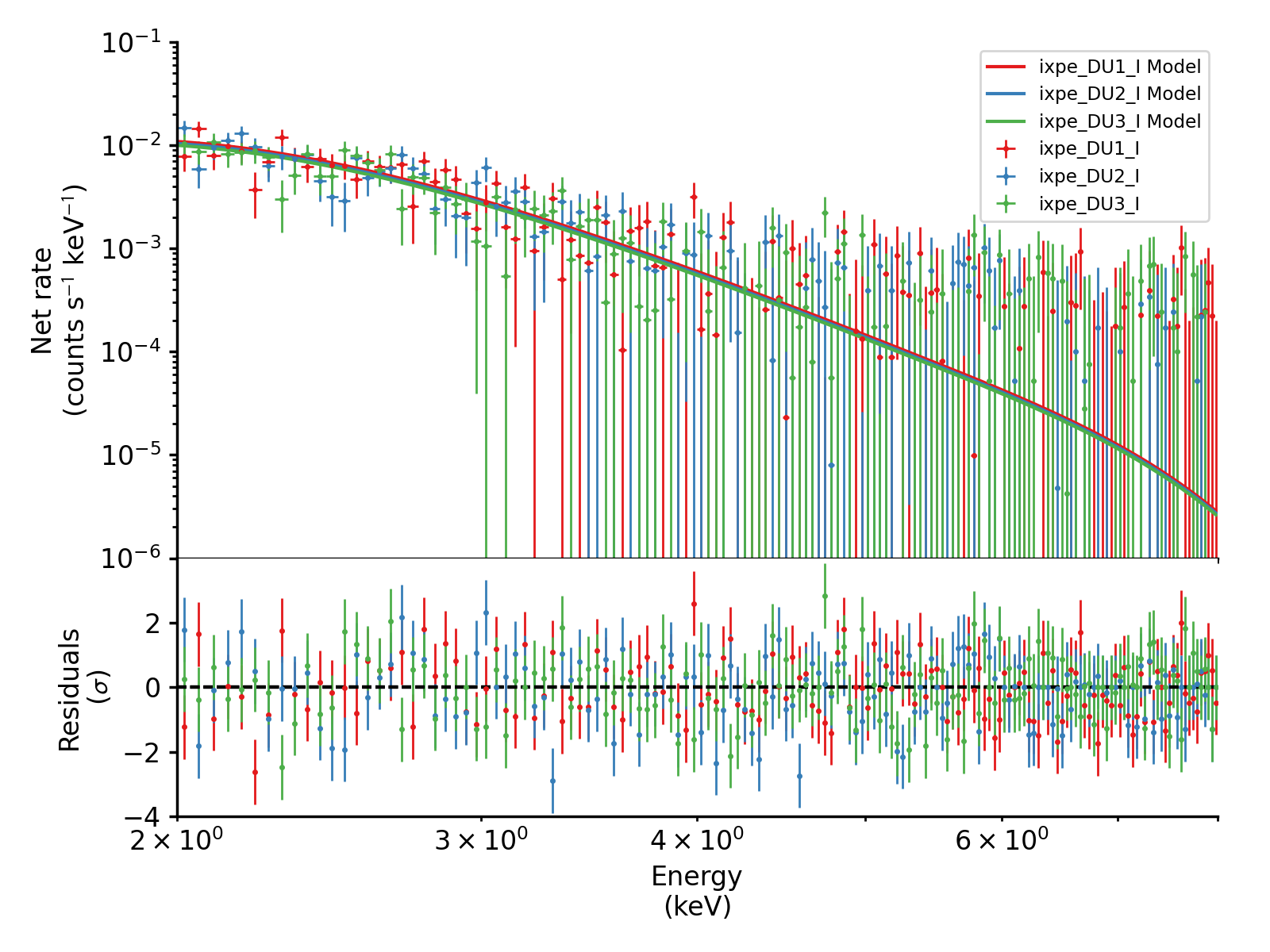}
    \includegraphics[width=0.31\textwidth,height=0.23\textwidth]{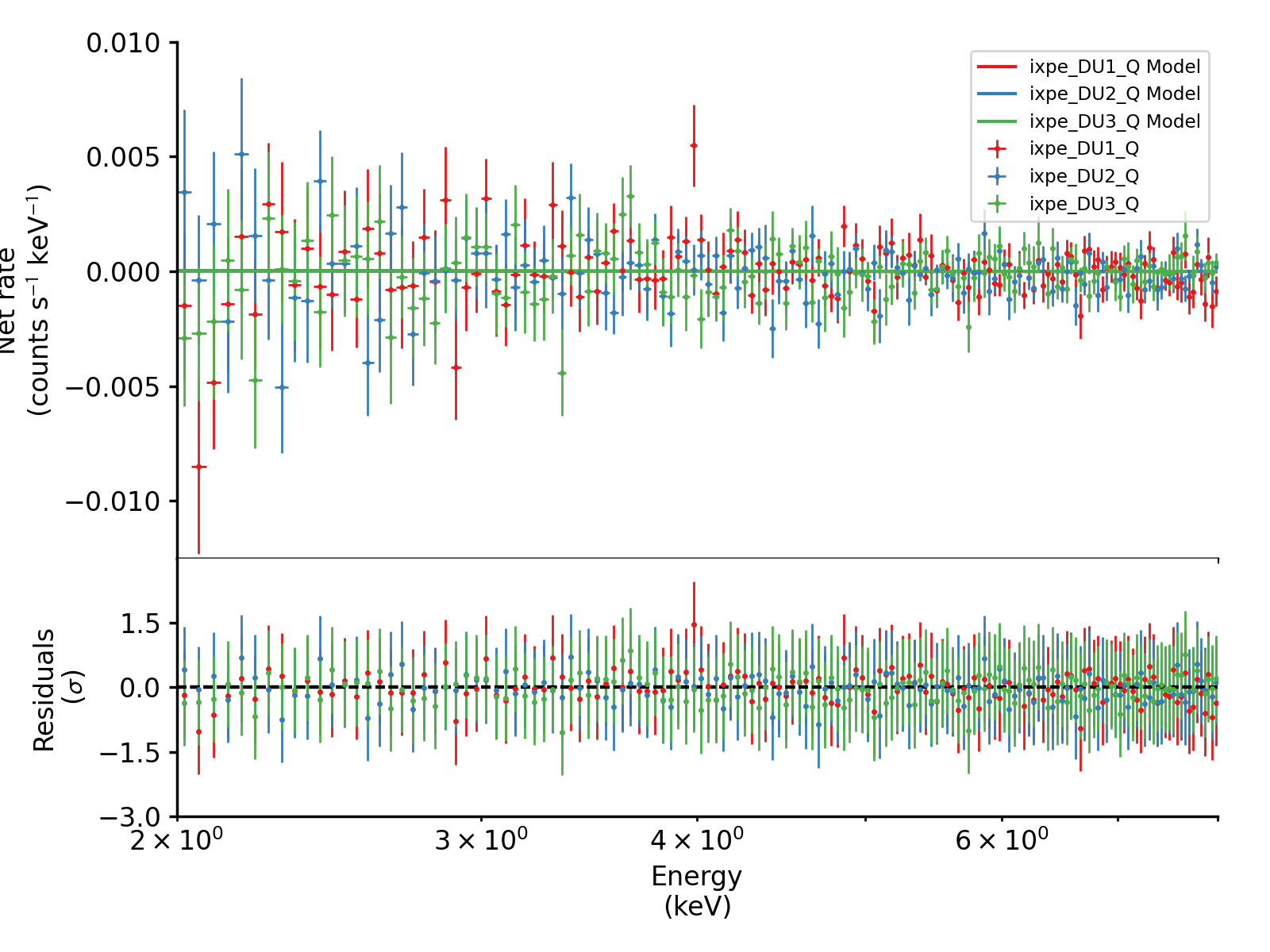}
    \includegraphics[width=0.31\textwidth,height=0.23\textwidth]{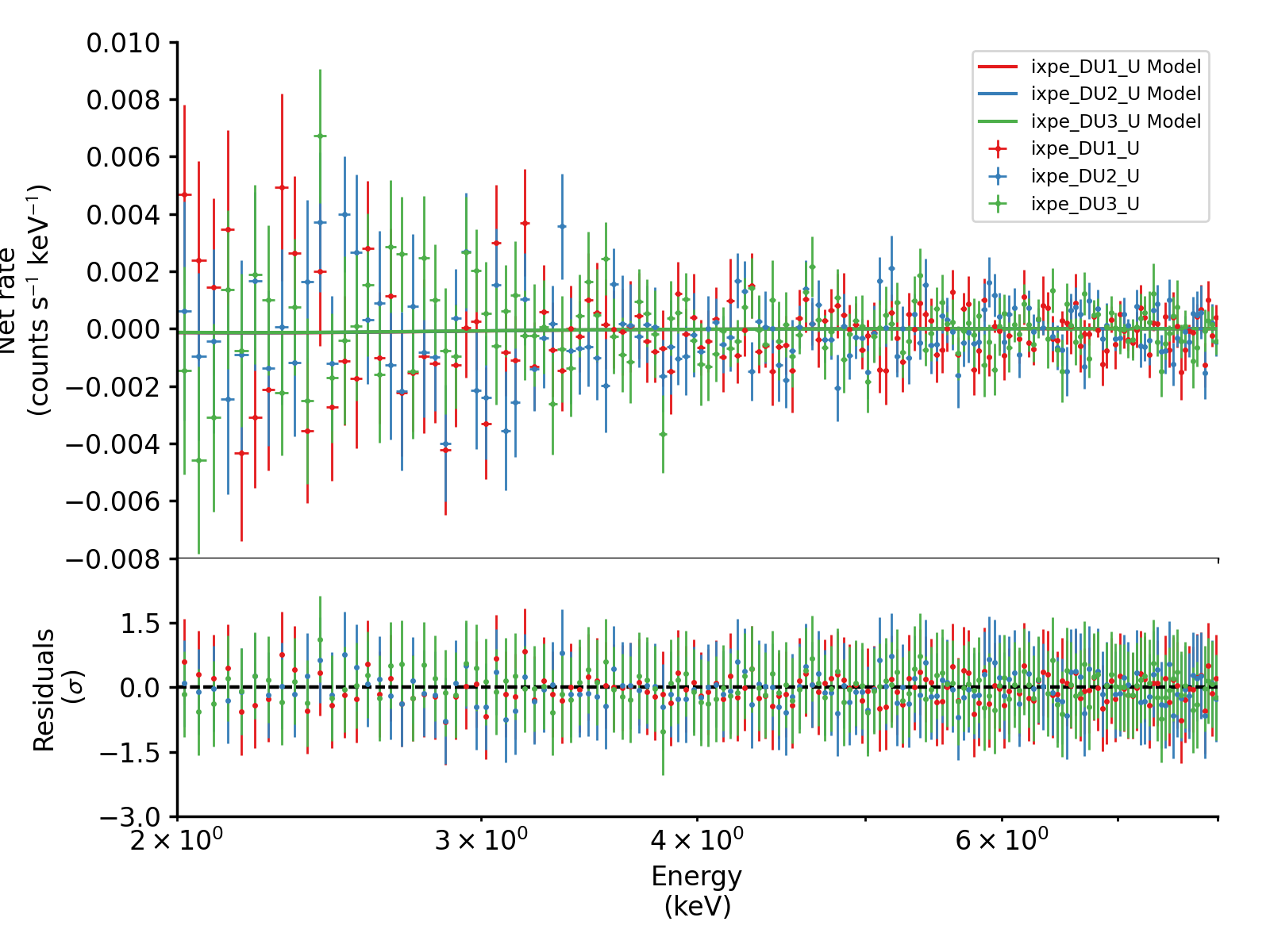}
    \caption{I, Q and U spectra (background rejected and subtracted) for the \core{} region (top row), \rone{} region (middle row), and \rtwo{} region (bottom row). All spectropolarimetric fits have a $\chi^2_{{\rm red}}\sim 1$.}
    \label{fig:spectra}
\end{figure}

\begin{figure}[ht]
    \centering
    \includegraphics[height=6cm]{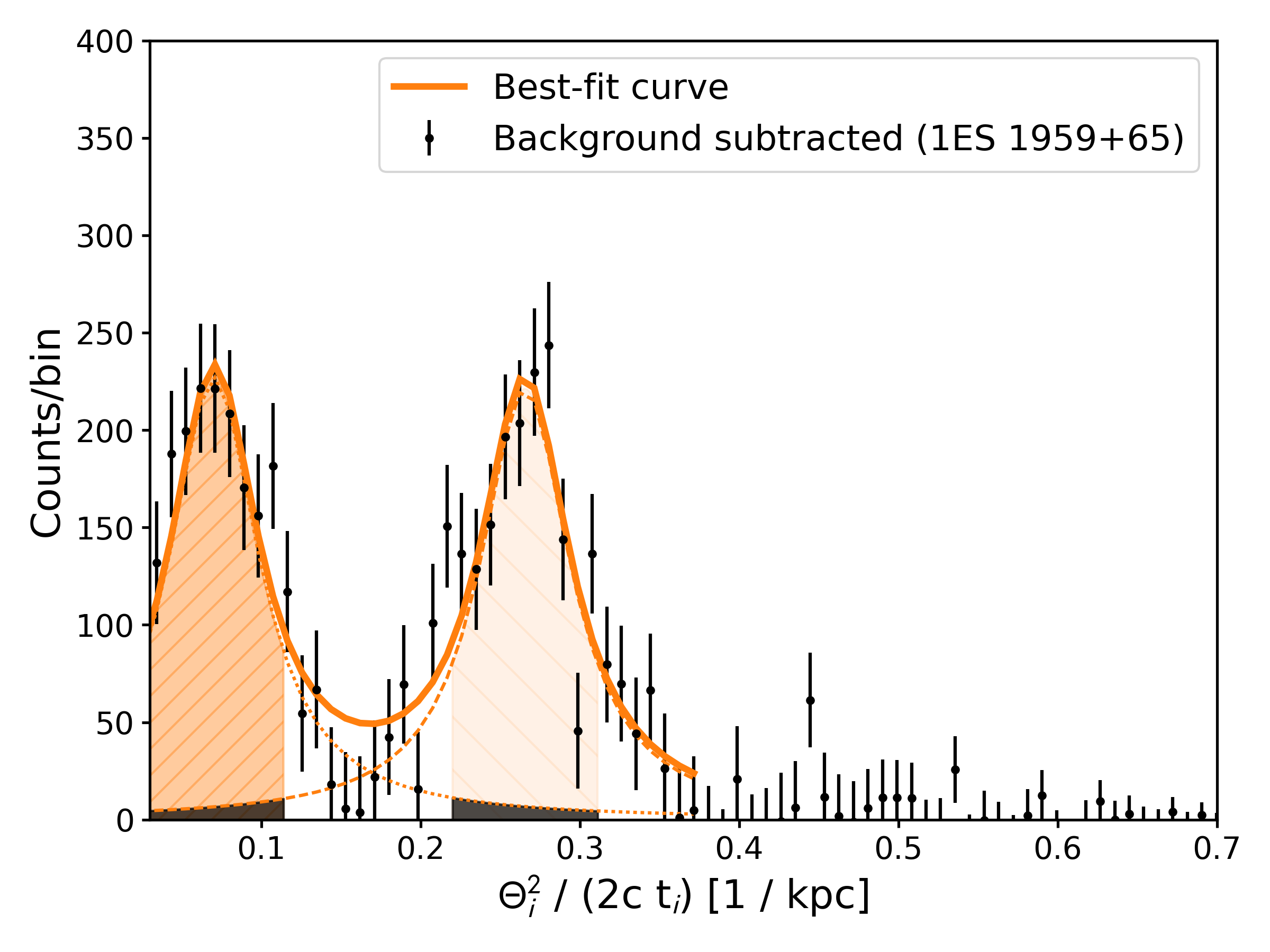}
    \includegraphics[height=6cm]{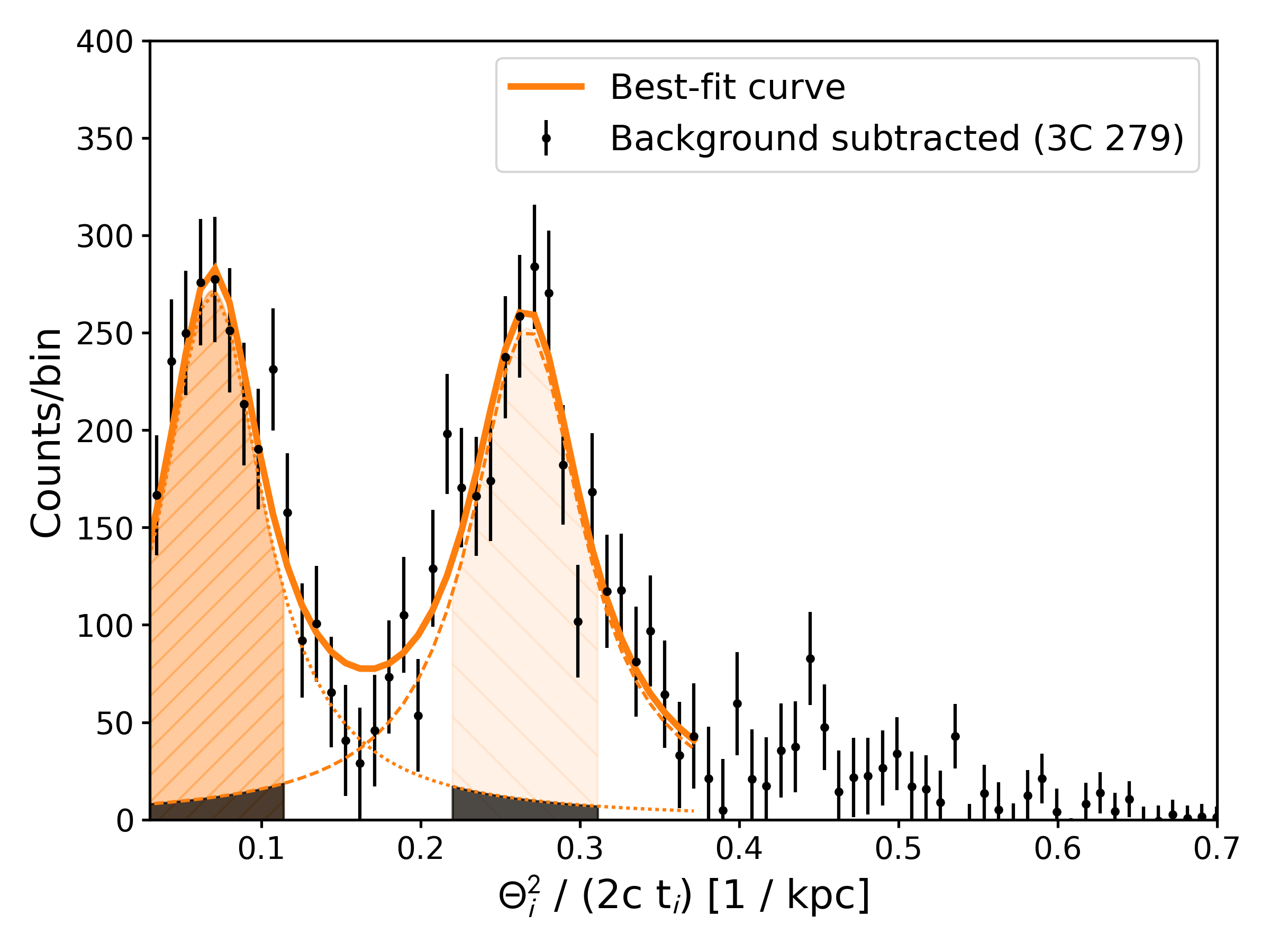}
    \caption{Equivalent plots to that in Figure~\ref{fig:ringevol} right panel, but subtracting a different background.}
    \label{fig:ring_fit}
\end{figure}

\begin{figure}[ht]
    \centering
    \includegraphics[height=6cm]{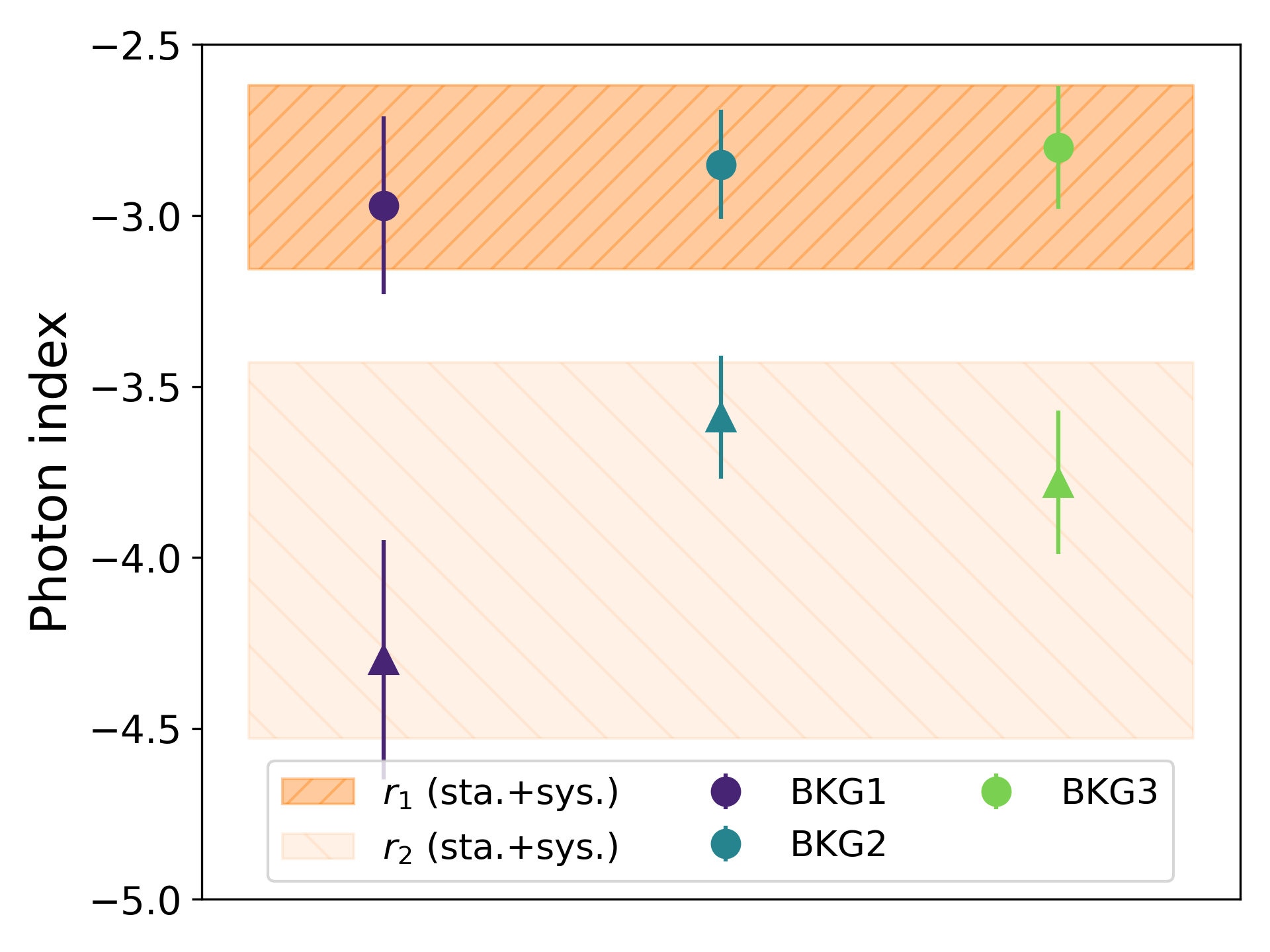}\quad\quad
    \includegraphics[height=6cm]{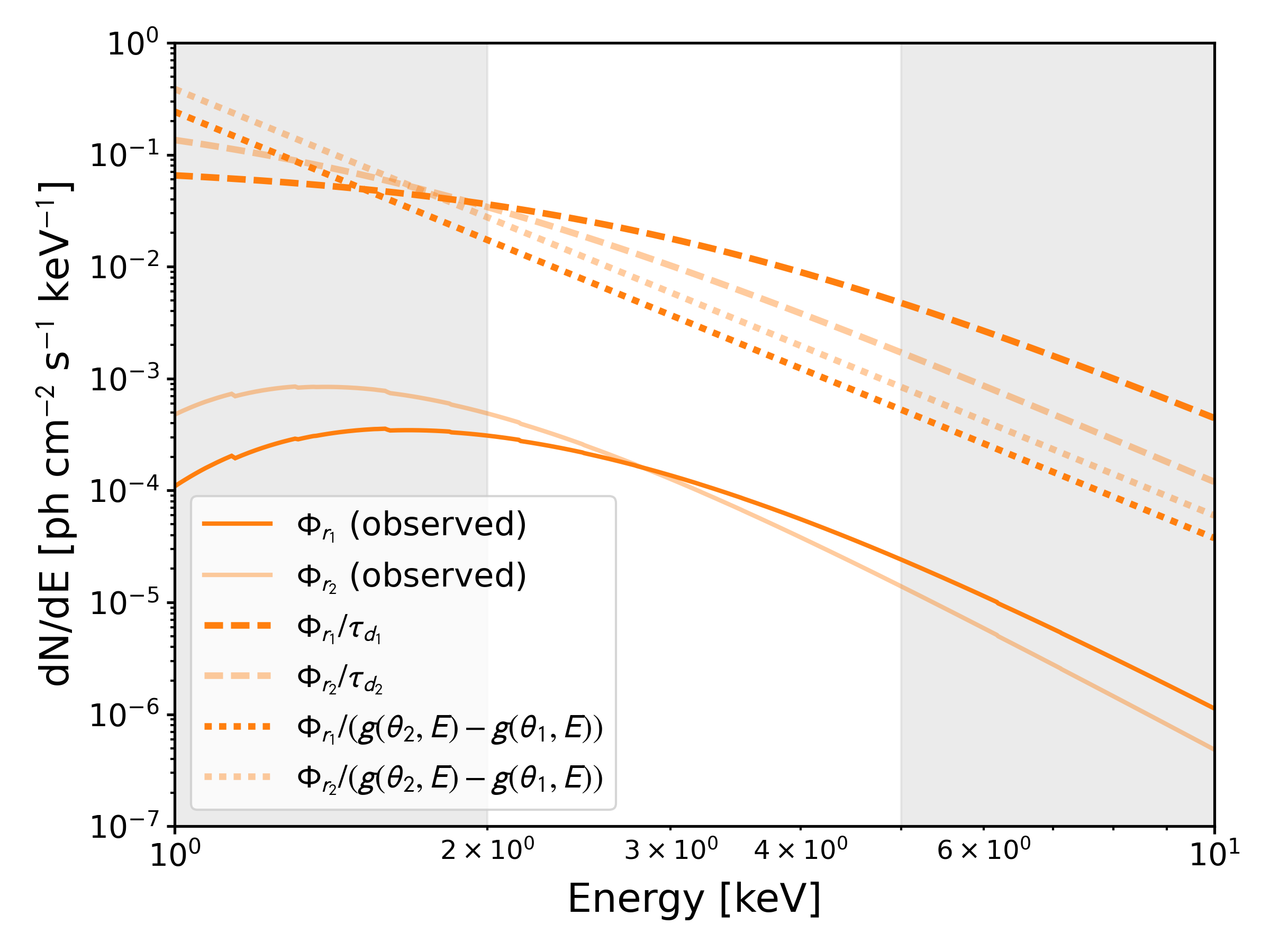}
    \caption{Left: Comparison of the best-fit spectral indices of the two rings emissions. Right: Effect of the corrections for the optical depth and for the energy-dependent scattering efficiency. The light-gray regions cover the energy ranges excluded in the fitting procedure. }
    \label{fig:ring_intrinsic_corrections}
\end{figure}

\begin{figure}[t]
    \centering    
    \includegraphics[height=0.3\textwidth]{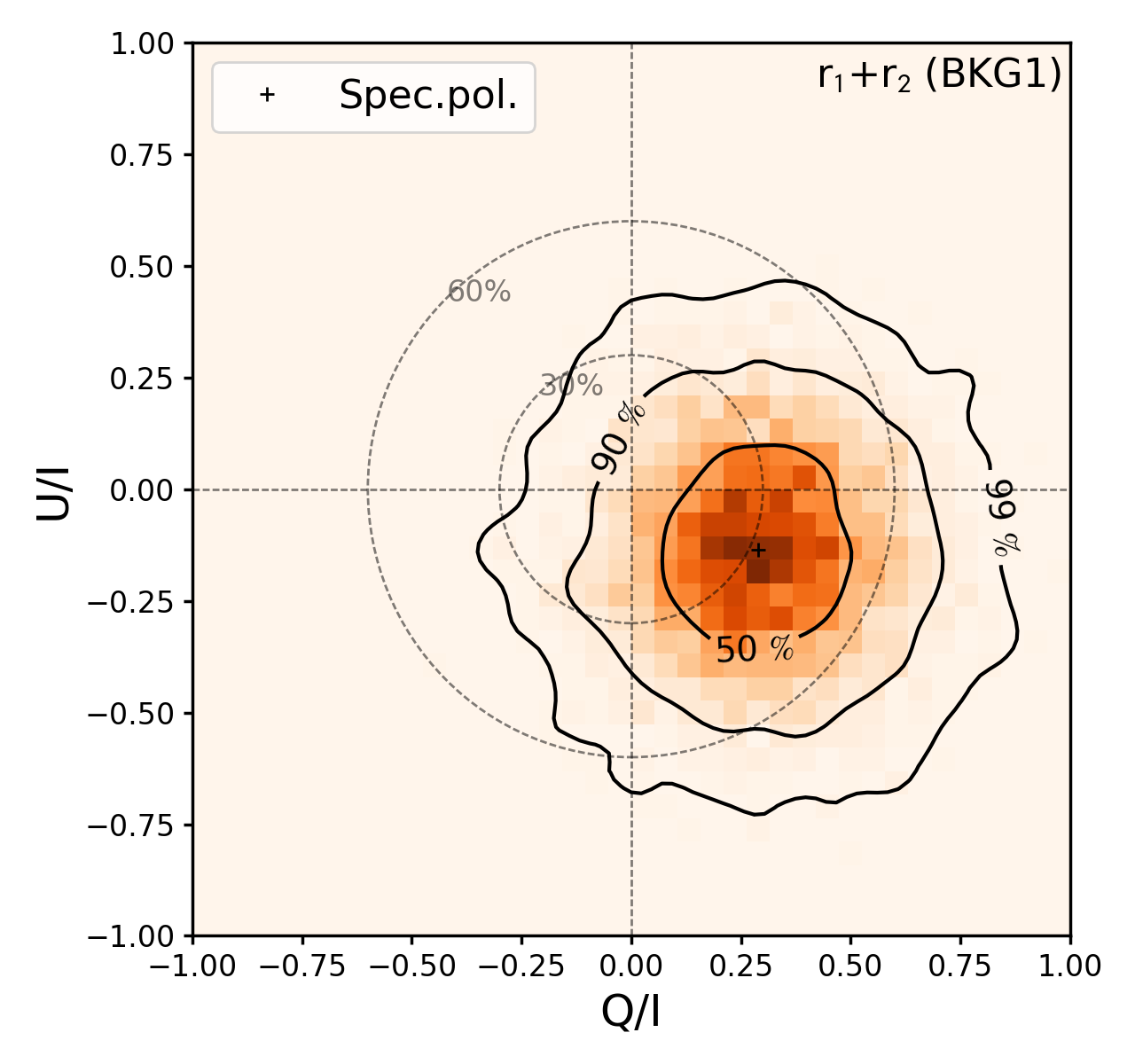}
    \includegraphics[height=0.3\textwidth]{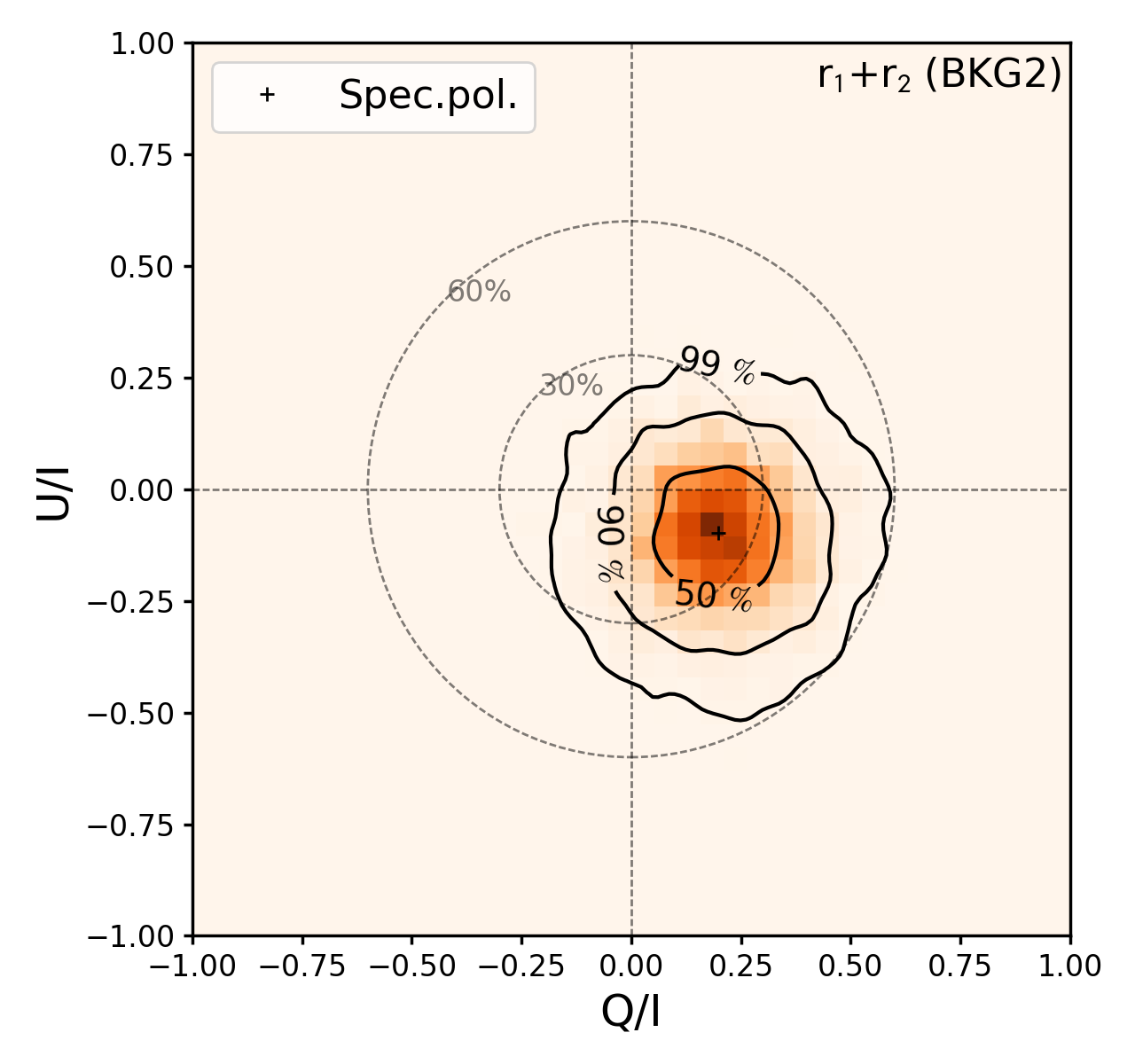}
    \includegraphics[height=0.3\textwidth]{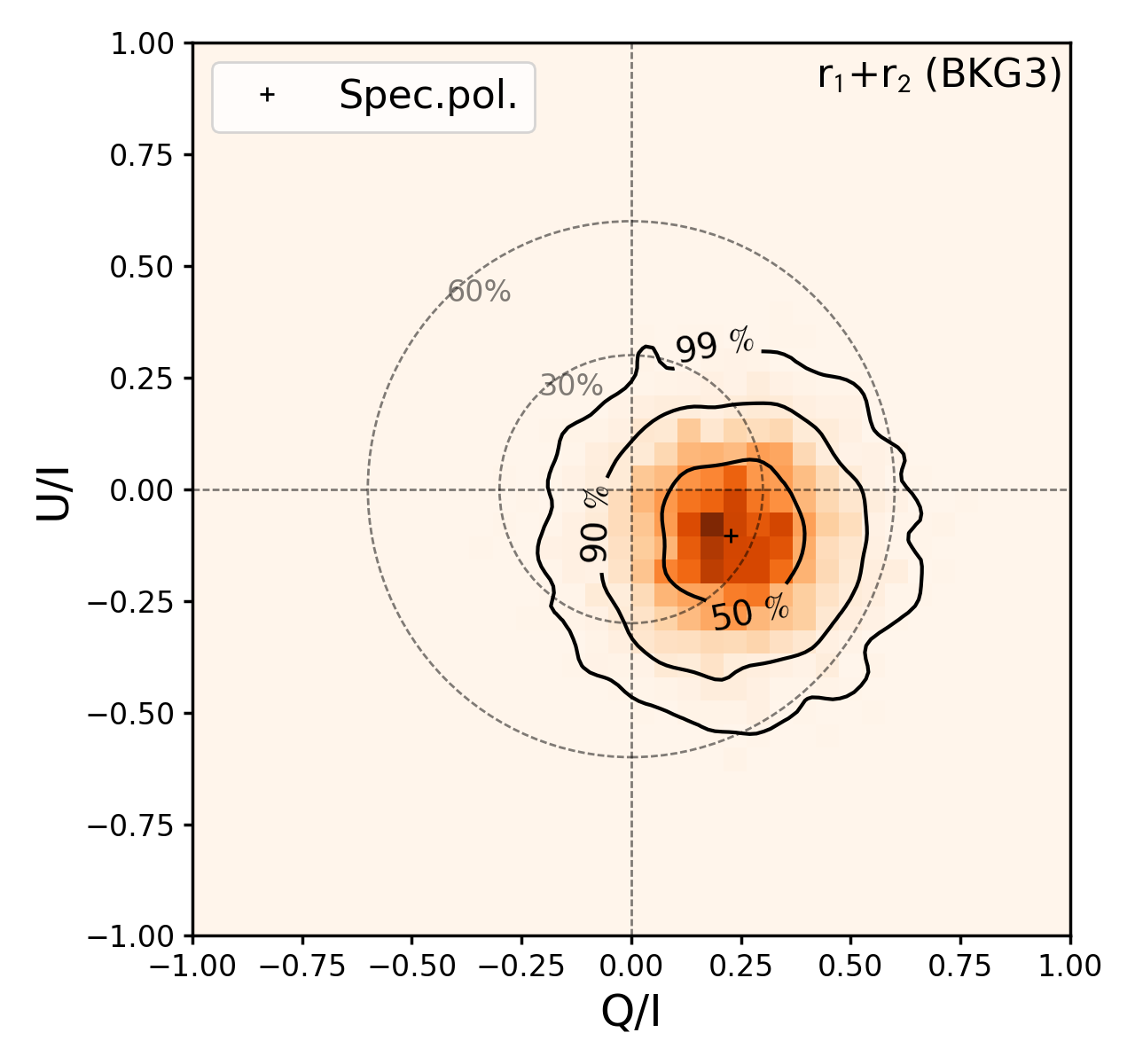}\\
    \includegraphics[height=0.3\textwidth]{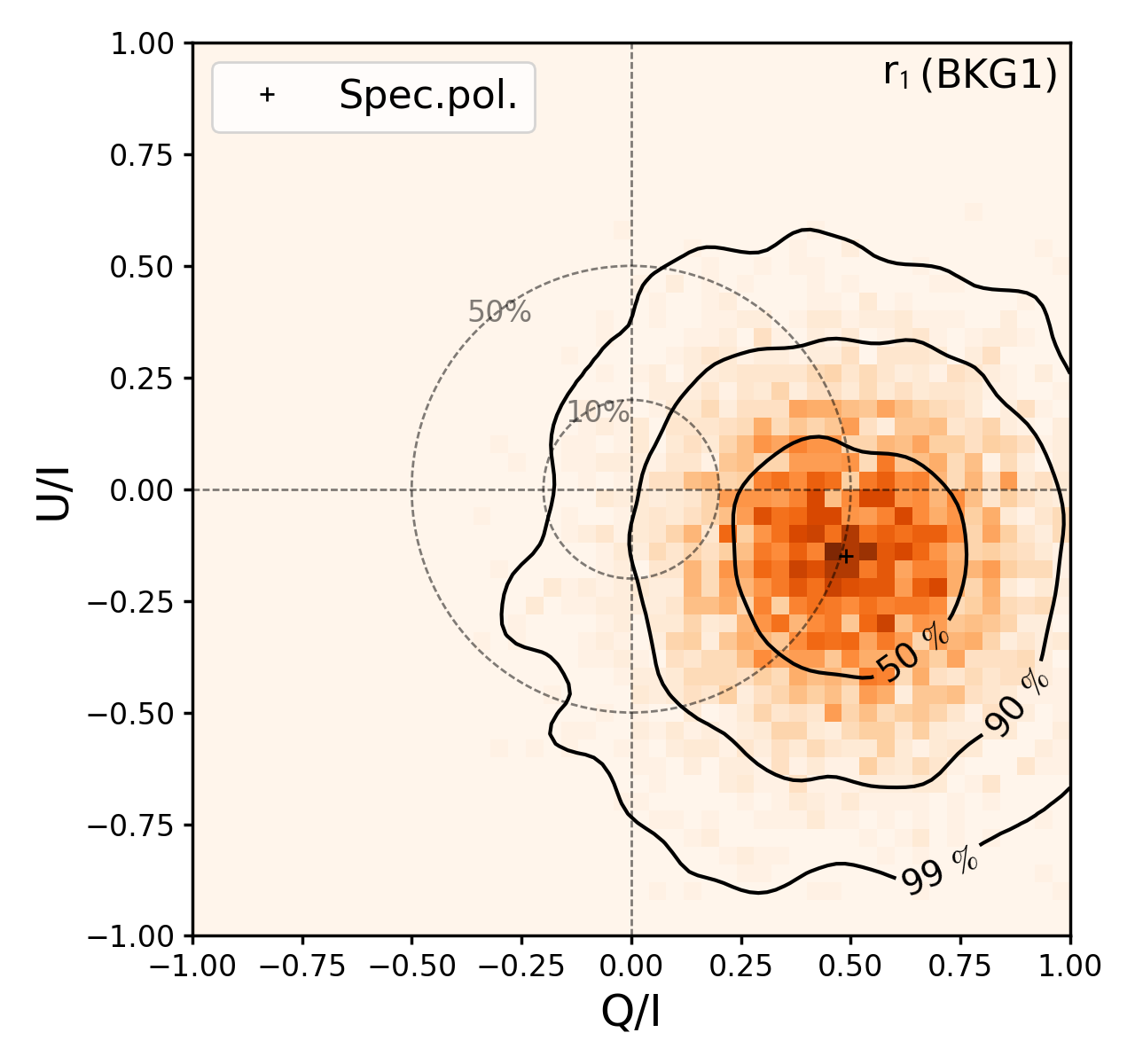}
    \includegraphics[height=0.3\textwidth]{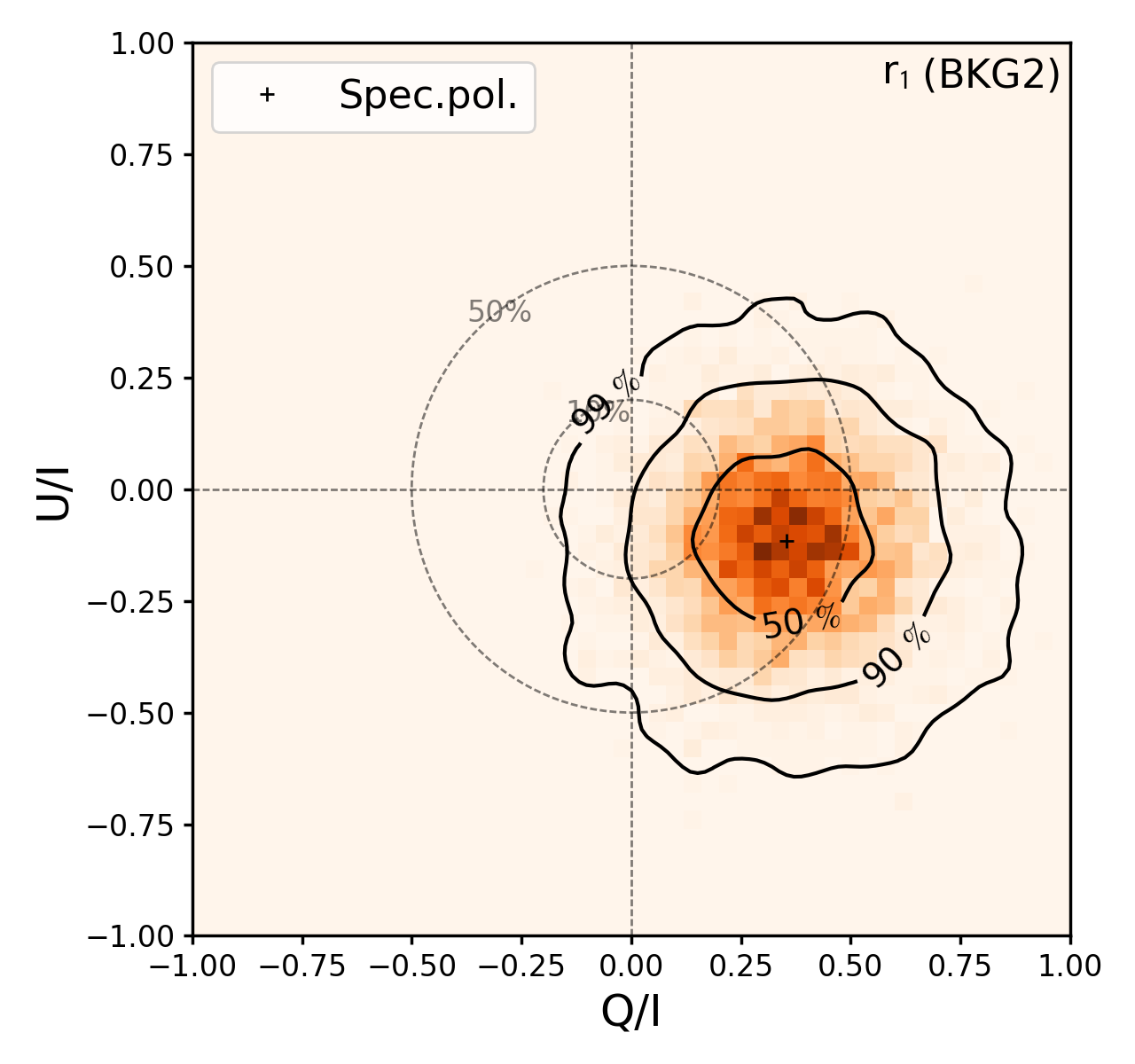}
    \includegraphics[height=0.3\textwidth]{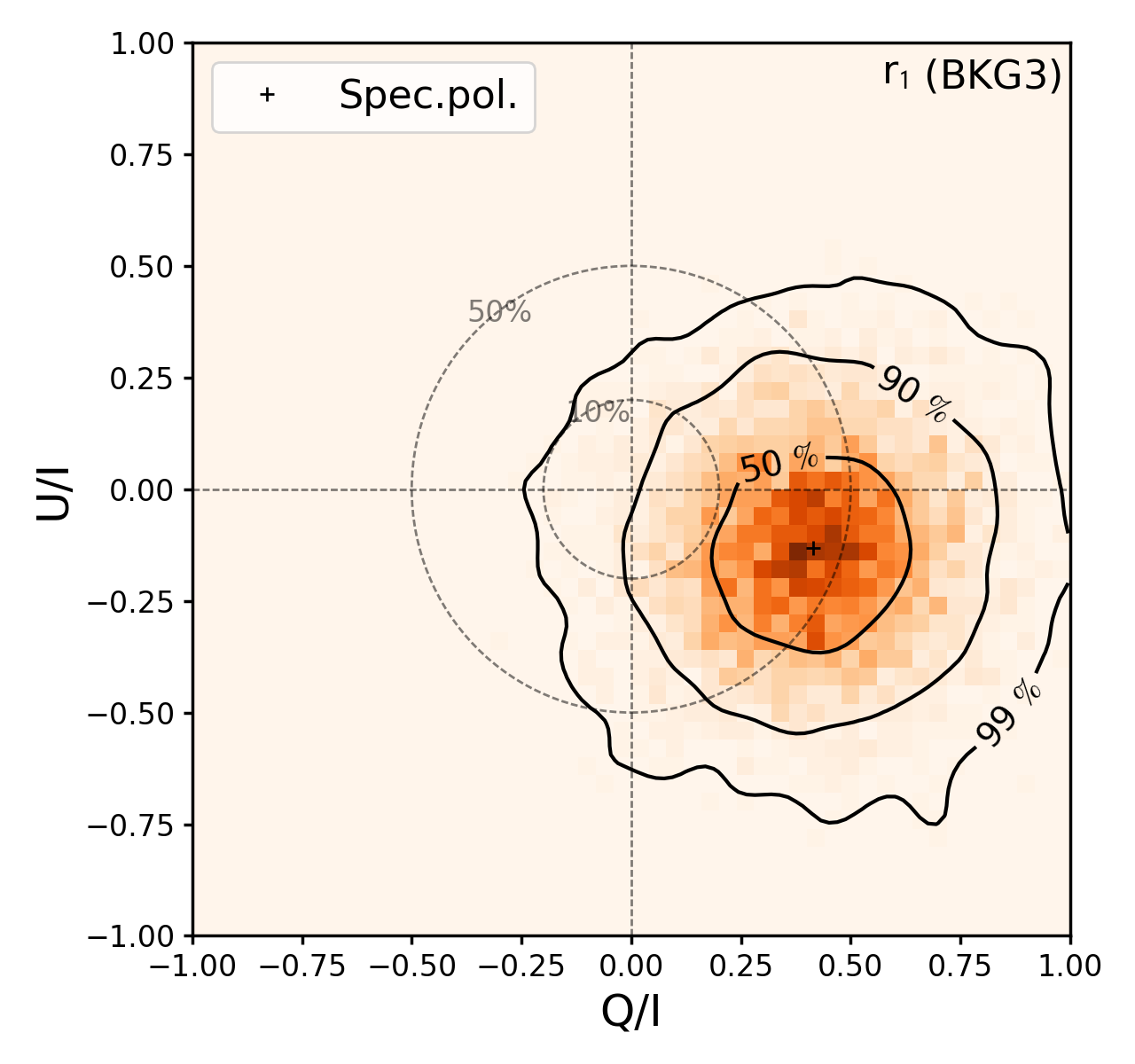}\\
    \includegraphics[height=0.3\textwidth]{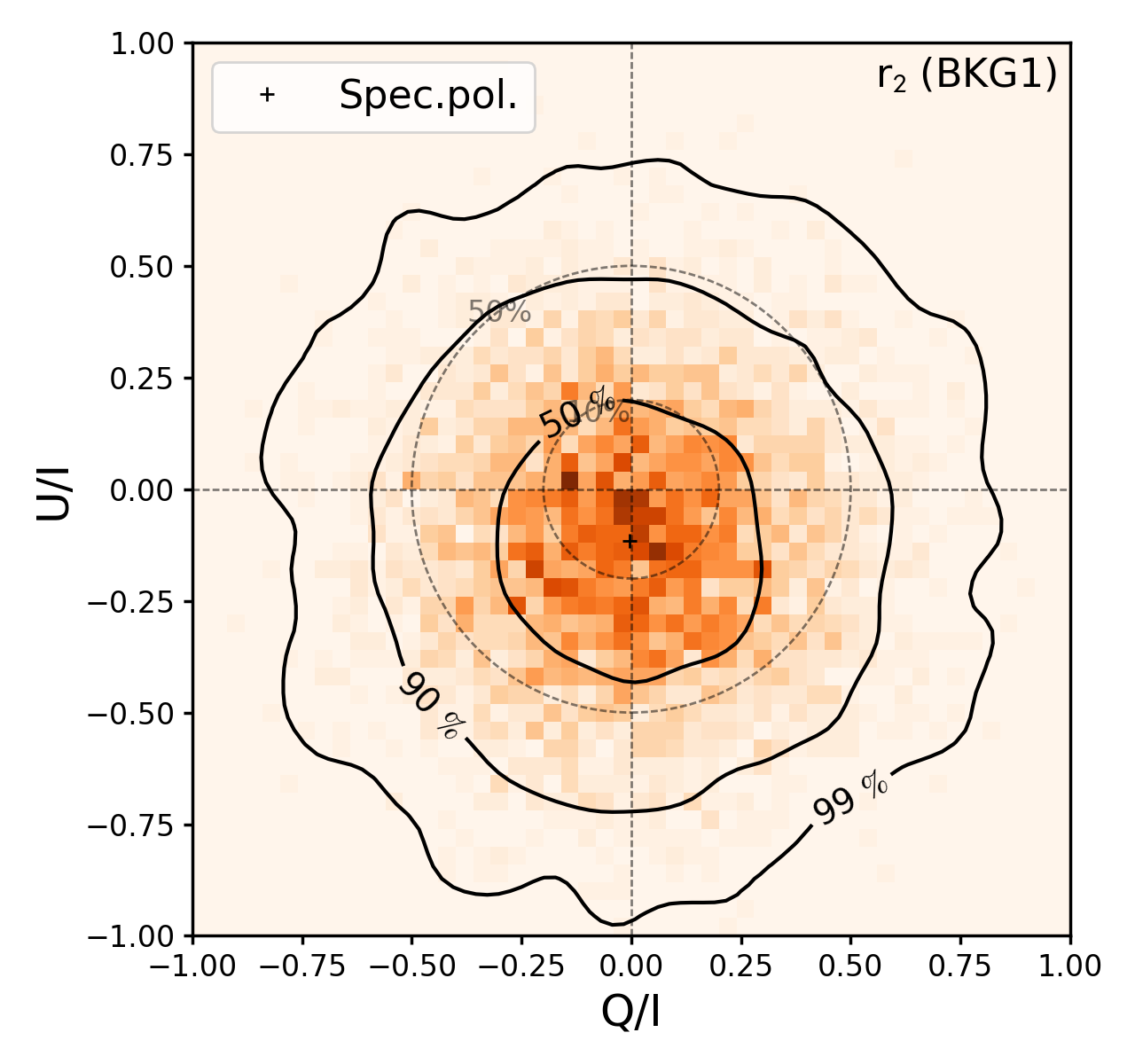}
    \includegraphics[height=0.3\textwidth]{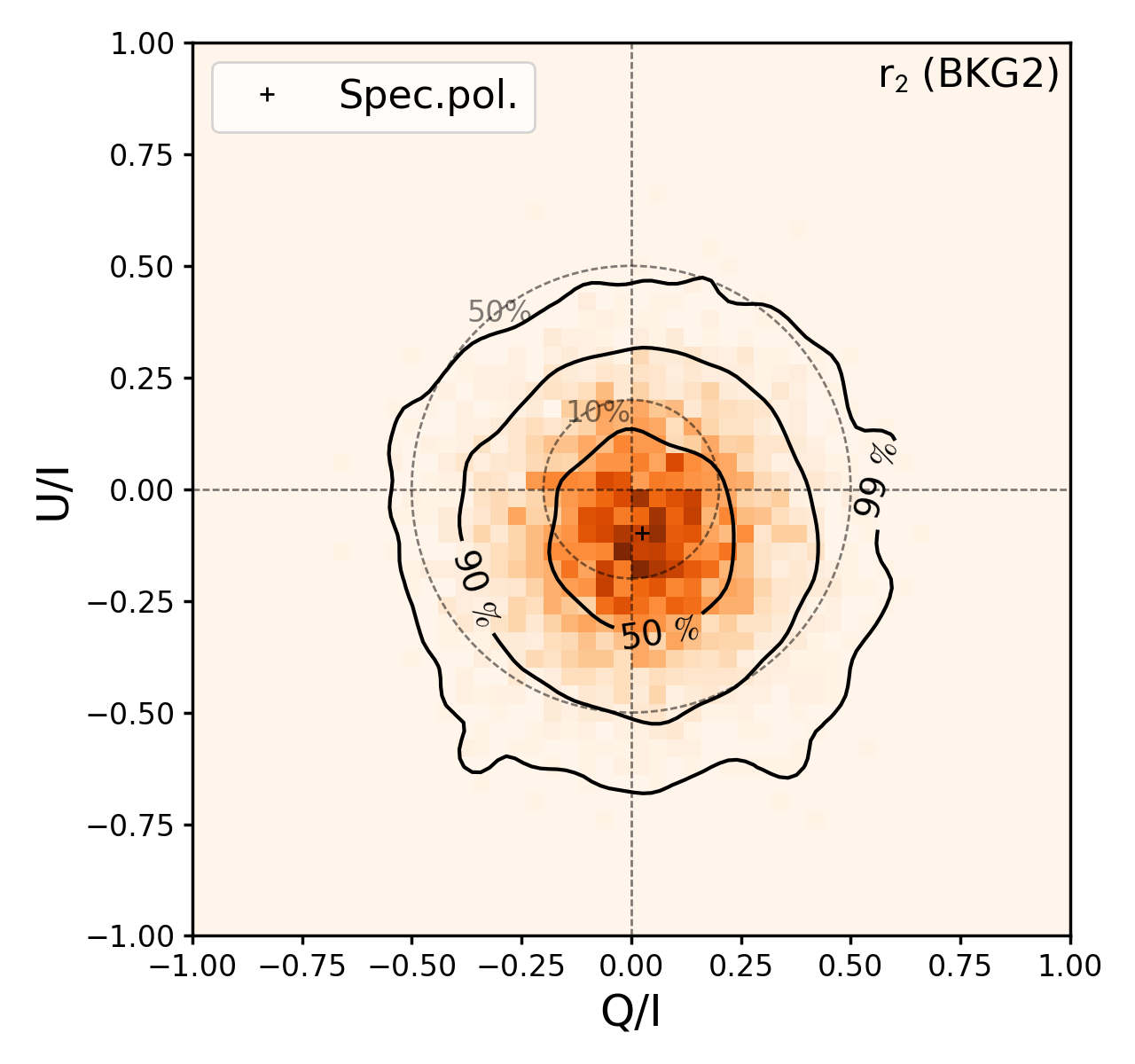}
    \includegraphics[height=0.3\textwidth]{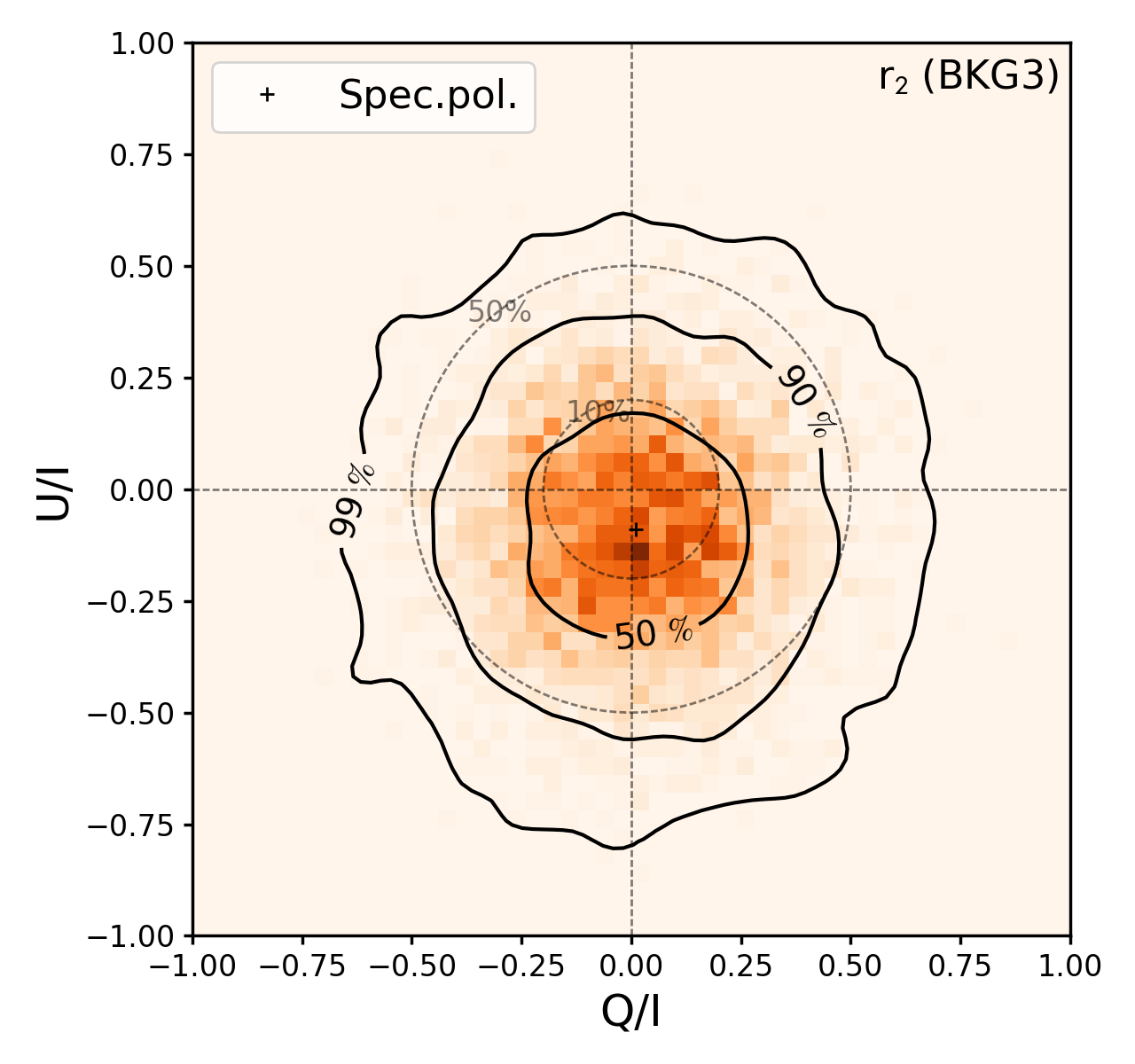}\\

    \caption{Q/I versus U/I plots resulting from the spectropolarimetric analysis for the combined analysis of \rone{} + \rtwo{} (top row) and for \rone{} and \rtwo{} individual analyses (middle and bottom row, respectively), shown for completeness. The three columns refer to different background assumptions: BKG1 (extracted from IXPE observations of 1ES~1959+65) on the left column, BKG2 (from BL Lac observation) on the central column, and BKG3 (from 3C~279 observation) on the right column. This shows that the result of the combined analysis is mostly driven by \rone{}, while \rtwo{} seems more unpolarized. However, within the 50\% contours \rone{} and \rtwo{} are compatible.}
    \label{fig:polarplots}
\end{figure}

\begin{figure}[ht]
    \centering
    \includegraphics[width=11.5cm]{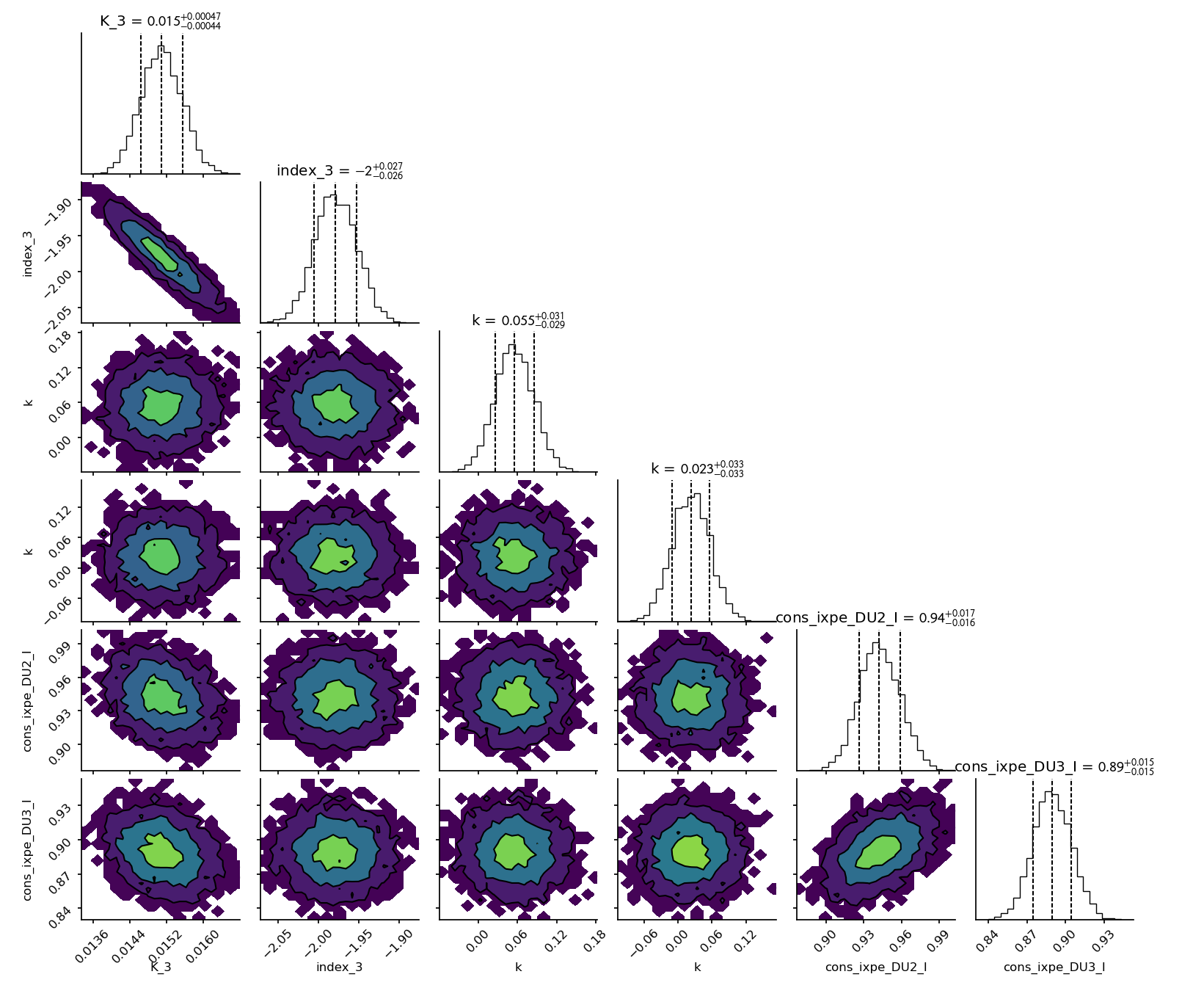}\\
    \includegraphics[width=11.5cm]{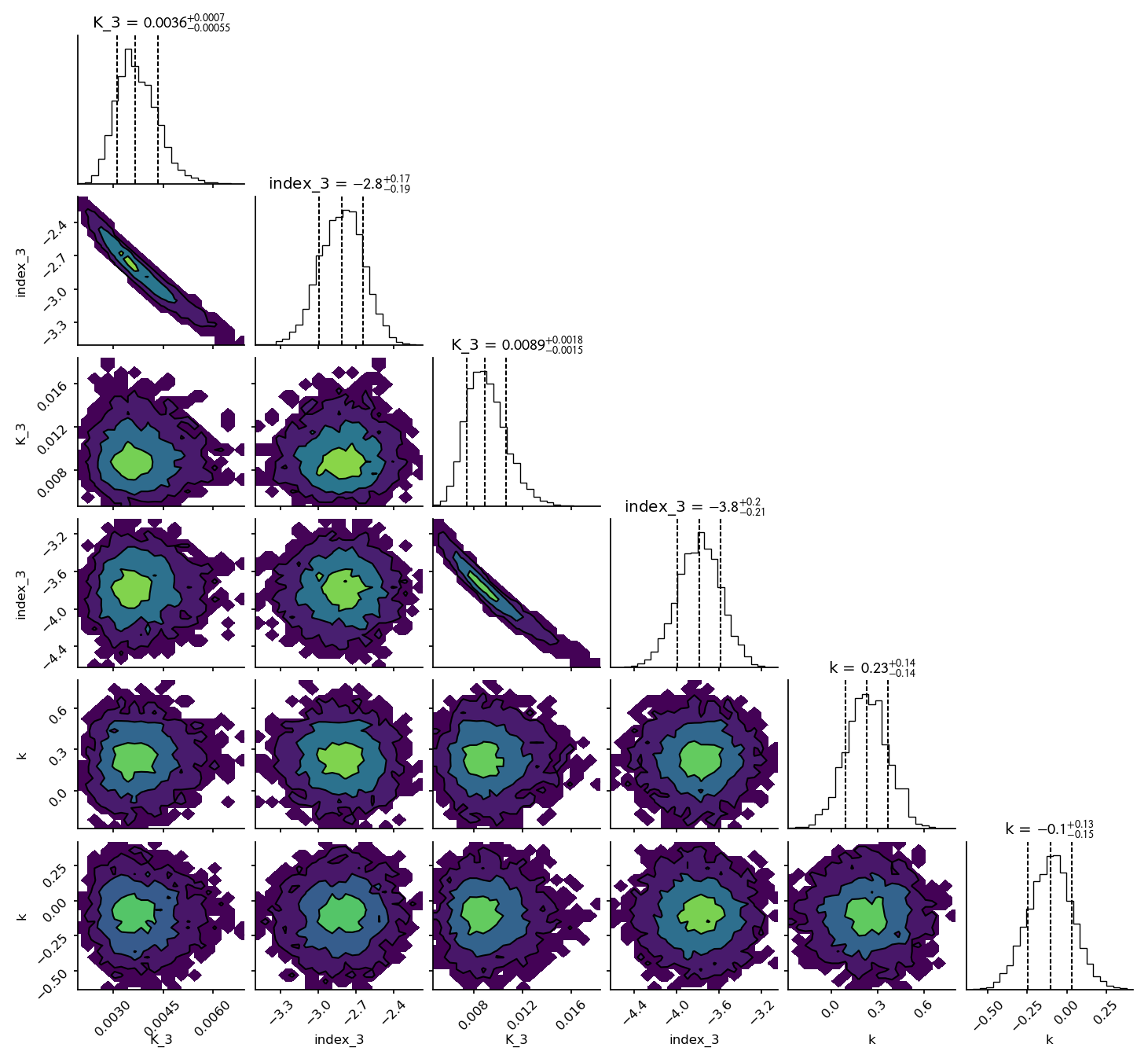}
    \caption{Corner plots resulting from the Bayesian spectropolarimetric fit performed with \threeml{} of the \core{} region (top) and the combined fit of \rone{} and \rtwo{} (bottom). This shows how the Bayesian approach yields results consistent with the frequentist approach adopted in this work. It also shows the nice convergence of the fit for all the parameters of interest. In both plots, we show the case of BKG2 assumed as background model for the background subtraction (the cases of BKG1 and BKG3 yield analogous results).}
    \label{fig:corner}
\end{figure}

\end{document}